\documentclass[twocolumn]{aastex701}
\usepackage[T1]{fontenc}

\newcommand{\ctht}{C$_2$H$_2$}

\begin{document}

\title{The JDISC Survey: Inner Disk Chemistry of Class I/FS Disks \\and Tentative Evidence for Early Pebble Drift}

\author[0000-0002-0661-7517]{Ke Zhang}
\affiliation{Department of Astronomy, University of Wisconsin-Madison, 
475 N Charter St, Madison, WI 53706, USA}
\email[show]{ke.zhang@wisc.edu}

\author[0000-0003-4335-0900]{Andrea Banzatti}
\affiliation{Department of Physics, Texas State University, 749 N Comanche Street, San Marcos, TX 78666, USA}
\email{}

\author[0000-0003-3682-6632]{Colette Salyk}
\affiliation{Vassar College, 124 Raymond Avenue, Poughkeepsie, NY 12604, USA}
\email{}

\author[0000-0002-1566-389X]{Abygail Waggoner}
\affiliation{Department of Astronomy, University of Wisconsin-Madison,
475 N Charter St, Madison, WI 53706, USA}
\email{}

\author[0000-0001-7552-1562]{Klaus Pontoppidan}
\affiliation{Jet Propulsion Laboratory, California Institute of Technology, 4800 Oak Grove Drive, Pasadena, CA 91109, USA}
\email{}

\author[0000-0002-5296-6232]{Mar\'{i}a Jos\'{e} Colmenares}
\affiliation{Department of Astronomy, University of Michigan, Ann Arbor, MI 48109, USA}
\email{}

\author[0000-0001-7962-1683]{Ilaria Pascucci}
\affiliation{Department of Planetary Sciences, University of Arizona, 1629 East University Boulevard, Tucson, AZ 85721, USA}
\email{}

\author[0000-0002-2828-1153]{Lucas A. Cieza}
\affiliation{Instituto de Estudios Astrof\'isicos, Universidad Diego Portales, Av. Ejercito 441, Santiago, Chile}
\affiliation{Millennium Nucleus on Young Exoplanets and their Moons (YEMS),
Ej\'ercito 441, Santiago, Chile}
\email{}

\author[0000-0002-4147-3846]{Miguel Vioque}
\affiliation{European Southern Observatory, Karl-Schwarzschild-Str. 2, 85748 Garching bei München, Germany}
\email{}

\author[0000-0001-8764-1780]{Paola Pinilla}
\affiliation{Mullard Space Science Laboratory, University College London, Holmbury St Mary, Dorking, Surrey RH5 6NT, UK}
\email{}

\author[0000-0003-0787-1610]{Geoffrey A. Blake}
\affiliation{Division of Geological and Planetary Sciences, California Institute of Technology, MC 150-21, Pasadena, CA 91125, USA}
\email{}

\author[0000-0002-5758-150X]{Joan Najita}
\affiliation{NSF’s NOIRLab, 950 N. Cherry Avenue, Tucson, AZ 85719, USA}
\email{}

\author[0009-0008-8176-1974]{Joe Williams}
\affiliation{Department of Physics and Astronomy, University of Exeter, Exeter, EX4 4QL, UK}
\email{}

\author[0000-0002-3291-6887]{Sebastiaan Krijt}
\affiliation{Department of Physics and Astronomy, University of Exeter, Exeter, EX4 4QL, UK}
\email{}

\author[0000-0001-8240-978X]{Till Kaeufer}
\affiliation{Department of Physics and Astronomy, University of Exeter, Exeter, EX4 4QL, UK}
\email{}

\author[0000-0001-6947-6072]{Jane Huang}
\affiliation{Department of Astronomy, Columbia University, 538 W. 120th Street, Pupin Hall, New York, NY 10027, USA}
\email{}

\author[0000-0002-7607-719X]{Feng Long}
\affiliation{Kavli Institute for Astronomy and Astrophysics, Peking University, Beijing 100871, CHINA}
\email{}

\author[0000-0001-8184-5547]{Chengyan Xie}
\affiliation{Lunar and Planetary Laboratory, University of Arizona, Tucson, AZ 85721, USA}
\email{}

\author[0000-0001-6218-2004]{Minjae Kim}
\affiliation{Mullard Space Science Laboratory, University College London, Holmbury St Mary, Dorking, Surrey RH5 6NT, UK}
\affiliation{Space Science Exploration Mission Directorate, Korea AeroSpace Administration, Sacheon-si, Gyeonsangnam-do, 52535, Republic of Korea}
\email{}

\author[0009-0002-2380-6683]{Eshan Raul}
\affiliation{Department of Astronomy, University of Wisconsin-Madison,
475 N Charter St, Madison, WI 53706, USA}
\email{}

\author[0000-0003-3573-8163]{Dary A. Ru\'iz-Rodr\'iguez}
\affiliation{National Radio Astronomy Observatory, 520 Edgemont Rd., Charlottesville, VA 22903, USA}
\email{}

\author[0000-0003-2631-5265]{Nicole Arulanantham}
\affiliation{Astrophysics \& Space Institute, Schmidt Sciences, New York, NY 10011, USA}
\email{}

\author[0000-0002-1103-3225]{Beno\^it Tabone }
\affiliation{Université Paris-Saclay, CNRS, Institut d'Astrophysique Spatiale, 91405 Orsay, France}
\email{}

\author[0000-0002-0554-1151]{Mayank Narang}
\affiliation{Jet Propulsion Laboratory, California Institute of Technology, 4800 Oak Grove Drive, Pasadena, CA 91109, USA}
\email{}

\author[0000-0001-8284-4343]{Karina Mauco}
\affiliation{European Southern Observatory, Karl-Schwarzschild-Strasse 2, 85748 Garching bei München, Germany}
\email{}

\begin{abstract}

We present the first chemical survey of Class I and Flat-Spectrum (I/FS) disks using JWST MIRI/MRS, targeting sixteen sources in the Ophiuchus star-forming region. Through empirical line luminosity measurements and multi-component slab modeling, we characterize the molecular reservoir of these young systems and compare them to twelve Class II disks of similar stellar mass. Water, HCN, \ctht, and CO$_2$ are frequently detected in I/FS sources with inclinations $i < 70^{\circ}$, whereas edge-on systems show significantly suppressed emission. Compared to Class II disks, I/FS sources show suggestive---though not yet statistically significant---evidence for elevated cold water ($\sim$200\,K) mass and lower CO$_2$ excitation temperatures. Statistical analyses identify accretion luminosity as the primary correlate of molecular mass across both evolutionary stages. Once this dependence is removed, cold water and CO$_2$ masses anti-correlate with mm-dust disk radius, while hot water remains insensitive to disk size.  These patterns are qualitatively consistent with pebble drift models that predict early water enrichment followed by delayed CO$_2$ delivery, suggesting an evolutionary progression from molecular-poor Class 0 sources, through water-rich Class I/FS disks, to Class II disks with reduced cold water excess. This work provides an initial evolutionary framework for disk chemistry that requires larger, multi-region samples to confirm.

\end{abstract}


\section{Introduction}
Understanding the chemical evolution of protoplanetary disks is critical for tracing the chemical origins of planetary building blocks \citep[e.g.,][]{Krijt2023,Oeberg2023,Zhang2024}. The chemical composition of the inner few au region of protoplanetary disks can be best characterized by emission of warm gas phase molecules and atoms at near-and mid-InfraRed (IR) wavelengths. H$_2$O, CO, HCN, \ctht, and OH have been commonly detected in near-IR ground observations and mid-IR spectra of \textit{Spitzer}/IRS \citep[e.g.,][]{Salyk2008,Carr2008,Pontoppidan2010,Salyk2011,Carr2011,Najita2013,Pascucci2013}. Over the past few years, the JWST Mid InfraRed Instrument/Medium-resolution spectrograph (MIRI/MRS), with its unprecedented sensitivity and spectral coverage from 5--28\,$\mu$m, have provided many new important insights into the inner disk chemistry of Class II disks, including the detections of complex organics and constraints on the radial distribution of water \citep[e.g.,][]{Grant2023,Tabone2023,Henning2024,Banzatti2023,Banzatti2025,Arulanantham2025,Arabhavi2026}. 

However, it remains unclear how the volatile compositions observed in Class II systems connect to the earlier, embedded stages of disk evolution when planets/planetesimals are likely beginning to form. Most existing JWST/MIRI studies have focused on Class II disks ($\geq$0.5\,Myr), which represent the later stages of disk evolution when envelope has dissipated \citep{Grant2023,Henning2024,Banzatti2023,Arulanantham2025,Temmink2025}. 
A growing body of evidence suggests that planet formation may begin much earlier than previous estimations. Many key processes ---such as pebble growth, planetesimal and planetary core formation--- may already be well under the way at the embedded disk stage ($<$1\,Myr). Evidence includes millimeter-sized pebbles in many embedded disks and radial dust substructures detected in some embedded disks indicating possible concentration of pebbles and/or planetary formation \citep{Tobin2012,Najita2014,ALMAPartnership2015,Harsono2018,Tobin2020,Tychoniec2020,SeguraCox2020,Cieza2021,Ohashi2023,Hsieh2024,Hsieh2025}. Furthermore, isotopic signatures in meteorites point to the rapid formation of planetesimals or Jupiter’s core, likely occurring within the first million years (Myr) of the solar system. \citep{Kruijer2017,Lichtenberg2021}.

A few MIRI studies have targeted deeply embedded Class 0/I sources around low-mass protostars and revealed gas-phase water and organic emission in a few cases \citep{Yang2022,Salyk2024,vanGelder2024,vanDishoeck2025}.  However, most of the existing MIRI/MRS spectra of embedded disks are Class 0 and only included six very young Class I sources (T$_{\rm bol} < 200$\,K). It is unclear whether rotationally supported disks have formed in most of these systems. Furthermore, the majority of detected molecular emissions in the existing embedded samples appear to be relatively low temperature ($<$250\,K), which originates from the cold outer disk or inner envelope rather than from the innermost disk region. Indeed, in some Class 0 sources, mid-infrared molecular line emission clearly show a wind/outflow component in MIRI images \citep{vanGelder2024,vanDishoeck2025}.
These deeply embedded, envelope-dominated sources represent only the earliest phase of the embedded stage. The bulk population of embedded disks are Class I (0.1-0.6\,Myr) and Flat spectrum (FS, 0.6-1\,Myr) sources \citep{Evans2009}, which represents a crucial yet underexplored phase in the chemical evolution of disks.

Characterizing the inner disk chemistry of Class I/FS systems offers a unique opportunity to test whether early pebble drift leaves detectable imprints on the volatile inventory of the planet-forming region. In the embedded phase, dust traps may not yet have formed or may be less developed \citep{SeguraCox2020,Ohashi2023,Hsieh2025}, allowing icy pebbles to drift inward largely unimpeded. Models predict that this efficient radial transport can deliver substantial amounts of volatile ices, such as H$_2$O, to the inner disk, potentially enriching its molecular reservoir beyond what is seen in Class II systems where dust traps are more prevalent \citep[e.g.,][]{Kalyaan2023,Easterwood2024,Sellek2025,Luo2026}. Additionally, the physical environment of embedded disks may further shape their chemistry: higher accretion rates and reprocessed radiation from the surrounding envelope---which absorbs stellar and accretion luminosity and re-emits it as thermal radiation back onto the disk---can produce warmer mid-plane temperatures \citep[e.g.,][]{Harsono2015,Woitke2018}, while less dust growth and settling may result in more flared geometries that enhance the exposure of disk surfaces to radiation and subsequently warmer disk.  

The goal of this paper is to provide the first detailed characterization of the chemical composition of Class I/FS disks, based on 16 disks observed in the JWST GO 3034 and 7135 programs (\citealt{Zhang2023,Zhang2025}).  We find that molecular line emissions are commonly detected in these Class I/FS disks. By comparing with 12 Class II disks within a similar stellar mass range, we explore possible chemical evolution from the embedded stage to Class II disks. Given the modest sample sizes, this study is exploratory in nature; our statistical analysis identifies the sample sizes needed for more definitive conclusions. Ongoing JWST programs will substantially increase the statistical power by providing larger, uniformly processed samples of Class I/FS disks in Ophiuchus and other star-forming regions.

The paper is organized as follows.
In \S2, we describe the sample selection, JWST/MIRI–MRS observations, and data calibration procedures.
\S3 outlines the extinction correction, continuum subtraction, and slab modeling methodology.
In \S4, we present a comparison of both empirical characteristics (line fluxes, flux ratios) and slab modeling results between the Class I/FS and Class II samples, along with statistical tests for correlations with stellar and disk properties.
\S5 discusses the possible chemical evolution from embedded disks to Class II disks and places the observed trends in the context of current pebble drift frameworks.
Finally, \S6 summarizes the main results of this study.

\section{Observations} \label{sec:obs}

\subsection{Class I/FS Sample} \label{subsec:sample}

Our sample consists of 16 Class I/FS disks in the Ophiuchus star-forming region, from two JWST GO programs (3034 and 7135, PI. Zhang). The evolutionary classifications are based on the infrared spectral index $\alpha_{\rm IR}$ measured between 2 and 20\,$\mu$m, following the scheme adopted by \citet{Cieza2019}. In this scheme, sources with $\alpha_{\rm IR} > 0.3$ are classified as Class~I, indicating young systems still deeply embedded within infalling envelopes. Those with $-0.3 < \alpha_{\rm IR} < 0.3$ are designated Flat Spectrum (FS), representing a transitional stage in which the envelope contribution to the SED is diminishing but not yet negligible. Sources with $-1.6 < \alpha_{\rm IR} < -0.3$ are classified as Class~II, where the infrared excess is dominated by the circumstellar disk.

We caution that $\alpha_{\rm IR}$-based classifications are inclination-dependent: an edge-on disk appears redder and cooler, and hence more embedded, than the same disk seen closer to face-on \citep{Robitaille2006,Robitaille2007}. The $L_{\rm submm}/L_{\rm bol}$ ratio is less inclination-sensitive and a more robust stage indicator \citep{Andre1993}, but we adopt $\alpha_{\rm IR}$ and $T_{\rm bol}$ because these are available uniformly across our sample while $L_{\rm submm}/L_{\rm bol}$ is not. To mitigate inclination effects, we exclude the highly inclined ($i>70^{\circ}$) disks from the chemical comparison and use $T_{\rm bol}$ only as a \emph{relative} age indicator.

Table~\ref{table:stars} shows the stellar and disk properties of the 16 Ophiuchus Class I/FS systems in this study. 

Ten sources in the 3034 program were originally selected as the Ophiuchus embedded disk sub-sample of the AGE-PRO ALMA large program \citep{Zhang2025_AGEPRO}. These ten Class I/FS sources are all $<$1\,Myr old \citep{Evans2009} and span a narrow stellar spectral type range of M3--K6 (M$_\star \sim$0.3--0.8\,M$_\odot$), enabling controlled comparisons with Class II samples in the same mass range. The ten disks were selected to cover the full spread of millimeter continuum disk luminosities between known M3-K6 sources in the Ophiuchus region \citep{Cieza2019,RuizRodriguez2025}. 

An additional six sources (Oph~11--16) are from the 7135 program, which is designed to survey all $\sim$40 Class~I/FS disks in the Ophiuchus region. These were among the first observed in Cycle~4. Five of the six (Oph~11--15) have literature spectral types of M4--M5.5 ($M_\star \sim 0.07$--$0.18$\,M$_\odot$; \citealt{McClure2010}, Ruiz-Rodriguez et al.\ in prep), making them generally lower-mass than the AGE-PRO subsample. While these sources have fewer constraints on stellar and disk gas properties compared to the 10 AGE-PRO sources, they all have high-resolution millimeter-wavelength ALMA observations that provide dust disk information (\citealt{Cieza2019,Cieza2021,Bhowmik_2026_ODISEA}).

\begin{deluxetable*}{llccccclcccc}
\tabletypesize{\scriptsize}
\tablecaption{Stellar and disk properties of the Ophiuchus Class I/FS disk sample \label{table:stars}}
\tablewidth{0pt}
\tablehead{
\colhead{Index} & \colhead{Source Name} & \colhead{Common Name} &
\colhead{SpT} & \colhead{Class} & \colhead{T$_{\rm bol}$} & \colhead{Av} & \colhead{M$_\ast$} & \colhead{M$_{\rm dust}$} &  \colhead{logM$_{\rm gas}$}& \colhead{incl} & \colhead{$R_{d,90\%}$}
\\
\colhead{} & \colhead{} & \colhead{} &
&  & \colhead{(K)}&  &
\colhead{($M_\odot$)}& \colhead{($M_\oplus$)} & \colhead{($M_\odot$)}  &
\colhead{(deg)} & \colhead{(au)}
}
\decimalcolnumbers
\startdata
Oph 1 & SSTc2dJ162623.6-242439 & GY 21 & K7 & FS & 670 & 16.1 & 0.61 & 61 & -0.63 & 73 & 95 \\
Oph 2 & SSTc2dJ162703.6-242005 & ISO-Oph 94 & M1.5 & FS & 750 & 10.1 & 0.36 & 17 & -2.12 & 76 & 69 \\
Oph 3 & SSTc2dJ162719.2-242844 & WL 3 & K7 & FS & 660 & 31.5 & 0.61 & 12 & -2.43 & 61 & 17 \\
Oph 4 & SSTc2dJ162728.4-242721 & Elias 2-32 & K6.5 & FS & 310 & 24.5 & 0.63 & 1 & -3.39 & 61 & 18 \\
Oph 5 & SSTc2dJ162737.2-244237 & GY 301 & K9 & I & 450 & 40.0 & 0.55 & 3 & -3.46 & 25 & 6 \\
Oph 6 & SSTc2dJ162738.9-244020 & GY 312 & M2.5 & I & 290 & 12.1 & 0.30 & 20 & -0.96 & 74 & 40 \\
Oph 7 & SSTc2dJ163135.6-240129 & IRAS 16285-2355 & K6 & I & 300 & 23.3 & 0.65 & 180 & -0.39 & 47 & 73 \\
Oph 8 & SSTc2d J162305.4-230257 & \nodata & M1 & I & 650 & 1.3 & 0.40 & 2 & -3.21 & 72 & 34 \\
Oph 9 & SSTc2dJ162627.5-244153 & GY 33 & M2 & FS & 750 & 5.0 & 0.33 & 5 & -1.99 & 73 & 32 \\
Oph 10 & SSTc2dJ162718.4-243915 & BBRCG 39 & M0 & I & 590 & 44.0 & 0.48 & 16 & -2.24 & 74 & 54 \\
Oph 11 & SSTc2dJ162652.0-243039 & BBRCG 13 & M5$^{a}$ & FS & 680 & 34.7 & 0.07$^{a}$ & 6 & \nodata & 37 & 13 \\
Oph 12 & SSTc2dJ162654.3-242438 & GY 152 & M5.5$^{a}$ & FS & 720 & 48.5 & 0.07$^{a}$ & 2 & \nodata & \nodata & \nodata \\
Oph 13 & SSTc2dJ162709.3-244022 & GY 213 & M4$^{a}$ & FS & 720 & 23.5 & 0.18$^{a}$ & 6 & \nodata & 24 & 12 \\
Oph 14 & SSTc2dJ162741.7-244336 & GY 323 & M5$^{a}$ & FS & 720 & 28.3 & 0.07$^{a}$ & 2 & \nodata & \nodata & \nodata \\
Oph 15 & SSTc2dJ163143.8-245525 & ISO-Oph 200 & M4$^{b}$ & I & 500 & 22.8 & 0.18$^{b}$ & 6 & \nodata & 31 & 7 \\
Oph 16 & SSTc2dJ162724.6-244103 & CRBR 2422.8-3423 & \nodata & I & 300 & 55 & \nodata & 3 & \nodata & 71 & 8 \\
\enddata
\tablecomments{Col.~(1) index used in this work, (2) Source name, (3) Common name,
(4) Spectral type, (5) Disk class based on SED, (6) Bolometric temperature, (7) V band extinction magnitude, (8) Stellar mass, (9) Dust disk mass, (10) Gas disk mass, (11) Disk inclination angle, (12) Dust disk radius, measured as the 90\% radius of 1.3\,mm continuum emission for Oph 1-10 and 0.7\,mm continuum emission for Oph 11-16.}
\tablerefs{T$_{\rm bol}$ from \citet{Evans2009};  Class information from \citet{Cieza2019}; For Oph 1-10, the Spectral type, A$_V$, stellar mass, dust disk mass are from \citet{RuizRodriguez2025};  inclination angle and dust disk radius are from \citet{Vioque2025}; \citet{Trapman2025_AGEPRO_mass} for gas disk mass. For Oph 11-16, the Spectral type, A$_V$, stellar mass $^{a}$\citet{McClure2010} and $^{b}$Ruiz-Rodriguez et al.\ in prep. Dust mass, inclination angle, and dust disk radius are from \citet{Bhowmik_2026_ODISEA}}
\end{deluxetable*}

\begin{deluxetable*}{lcccccccccccc}
\tabletypesize{\scriptsize}
\tablecaption{Stellar and disk properties of the Class II disk sample \label{table:classII}}
\tablewidth{0pt}
\tablehead{
\colhead{Source} &
\colhead{Dist} & \colhead{SpT} & \colhead{$T_{\rm eff}$} & \colhead{$L_\star$} & \colhead{$M_\star$} & \colhead{log$\dot{M}_{\mathrm{acc}}$} & \colhead{log$L_{\mathrm{acc}}$} & \colhead{$M_{d}$} & \colhead{$R_{d,90\%}$} & \colhead{incl} & \colhead{$n_{13-26}$} & \colhead{Age}
\\
\colhead{} & \colhead{(pc)}
& & \colhead{(K)} &
\colhead{($L_\odot$)} & \colhead{($M_\odot$)} & \colhead{($M_\odot\,\mathrm{yr}^{-1}$)} & \colhead{($L_\odot$)} & \colhead{($M_\oplus$)} & \colhead{(au)} & \colhead{(deg)} & & \colhead{(Myr)}
}
\decimalcolnumbers
\startdata
CI Tau & 160 & K5.5 & 4162 & 1.65 & 0.70 & $-7.5$ & $-0.19$ & 103.4 & 174 & 50 & $-0.18$ & $0.62^{+0.23}_{-0.12}$ \\
GQ Lup & 158 & K6 & 4115 & 1.60 & 0.67 & $-7.4$ & $-0.80$ & 25.6 & 22 & 61 & $-0.18$ & $0.60^{+0.21}_{-0.10}$ \\
IQ Tau & 141 & M1.1 & 3704 & 0.86 & 0.48 & $-7.9$ & $-0.97$ & 35.5 & 96 & 62 & $-0.40$ & $0.56^{+0.18}_{-0.06}$ \\
RU Lup & 159 & K7 & 4074 & 1.45 & 0.64 & $-7.1$ & $0.01$ & 155.0 & 55 & 19 & $0.07$ & $0.63^{+0.22}_{-0.13}$ \\
FZ Tau & 129 & M0.5 & 3810 & 1.19 & 0.48 & $-6.5$ & $-0.04$ & 5.3 & 15 & 13 & $-0.85$ & $0.50$ \\
Elias 27 & 116 & M0 & 3890 & 0.91 & 0.58 & $-7.2$ & $-0.65$ & 134.1 & 203 & 56 & $-0.42$ & $0.74^{+0.41}_{-0.24}$ \\
GO Tau & 142 & M2.3 & 3515 & 0.27 & 0.35 & $-9.5$ & $-1.70$ & 31.4 & 144 & 54 & $0.19$ & $1.58^{+1.30}_{-0.71}$ \\
Elias 20 & 138 & M0 & 3890 & 2.24 & 0.52 & $-6.9$ & $-0.90$ & 59.8 & 58 & 49 & $-0.70$ & $0.50$ \\
WSB 52 & 136 & M1 & 3715 & 0.71 & 0.47 & $-7.6$ & $-0.77$ & 37.4 & 28 & 54 & $-0.22$ & $0.67^{+0.32}_{-0.17}$ \\
GK Tau & 129 & K6.5 & 4067 & 1.48 & 0.64 & $-8.3$ & $-1.62$ & 2.4 & 12 & 40 & $-0.08$ & $0.62^{+0.21}_{-0.12}$ \\
SR 4 & 134 & K7 & 4074 & 1.17 & 0.65 & $-6.9$ & $-0.22$ & 37.4 & 29 & 22 & $0.77$ & $0.79^{+0.38}_{-0.29}$ \\
Sz 114 & 162 & M5 & 3162 & 0.20 & 0.17 & $-9.1$ & $<-2.21$ & 30.6 & 52 & 21 & $0.03$ & $1.04^{+1.64}_{-0.31}$ \\
\enddata
\tablecomments{Spectral types, $T_{\rm eff}$, $L_\star$, and mass accretion rates are adapted from compiled results of \citet{Andrews2018a}. The $\log L_{\rm acc}$ values are derived from H\,\textsc{i} (10--7) line luminosity (updated for JDISC v9.1 this work). $R_{d,90\%}$ are measured from radial profiles of public 1.3\,mm continuum DSHARP profiles and private communication. Inclination are from values compiled by \citet{Arulanantham2025}. Ages are estimated by this work. The original MIRI data were taken from the following GO 1 programs: PID 1584 (PI: C. Salyk; co-PI: K. Pontoppidan; \citealt{Arulanantham2025}), PID 1549 (PI: K. Pontoppidan; \citealt{Pontoppidan2024}) and PID 1640 (PI: A. Banzatti; \citealt{Banzatti2023,RomeroMirza2024}).}
\end{deluxetable*}

\subsection{Class II sample for comparison}\label{subsec:classII_sample}

To enable a comparison with the Class I/FS sample, we selected a sample of Class II disks from publicly available JDISCS JWST/MIRI GO 1 data \citep{Pontoppidan2024,Banzatti2025,Arulanantham2025}. The parent sample presented in \citet{Arulanantham2025} includes 31 disks, spanning a stellar spectral type between M5 and A1. For a similar stellar mass comparison, we limit our sample to sources with $M_\star \le$0.7\,M$_\odot$. The final comparison sample consists of 12 Class II disks. The original MIRI spectra of the Class II sample have been published in \citet{Banzatti2025,Arulanantham2025}.

\begin{figure*}[!htbp]
\centering
\vspace{-0.cm}
\includegraphics[width=0.9\textwidth]{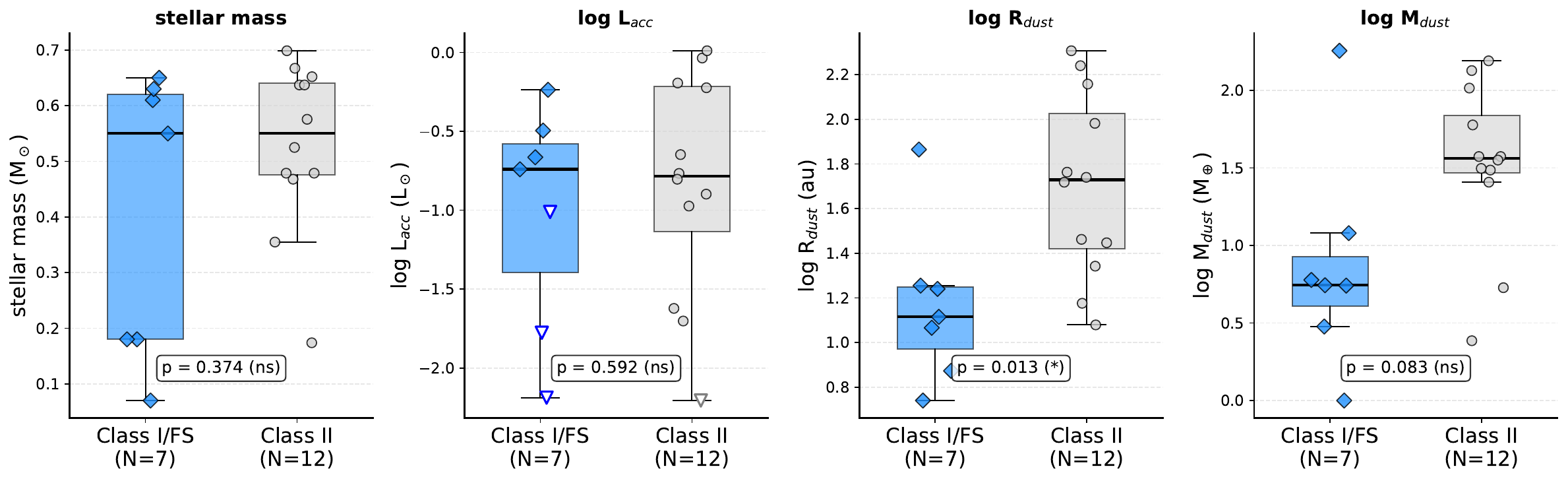}
\vspace{-0.cm}
\caption{ Comparison of stellar and disk properties between the Class I/FS ($i<70^\circ$) and Class II samples. The horizontal bars show the median values of each group. The p-values shown are from a Mann-Whitney U test (also known as the Wilcoxon rank-sum test) comparing the distributions of each property between the two samples. This non-parametric test assesses whether the two classes are drawn from the same underlying distribution, with ``ns" indicating non-significant (p $>$ 0.05) and ``*" indicating significant (p $<$ 0.05) differences—only the dust disk radius ($R_{\rm dust}$) shows a statistically significant difference between the two samples. \label{fig:disk_comparison}
}
\vspace{-0.cm}
\end{figure*}

Table~\ref{table:classII} lists the stellar and disk properties of the comparison Class II sample. Figure~\ref{fig:disk_comparison} shows the comparison between the two samples for the Class~I/FS sources with $i<70^{\circ}$, including stellar mass, accretion luminosity, dust disk size and mass. We restrict this comparison to the low-inclination subsample because edge-on systems show strongly suppressed molecular emission (Section~\ref{sec:results}), making them unsuitable for the chemical comparison that follows. The median stellar masses are 0.62 and 0.58\,M$_\odot$ for Class I/FS and Class II samples, respectively. A Mann-Whitney U test shows the stellar mass, accretion luminosity, and dust disk mass cannot be statistically distinguished between the two samples, while the difference in dust disk size is significant. The dust size difference is likely due to both evolution and sample bias, as the dust disk sizes of embedded sources tend to be smaller than Class II disks \citep{Tobin2020}, and the GO 1 sample of Class II disks is on the more massive and larger side of the disk size distribution \citep{Arulanantham2025,Manara2023}.     

\subsection{MIRI Observations and Data Reduction} \label{subsec:miri_obs}
The MIRI-MRS observations of 16 Ophiuchus Class I/FS disks were carried out as part of the Cycle 2 and 4 GO programs 3034 and 7135 (PI: K. Zhang) between 2023-2025, with a four-point dither pattern. MIRI-MRS provides integral-field spectroscopy covering 4.9--27.9\,$\mu$m across four channels, with spectral resolving power $R\sim$1500--3500 \citep{Wells2015,Argyriou2023}. Appendix Table~\ref{tab:obs_log} shows the observational dates and integration times.

All data were calibrated with the JWST Disk Infrared Spectral Chemistry Survey (JDISCS; \citealt{Pontoppidan2024,Arulanantham2025}) pipeline version 9.1.
This pipeline builds on the standard JWST reduction (version 1.18.0, CRDS 12.1.5; \citealp{Bushouse2025}) through stage 2b, followed by dedicated background subtraction and fringe correction using both stellar and asteroid calibrators \citep{Pontoppidan2024,Arulanantham2025}. The fringe calibration has been further refined by replacing the blackbody assumption for asteroids with a parametric emissivity model that accounts for silicate emission features \citep{Humes2024}, benchmarked against CALSPEC standard star spectra \citep{Bohlin2020}. This approach yields signal-to-noise improvements of up to a factor of $\sim$6 in MIRI channel 4. The Class~II spectra were originally reduced with JDISCS pipeline v8.4 \citep{Arulanantham2025}; we reprocessed them with v9.1 for homogeneous calibration across both samples. The updated pipeline yields 10--20\% higher fluxes at 18--27\,$\mu$m, owing to revised emissivity models.

\begin{figure*}[!htbp]
\centering
\vspace{-0.cm}
\includegraphics[width=0.95\textwidth]{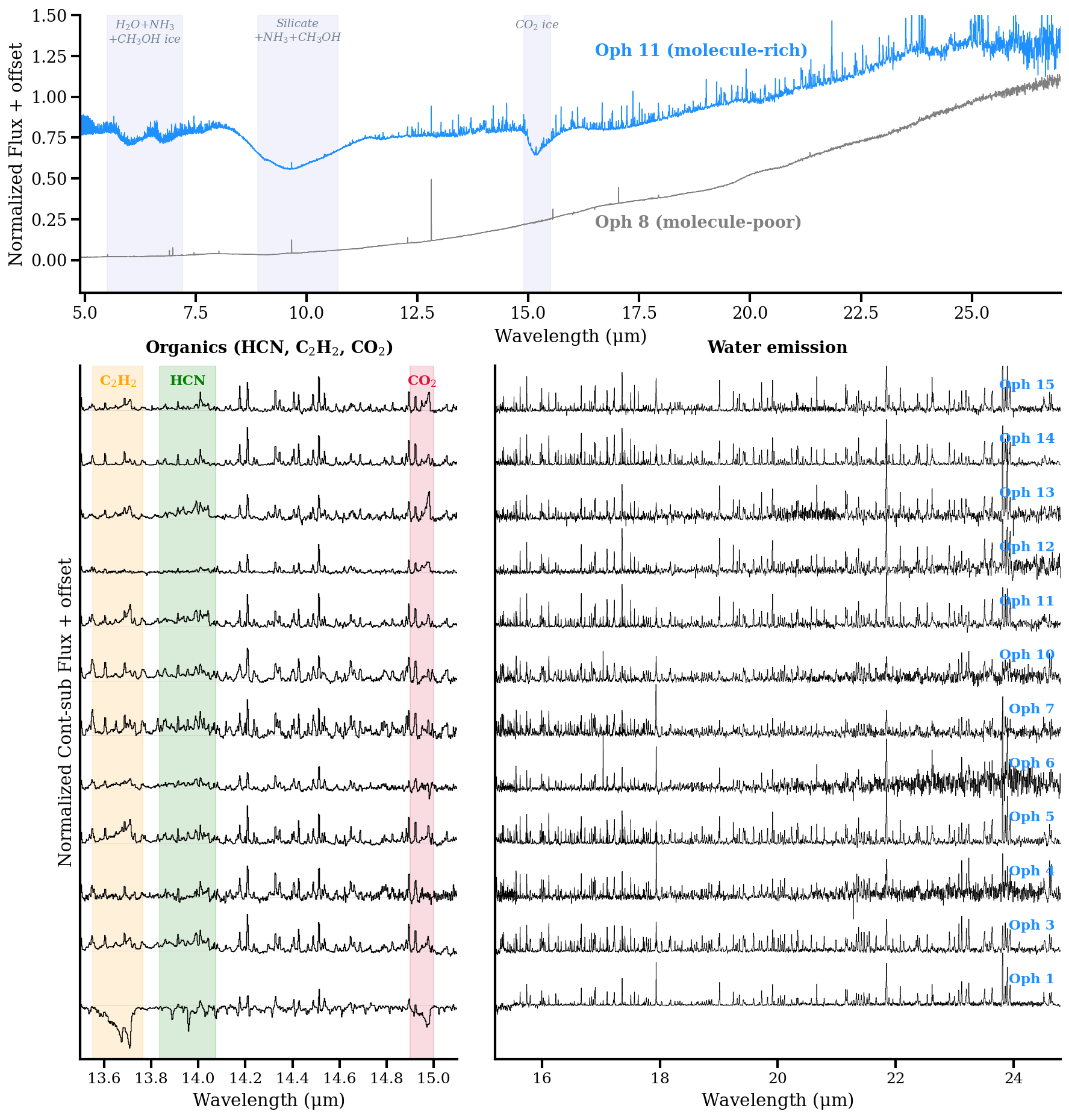}
\vspace{-.3cm}
\caption{ Top panel: JWST MIRI/MRS spectra of two representative Ophiuchus Class~I/FS sources --- Oph~11 (molecule-rich, blue) and Oph~8 (molecule-poor, grey) --- with ice absorption features highlighted (shaded bands). Bottom panel: extinction-corrected, continuum-subtracted spectra of the twelve sources with water emission detections (13--25\,$\mu$m).  \label{fig:obs_spec}
}
\vspace{-0.cm}
\end{figure*}

For embedded sources, the water emission can potentially originate from regions outside the Keplerian disk, such as the inner envelope or outflow. To check the spatial origin of the line emission, we constructed continuum-subtracted line images following the procedure of \citet{Narang_2026}. Appendix Figure~\ref{fig:line_images} shows the resulting images of H$_2$O (23.8\,$\mu$m), H$_2$ S(5) (6.91\,$\mu$m), and [Ne\,II] (12.81\,$\mu$m) for all 16 Class~I/FS sources. The water emission is spatially unresolved in all sources, while some sources show extended H$_2$ or jet-like [Ne\,II] emission whose morphology does not correlate with the cold water excess.  In addition, we measured the radial velocities of isolated H$_2$O lines in the 13--17\,$\mu$m range using single-Gaussian fits with iSLAT, following \citet{Banzatti2025}, and find no evidence of velocity shifts larger than 15\,km\,s$^{-1}$ relative to the systemic velocity in any of the seven ($i<70^{\circ}$) sources, ruling out high-velocity outflow components. Figure~\ref{fig:obs_spec} shows the MIRI spectra of the 16 Class I/FS sources.

\section{Method} \label{sec:method}
\subsection{Extinction correction}\label{subsec:extinction}
For the Class I/FS disks, we first apply extinction correction to the spectra to account for envelope and/or foreground extinction before measuring the line fluxes.
We adopt the extinction correction treatment of \citet{vanGelder2024a} for prestellar systems, which used a modified version of the extinction law of \citet{McClure2009} to account for the additional extinction from ice species and silicates. Basically, the total extinction (in unit of optical depth) is decomposed into two components: $\tau_{\rm tot}(\lambda) = \tau_{\rm ice,silicate} (\lambda) + \tau_{\rm ext} (\lambda)$, where $\tau_{\rm ice,silicate} (\lambda)$ is the extinction caused by the ice and silicate absorption features, and $\tau_{\rm ext}$ is the absolute extinction. The optical depth is computed using a third-order polynomial fitting based on relatively absorption-free wavelength regions (5.1-5.3, 7.7-7.9, and 23-26\,$\mu$m). The $\tau_{\rm ice,silicate} (\lambda)$ is estimated as $\tau_{\rm ice,silicate} (\lambda)=-ln \left(\frac{F_\lambda}{F_{cont}}\right)$. The absolute extinction $\tau_{\rm ext} = 0.085 \lambda^{-0.25} A_V$ based on \citet{McClure2009}, assuming $A_K= A_V/7.75$ mag. This \citet{McClure2009} curve is appropriate for the moderate extinctions of these relatively evolved, less deeply embedded Class~I/FS sources ($T_{\rm bol}=200$--750\,K); we discuss the choice of extinction curve, the determination of $A_V$, and why an H$_2$-based extinction estimate is not feasible for our sample in Appendix~\ref{sec:extinction_appendix}. We take the A$_V$ values listed in Table~\ref{table:stars}, adopted from the literature and works in preparation. An example of extinction correction is shown in Appendix Figure~\ref{fig:extinction_example}. For Class II disks, we do not apply any extinction correction, as the A$_V$ is estimated to be less than 1 for this sample.

\subsection{Continuum subtraction}
To estimate the continuum level, we use the \texttt{ctool} python tool \citep{Pontoppidan2026}. It estimates the continuum of a spectrum through an iterative process that alternates between sigma-clipping and Savitzky-Golay filtering of the flux array. At each iteration, outliers—typically due to lines—are masked, and the smoothed baseline is updated, yielding a robust continuum estimate that preserves broad spectral features while suppressing narrow line contributions. We also exclude the organic region between 13.4-14.1\,$\mu$m to avoid over-subtracting broad emission features from ro-vibrational lines of HCN and \ctht. The original \texttt{ctool} was designed for emission-line spectra, but some wavelength regions in our data clearly exhibit absorption lines. To address this, we divided the fitting into multiple wavelength intervals, depending on whether they are dominated by emission or absorption features by eye evaluation. For emission-dominated regions, we clip positive outliers above a chosen threshold to estimate continuum, whereas for absorption-dominated regions, we clip negative outliers below the threshold. The continuum subtracted data are shown in Figure~\ref{fig:obs_spec}.

\subsection{Line fluxes and Accretion luminosities}

After the extinction correction and continuum subtraction, we measure line fluxes of H$_2$O and organics for empirical comparison between the Class I/FS and II samples. Our Class I sample has abundant ice absorption features, and a few sources also show absorption lines of gas species, which are analyzed by Kim et al. (in prep), and \citet{Colmenares_2026_Oph1}. This work focuses on the emission lines. 

For H$_2$O, we measure fluxes for a suite of diagnostic lines recommended by \citet{Banzatti2025} (their Table 1), which span a wide range of upper-level energies and were selected to probe the relative contributions of hot, warm, and cold water reservoirs. The ratios of these lines (e.g., hot-to-cold, warm-to-hot) provide a model-independent diagnostic of the water excitation conditions and complement the slab modeling results presented in Section~\ref{sec:slab}. The line fluxes are measured with iSLAT \citep{Jellison2024}\footnote{\url{https://github.com/spexod/iSLAT}}, which performs single-Gaussian fits to measure the line centroid, full width at half maximum, flux, and their uncertainties. For non-detections, 3$\sigma$ upper limits are reported. For conversion from line fluxes to luminosities, we adopt 138.4\,pc as the average distance to Ophiuchus sources \citep{OrtizLeon2018}. 

We also measure the line fluxes of the vibrational bands of HCN, \ctht, and CO$_2$. These features consist of highly blended individual lines and are often contaminated by overlapping water emission at similar wavelengths. To account for this, we use best-fitting slab models of each species to extract their line fluxes (see \S\ref{sec:slab} for details on the slab modeling). The wavelength ranges used for integrating the fluxes are as follows: \ctht: 13.553–13.764\,$\mu$m, HCN: 13.837–14.075\,$\mu$m, and CO$_2$: 14.847–15.014\,$\mu$m. In cases of non-detections, we adopt the median temperature and column density of our samples of Class I/FS or Class II disks, and scale the emitting area until the peak of the model spectrum within the relevant wavelength range reaches the 1$\sigma$ noise level per resolution element. This lower threshold is adopted because the broad rovibrational bands of HCN, C$_2$H$_2$, and CO$_2$ extend over many resolution elements, so a per-element 1$\sigma$ peak still yields a detectable integrated signal.

Previous works have found that water line fluxes are tightly correlated with accretion luminosity (L$_{\rm acc}$) \citep{Salyk2011,Banzatti2020,Banzatti2025}. However, the traditional method of L$_{\rm acc}$ from UV continuum or H$\alpha$ line is not available for majority of our Class I/FS sources. Instead, we estimate the accretion luminosity of both our Class I/FS and II samples, by measuring the relatively isolated HI (10-7) line flux at 8.76\,$\mu$m and converting the average fluxes to L$_{\rm acc}$, based on the empirical relation found by \citet{Tofflemire2025,Shridharan2026} and updated in Hyden et al. 2026 (in prep). The H\,\textsc{i} (10--7) line is blended with water emission at this wavelength, so we measure its flux after subtracting the best-fitting water slab model; the subtracted water contribution is minor in most sources. We tabulate the resulting H\,\textsc{i} (10--7) line luminosities alongside the derived $L_{\rm acc}$ (Appendix Tables~\ref{tab:lineflux} and \ref{tab:lineflux_classII}), so that our accretion luminosities can be reproduced or recomputed with any adopted H\,\textsc{i}--$L_{\rm acc}$ relation.

Finally, we measure the infrared spectral index $n_{13-26}$ by taking the MIRI continuum flux measured at 13 and 26~$\mu$m, following the same approach as \cite{Banzatti2023,Arulanantham2025}. The continuum fluxes are measured from relatively line-free/weak regions: at 13.095--13.113 and 26.3--26.4~$\mu$m. 

The line luminosities and L$_{\rm acc}$ for the Class I/FS sample are listed in Appendix Table~\ref{tab:lineflux}. 
The line luminosities and L$_{\rm acc}$ of the Class II sample with JDISCS v9.1 version have been reported in \citet{Banzatti2023,Mallaney2026}.

\subsection{Slab models}\label{sec:slab}
We characterize the excitation conditions of key molecular species by fitting local-thermodynamic-equilibrium (LTE) slab models to the continuum-subtracted spectra. Each slab model approximates the emission of a molecule as an isothermal, plane-parallel column of gas characterized by a single excitation temperature $T$, column density $N$, and emitting area $A$, a standard approach for interpreting molecular emission in mid-infrared spectra \citep[e.g.,][]{Salyk2011,Tabone2023,Grant2023,Temmink2025}. 

We model all species simultaneously. Water is represented by three temperature components: hot (500–1500\,K), warm (200–800\,K), and cold (100–400\,K), following the three-component water decomposition of \citet{Temmink2024}, who showed that it reproduces the spectral features as well as models with additional components. For the rest of the molecules (HCN, C$_2$H$_2$, CO$_2$, and OH), each is modeled with a single-temperature component. OH line emission is often found to be in non-LTE \citep{Tabone2024}, and we only include it as a component to remove its contribution in MIRI spectra.

Each component's slab model is evaluated over a wavelength range appropriate to its spectral features: hot and warm water over 13–27\,$\mu$m, cold water over 18–27\,$\mu$m (where it contributes meaningfully), HCN, C$_2$H$_2$, and CO$_2$ over 13–18\,$\mu$m (where their primary rovibrational bands reside), and OH over 13–27\,$\mu$m. These component models are summed into a single composite spectrum over 13–27\,$\mu$m, and all seven are fitted together in a single MCMC run by comparing the composite to the observed spectrum at each step.

Synthetic spectra are generated using the \texttt{spectools\_ir}\footnote{\url{https://github.com/csalyk/spectools_ir/tree/main}} package \citep{Salyk2022}. We assume a Gaussian line profile, with instrumental FWHM values of 110\,km/s for 13–18\,$\mu$m and 130\,km/s for 18–27\,$\mu$m, consistent with resolving power measured by \citet{Pontoppidan2024}. In addition to instrumental broadening, we include thermal broadening based on the excitation temperature of each molecular component. We ignore turbulent broadening contribution as previous disk studies showed that turbulence broadening is only up to 10--20\% of the thermal broadening in Class II disks \citep{Flaherty2015,Flaherty2020}. The emission also arises from a range of disk radii and hence a range of Keplerian velocities, but MIRI's resolving power (velocity FWHM of $\sim$85--200\,km\,s$^{-1}$) is generally too coarse to resolve this broadening; only marginally resolved widths in some short-wavelength water lines \citep{Banzatti2025}. We verified that varying the velocity broadening up to $\delta v\sim2$\,km\,s$^{-1}$ (as adopted in some previous studies) slightly shifts the retrieved column density and emitting area, which partly compensate each other, while leaving the excitation temperature and total emitting mass essentially unchanged. Therefore, our temperature- and mass-based results are robust to the treatment of velocity broadening.

The per-channel flux uncertainties $\sigma_i$ are estimated following the same method as in \citet{RomeroMirza2024} and combined with continuum-subtraction uncertainties of 5\% of the continuum level for both the 13--18\,$\mu$m and 18--27\,$\mu$m regions. We fit the models in a Bayesian framework, adopting a Gaussian likelihood,
\begin{equation}
\ln\mathcal{L} = -\frac{1}{2}\sum_i \frac{\left(F_{{\rm obs},i}-F_{{\rm model},i}\right)^2}{\sigma_i^2},
\end{equation}
where the sum runs over the spectral channels $i$ and $F_{\rm model}$ is the composite slab model. We adopt uniform priors on all parameters: the excitation temperature spans 500--1500, 200--800, and 100--400\,K for the hot, warm, and cold water components, respectively, and 100--2000\,K for HCN, C$_2$H$_2$, CO$_2$, and OH; the log column density $\log N\,({\rm cm}^{-2})$ spans 12--22; and the log emitting area $\log A\,({\rm au}^2)$ spans $-4$ to 4, for all components. We sample the resulting posterior distribution using the affine-invariant MCMC ensemble sampler \texttt{emcee} \citep{ForemanMackey2013}, typically with 70 walkers and 3000–5000 steps. The chains generally converge within the first 2000 steps, and the remaining samples are used to compute the posterior parameter distributions and their uncertainties.

In cases of non-detections, we estimate the upper limits of emitting mass by adopting the median temperature and column density of our samples of Class I/FS or Class II disks, and scale the emitting area until the peak of the model spectrum within the relevant wavelength range reaches the 3$\sigma$ noise level for individual water lines, or the 1$\sigma$ noise level for the broadband rovibrational features of HCN, C$_2$H$_2$, and CO$_2$, as these features are spread over many resolution elements.

Several caveats apply to these 0D LTE slab retrievals. Each component is described by a single temperature and column density, whereas real disks have radial and vertical gradients, so the retrieved values are emission-weighted averages over the line-emitting region. The lines probe only the gas above the $\tau_{\rm dust}+\tau_{\rm line}\!\approx\!1$ surface, so the derived masses are lower limits to the true molecular content \citep[e.g.,][]{Tabone2026}; we refer to them as \emph{observable} masses throughout.

In the optically thin limit, the total number of emitting molecules (area $\times$ column density) is well constrained, though area and column density remain individually degenerate. Optically thin isotopologues such as H$_2^{18}$O and $^{13}$CO$_2$ can break this degeneracy and probe deeper reservoirs \citep[e.g.,][]{Grant2023,Salyk2026,Vlasblom_2025b}, but they are detected only in a few bright disks and are not available for most of our sample. Recent slab retrievals on synthetic spectra of thermochemical disk models nonetheless show that the retrieved parameters reliably trace conditions in the line-emitting layer \citep{Kaeufer_2024,Vlasblom_2025b}.

The LTE assumption may also break down for the rovibrational bands of HCN and CO$_2$, whose upper vibrational levels have high critical densities and can be sub-thermally populated in the upper disk layers \citep{Bruderer2015,Bosman2017}, adding further uncertainty to the inferred masses.

\subsection{Age estimation}

The ages of Class I/FS are highly uncertain. The Spitzer C2D large survey derived a statistical age for Class I 0.1-0.64\,Myr and 0.64-1.04\,Myr for FS \citep{Evans2009}. We adopt a statistical age of 0.5\,Myr for our Class I/FS sample \citep{Evans2009,RuizRodriguez2025}. Bolometric temperature (T$_{\rm bol}$) is used as an indicator of relative ages among Class I/FS.

For Class II sources, we derive isochronal ages, using effective temperatures and stellar luminosities from \citet{Andrews2018a,Manara2023} and comparing these with evolutionary tracks. We use the \texttt{Python} package \texttt{ysoisochrone}\footnote{\url{https://github.com/DingshanDeng/ysoisochrone}} \citep{Pascucci2016,Deng2025a}, which uses evolutionary tracks of \citet{Feiden2016} for sources with $T_{\rm eff}>3900\mathrm{~K}$ and that of \citet{Baraffe2015} for $T_{\rm eff} \le 3900\mathrm{~K}$. The typical uncertainty is ${\sim}0.5~\mathrm{Myr}$. In the majority of cases, the age estimates are in good agreement with values reported in the literature \citep[e.g.,][]{Andrews2018a,Long2019,RomeroMirza2024}. The resulting ages are listed in Table~\ref{table:classII}; the majority of our Class~II sources span 0.5--1.6\,Myr, with only GO~Tau ($\sim$1.6\,Myr) significantly older than 1\,Myr.  Appendix Figure \ref{fig:age} shows the H-R diagram locations and comparison with evolutionary tracks for the Class II sources. We adopt our homogeneous approach to ensure internal consistency across the Class II sample. 

\section{Results} \label{sec:results}

 In this section, we compare the molecular emission properties of embedded Class I/FS disks with those of more evolved Class II disks.
We first examine the impact of disk inclination on molecular detections in the Class~I/FS sample, which reveals that edge-on systems ($i>70^{\circ}$) have significantly suppressed line detections. We therefore restrict the subsequent Class~I/FS vs.\ Class~II comparison to the seven Class~I/FS sources with $i<70^{\circ}$ and compare them with the twelve Class~II disks described in \S\ref{subsec:classII_sample}. We analyze detection rates, line fluxes, and line ratios to identify empirical differences, then compare slab model results to quantify differences in excitation temperature and emitting mass of key molecular species. Finally, we search for correlations between slab model parameters and disk properties.

\subsection{Inclination dependence of molecular detections}

\begin{figure}[!htbp]
\centering
\vspace{-0.cm}
\includegraphics[width=0.48\textwidth]{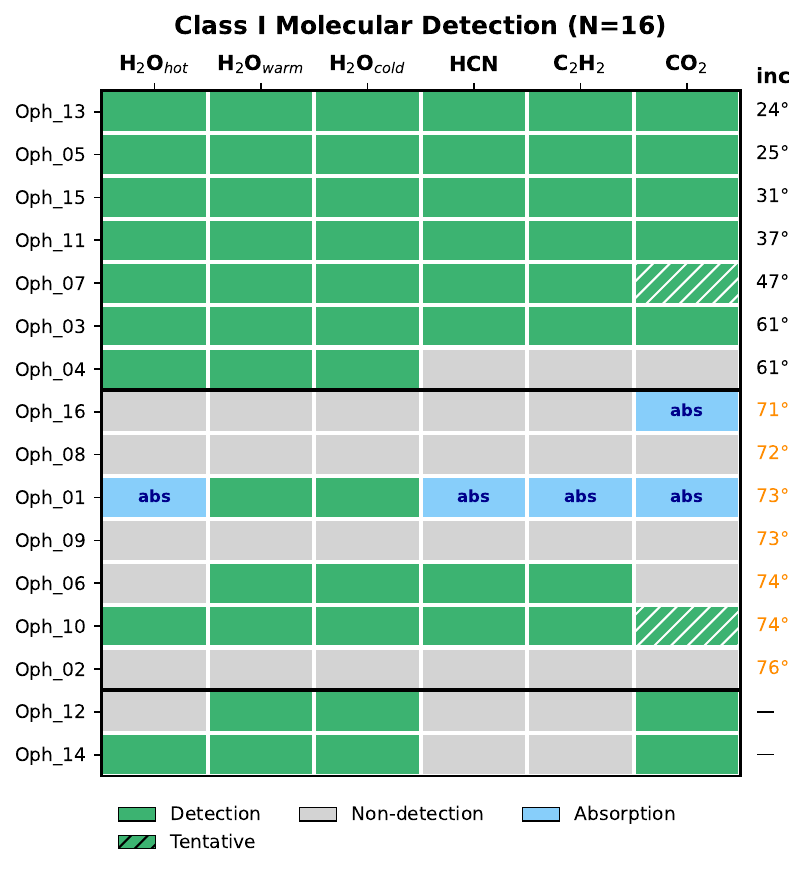}
\vspace{-0.2cm}
\caption{ Detection rate comparison among the Ophiuchus Class I/FS disks. \label{fig:detection_rate}
}
\vspace{-0.cm}
\end{figure}

Among the sixteen Class~I/FS sources, we find a strong dependence of molecular detection rates on disk inclination, as shown in Figure~\ref{fig:detection_rate}. For the seven sources with $i<70^{\circ}$ (Oph~3, 4, 5, 7, 11, 13, 15), H$_2$O is detected in all seven (100\%), and HCN, \ctht, and CO$_2$ are each detected in six out of seven (86\%). In contrast, the seven sources with $i>70^{\circ}$ (Oph~1, 2, 6, 8, 9, 10, 16) show markedly lower detection rates. Here and in Figure~\ref{fig:detection_rate} a ``detection'' refers to a constrained slab-model component (Appendix Tables~\ref{tab:slab_h2o} and \ref{tab:slab_organics}), rather than a measurable flux in any single line: only Oph~10 has a hot H$_2$O component, warm or cold H$_2$O is recovered in only three out of seven (43\%), the detection rates of HCN and \ctht~drop to 28.6\%, and CO$_2$ emission is only tentatively detected in 1/7 sources. Two sources (Oph~12 and 14) do not have inclination constraints because their mm images are spatially unresolved.
The suppressed emission in high-inclination systems likely reflects geometric effects \citep[e.g.,][]{Somigliana_2026_MINDS_HKTau}: at $i>70^{\circ}$, the optically thick outer disk and residual envelope obscure the line-emitting inner disk, while the continuum further dilutes the line-to-continuum ratio. To isolate intrinsic chemical differences from viewing geometry effects, we restrict the Class~I/FS sample to sources with $i<70^{\circ}$. All twelve sources in our Class~II sample have $i<70^{\circ}$ (Table~\ref{table:classII}) and are therefore included in the comparison.

\subsection{Molecular Detection Rates and line luminosities \\ Class~I/FS ($i<70^{\circ}$) vs.\ Class~II}

Comparing the seven $i<70^{\circ}$ Class~I/FS sources with the twelve Class~II disks, the detection rates are similar for H$_2$O (100\% vs.\ 100\%), HCN (86\% vs.\ 100\%), and CO$_2$ (86\% vs.\ 100\%). The only notable difference is in \ctht, which is detected in 86\% of the Class~I/FS sources but only 67\% (8/12) of the Class~II sample. Given the small size of both samples, the difference in the detection percentage may not be significant. In addition, GO~Tau---the oldest Class~II source in our sample ($\sim$1.6\,Myr)---shows emission from rarer isotopologues such as $^{13}$\ctht\ and more complex organics including C$_4$H$_2$ and HC$_3$N; none of these species are detected in our Class~I/FS sample.

\vspace{0.2cm}
\noindent\textbf{Line flux vs. accretion luminosity:} The line luminosities of diagnositic water lines and other molecules are listed in Table\,\ref{tab:lineflux} for the Class~I/FS sample and Table\,\ref{tab:lineflux_classII} for the Class~II sample. Previous studies have shown that water and other molecular line fluxes correlate strongly with accretion luminosity (L$_{\rm acc}$) \citep{Salyk2011,Banzatti2020,Banzatti2025}. Motivated by this, we also examine whether the correlation differs between the Class I/FS and Class II samples.

\begin{figure*}[!htbp]
\centering
\vspace{-0.cm}
\includegraphics[width=0.9\textwidth]{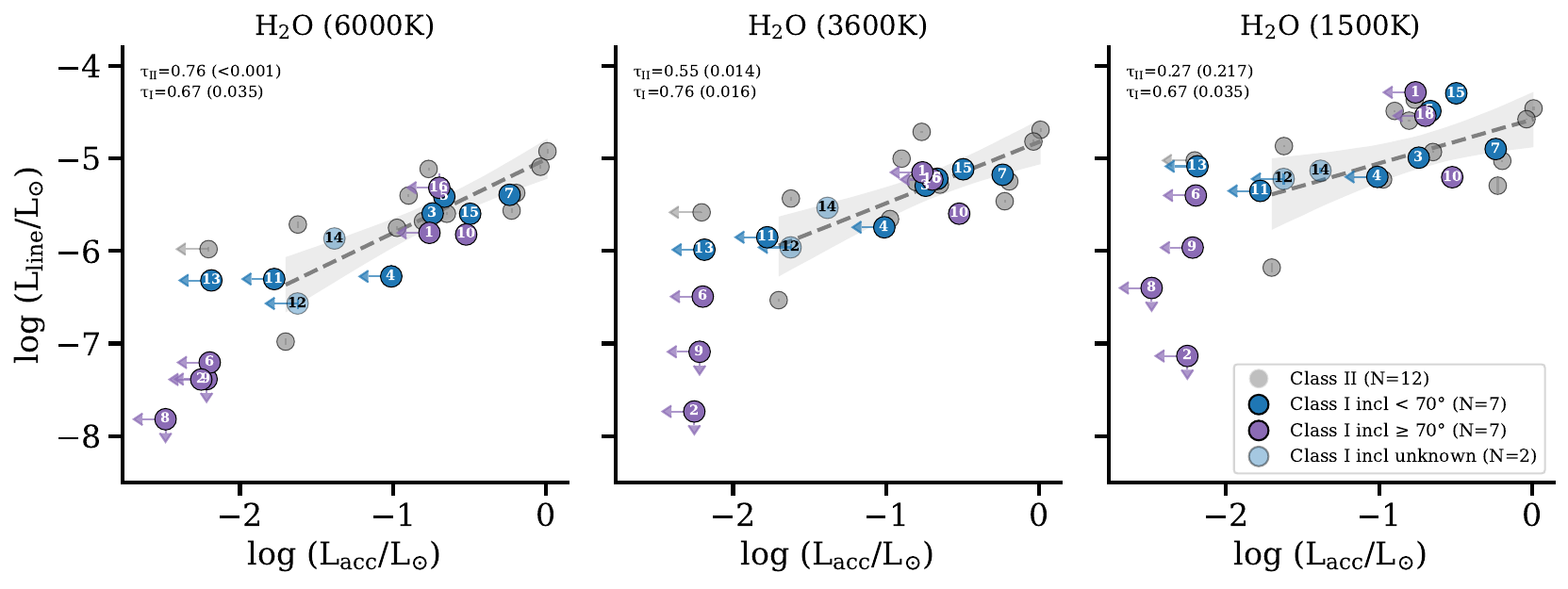}
\vspace{-0.1cm}
\includegraphics[width=0.9\textwidth]{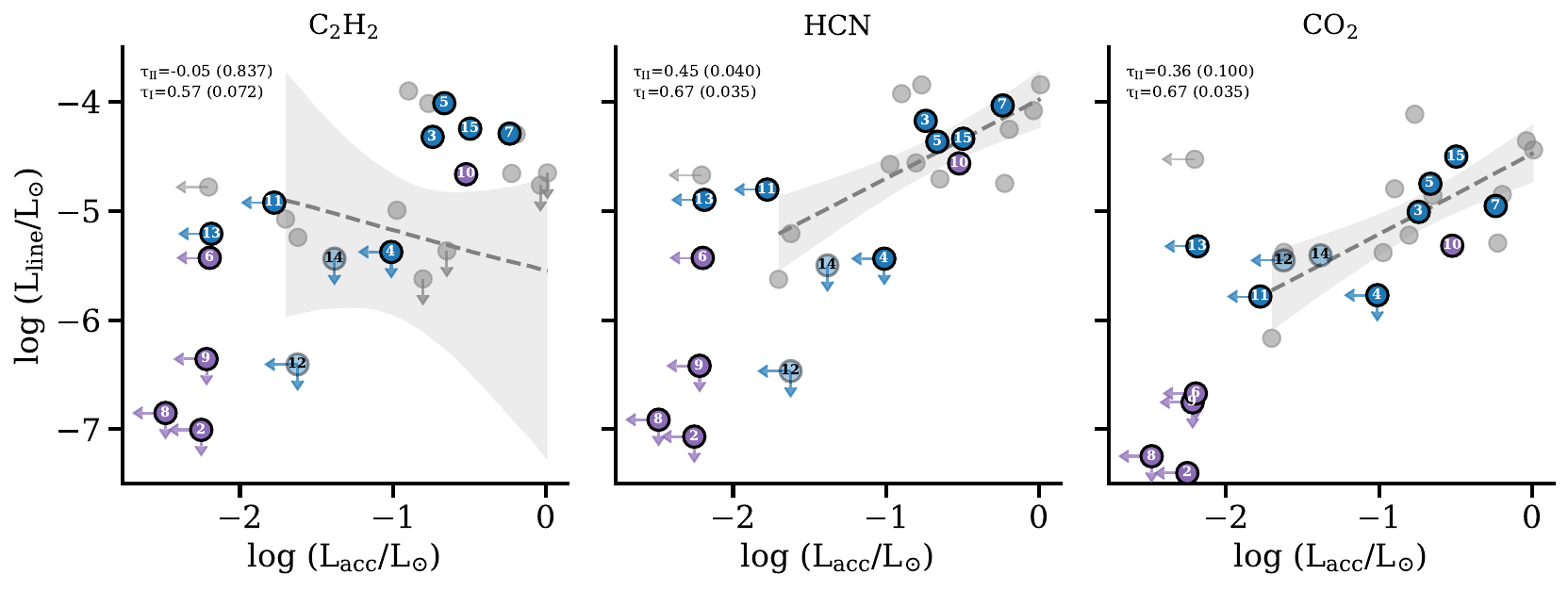}
\vspace{-0.2cm}
\caption{Line luminosities as a function of accretion luminosity. Top: diagnostic water lines. Bottom: HCN, \ctht, and CO$_2$. The Class I/FS and II samples generally follow the same trends, but Class I/FS shows larger scatter in the water lines. The grey dashed line and shaded band show the censored \texttt{linmix} fit \citep{Kelly_2007} to the Class~II reference sample (16th--84th percentile of the posterior), which includes the upper limits; a separate Class~I/FS \texttt{linmix} fit is not shown because too few Class~I/FS sources have $L_{\rm acc}$ detections. The generalized (censored) Kendall $\tau$ and its $p$-value are annotated in each panel for the Class~II ($\tau_{\rm II}$) and Class~I/FS\,($i<70^{\circ}$; $\tau_{\rm I}$) samples, computed separately so the two trends can be compared. \label{fig:flux_Lacc}
}
\vspace{-0.cm}
\end{figure*}

To compare the two samples while including the upper limits, we apply the generalized (censored) Kendall $\tau$ test---the same method used in Section~\ref{sec:correlation} (see details of the method in Appendix~\ref{sec:stat_methods})---separately to the Class~I/FS\,($i<70^{\circ}$) and Class~II samples (Figure~\ref{fig:flux_Lacc}). We find that the two samples generally follow the same water line luminosity--$L_{\rm acc}$ correlations, although the scatter is larger in the Class I/FS sample, in particular for the cold (E$_{\rm up}\sim$1500\,K) lines. HCN follows a similar trend in both samples (Figure~\ref{fig:flux_Lacc}, bottom). In contrast, neither \ctht\ nor CO$_2$ shows a clear correlation with $L_{\rm acc}$ in the Class~II sample, although both cover luminosity ranges comparable to the Class~I/FS sources. The censored Kendall $\tau$ coefficients are annotated in each panel.

\begin{figure*}[!htbp]
\centering
\vspace{0.1cm}
\includegraphics[width=0.9\textwidth]{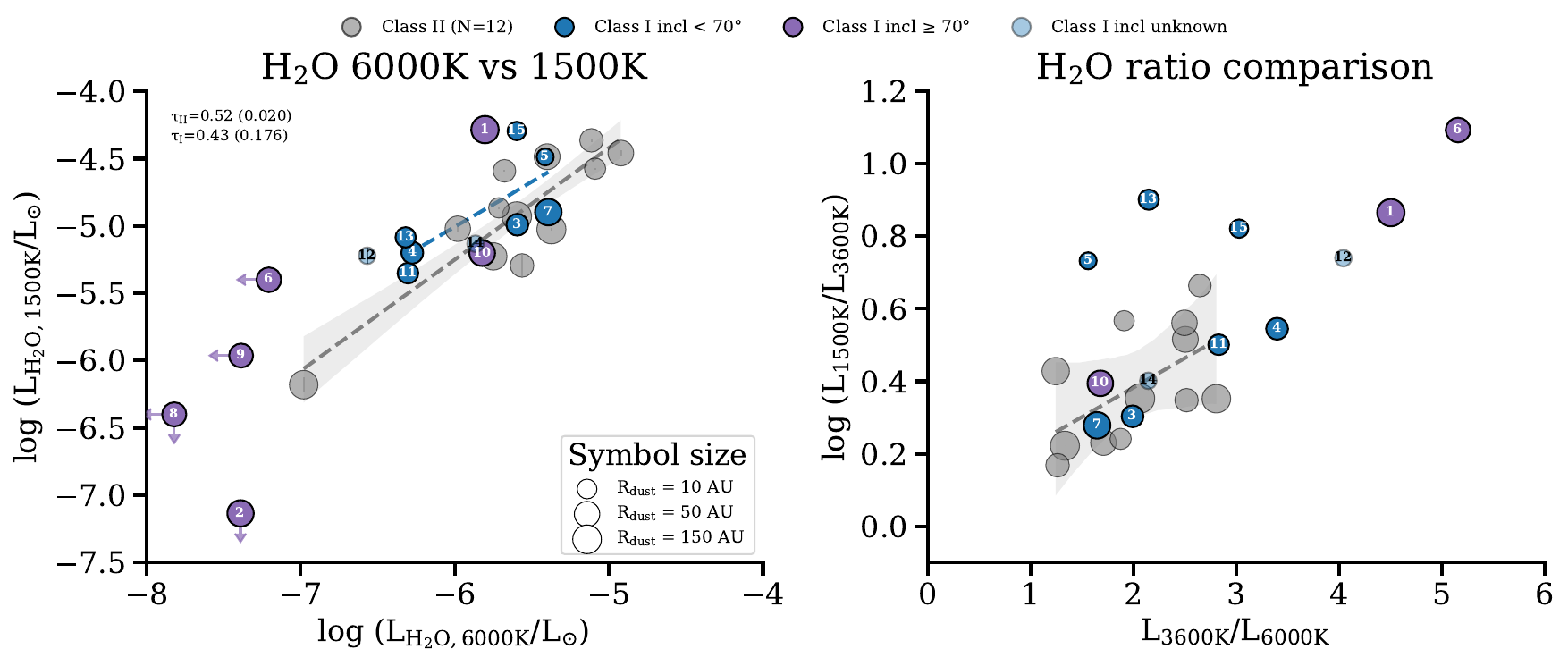}
\vspace{-0.2cm}
\caption{ Water line luminosity diagnostic plot. Left: the line luminosities of hot and cold water transitions cover similar luminosity range and trend as the Class II sample. The grey dashed line and shaded band show the censored \texttt{linmix} fit to the Class~II reference sample (16th--84th percentile), and the blue dashed line shows the corresponding Class~I/FS\,($i<70^{\circ}$) \texttt{linmix} fit (both hot-water axes are fully detected here, so both fits are constrained); the two slopes ($0.83\pm0.18$ and $0.67\pm0.56$) agree within the uncertainties. The censored Kendall $\tau$ for the Class~II ($\tau_{\rm II}$) and Class~I/FS ($\tau_{\rm I}$) samples are annotated. Right: Line luminosity ratios among water lines: the majority of the Class I/FS sources cover the upper half of the plot, suggesting these have excess of cold or warm water fluxes compared to many Class II sources. \label{fig:water_diagnostic}
}
\vspace{-0.cm}
\end{figure*}

We caution that a single water line luminosity is not a clean tracer of a single temperature component: particularly for the warm and cold lines, the measured flux can blend contributions from more than one component. We therefore regard the line luminosity--$L_{\rm acc}$ relations in Figures~\ref{fig:flux_Lacc} and \ref{fig:water_diagnostic} primarily as empirical consistency checks between the two samples, whereas our quantitative, physically interpreted correlations are based on the slab-model component masses, which separate the hot, warm, and cold water reservoirs (Section~\ref{sec:correlation_analysis}).

\noindent\textbf{Line flux  ratios:} To compare the relative contributions of hot, warm, and cold water reservoirs between the two samples in a model-independent way, we use the ratios of diagnostic water lines with different upper-level energies. Figure~\ref{fig:water_diagnostic} shows the line fluxes and diagnostic ratios for the Class~I/FS and Class~II samples. The hot and cold water lines (E$_{\rm up}\sim$6000\,K and 1500\,K, respectively) exhibit a similar flux range in both samples, spanning over two orders of magnitude.  
In contrast, the temperature diagnostic plots reveal a significant difference between the two samples. Of the seven $i<70^{\circ}$ Class I/FS sources, three (Oph 5, 13, 15) lie well above the Class II trends, suggesting exceptionally high cold-to-hot line fluxes. Oph 4 shows slightly higher warm and cold water line flux compared to Class II. The remaining three Class I/FS disks (Oph 3, 7, and 11) are located in a similar region as the bulk of Class II disks. Notably, the three cold-water-excess sources are among the most compact disks in the sample ($R_{\rm d, 90\%} = 6$, 12, and 7\,au for Oph~5, 13, and 15, respectively), while the three Class II-like sources have larger dust disks ($R_{\rm d, 90\%} = 17$, 73, and 13\,au for Oph~3, 7, and 11).

\subsection{Comparison of slab model results}

\begin{figure*}[!htbp]
\centering
\vspace{-0.cm}
\includegraphics[width=0.92\textwidth]{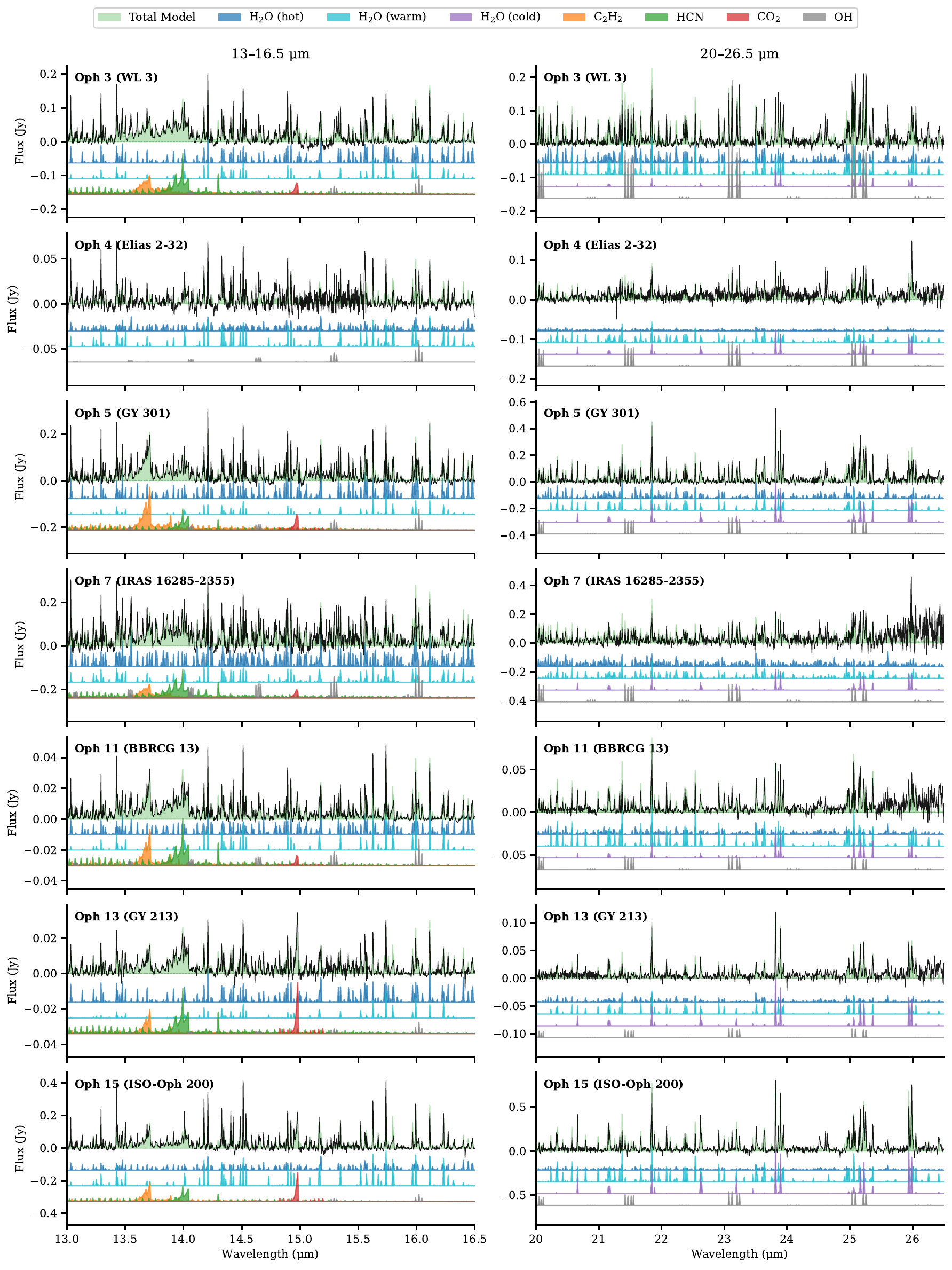}
\caption{ Slab models of i$<70^{\circ}$ Class I/FS sample, for the 13-16.5 and 20-26.5\,$\mu$m wavelength ranges. \label{fig:slab_model_classI}
}
\vspace{-0.cm}
\end{figure*}

The best-fitting slab model parameters for individual sources are listed in Appendix Tables~\ref{tab:slab_h2o}, \ref{tab:slab_organics}, \ref{tab:slab_classII_h2o}, and \ref{tab:slab_classII_organics}, and the median values for each sample are listed in Table~\ref{tab:slab_median}. The water slab parameters for our Class II sources are part of the larger 22-disk sample presented in \citet{Salyk2026}; here we additionally report slab parameters for HCN, C$_2$H$_2$, and CO$_2$. The best-fitting slab models of Class I/FS are shown in Figure~\ref{fig:slab_model_classI} for the $i<70^{\circ}$ sources and Appendix Figure~\ref{fig:ClassI_slab_inclined} for the rest. The Class~II slab models are shown in Appendix Figures~\ref{fig:ClassII_13um_slab} and \ref{fig:ClassII_20um_slab}. 

Table \ref{tab:slab_median} lists the median and standard deviation for each parameter. Uncertainties on the medians are estimated via 10,000 bootstrap iterations; in each, we resample sources with replacement and perturb their values using MCMC posterior errors. This process simultaneously accounts for sampling and measurement uncertainty, with the final error defined as the standard deviation of the bootstrap distribution.

\begin{deluxetable}{lcccccc}
\tablecaption{Median slab model parameters for Class I/FS and Class II samples. \label{tab:slab_median}}
\tablewidth{0pt}
\tablehead{
\colhead{Molecule} & \colhead{Class} & \colhead{$N_\mathrm{det}$} & \colhead{$T$} & \colhead{$\log N$} & \colhead{$\log A$} & \colhead{$\log M$} \\
 & & & (K) & (cm$^{-2}$) & (au$^2$) & ($M_\oplus$)
}
\startdata
H$_2$O (hot) & I & 7 & $909 \pm 95$ & $18.38 \pm 0.22$ & $-0.82 \pm 0.29$ & $-6.19 \pm 0.27$ \\
 & II & 12 & $894 \pm 58$ & $18.06 \pm 0.09$ & $-0.58 \pm 0.13$ & $-6.43 \pm 0.25$ \\
\hline
H$_2$O (warm) & I & 7 & $454 \pm 38$ & $18.05 \pm 0.26$ & $0.12 \pm 0.20$ & $-5.69 \pm 0.29$ \\
 & II & 12 & $469 \pm 35$ & $17.81 \pm 0.14$ & $0.38 \pm 0.23$ & $-5.74 \pm 0.25$ \\
\hline
H$_2$O (cold) & I & 7 & $196 \pm 19$ & $16.09 \pm 0.42$ & $2.26 \pm 0.32$ & $-5.15 \pm 0.41$ \\
 & II & 12 & $204 \pm 12$ & $15.97 \pm 0.26$ & $2.47 \pm 0.25$ & $-5.55 \pm 0.17$ \\
\hline
HCN & I & 6 & $798 \pm 88$ & $15.74 \pm 0.58$ & $-0.15 \pm 0.53$ & $-7.66 \pm 0.24$ \\
 & II & 12 & $783 \pm 123$ & $16.07 \pm 0.45$ & $-0.34 \pm 0.55$ & $-7.90 \pm 0.20$ \\
\hline
C$_2$H$_2$ & I & 6 & $790 \pm 148$ & $15.28 \pm 0.85$ & $0.07 \pm 0.70$ & $-8.19 \pm 0.26$ \\
 & II & 8 & $974 \pm 133$ & $16.03 \pm 0.69$ & $-0.47 \pm 0.53$ & $-8.71 \pm 0.26$ \\
\hline
CO$_2$ & I & 6 & $329 \pm 92$ & $16.00 \pm 0.65$ & $-0.09 \pm 0.54$ & $-7.45 \pm 0.55$ \\
 & II & 12 & $482 \pm 51$ & $16.62 \pm 0.27$ & $-0.57 \pm 0.25$ & $-7.43 \pm 0.26$
\enddata
\tablecomments{Median values of $T$, $\log N$, and $\log A$ are computed from detections only. Median $\log M$ is computed using the Kaplan-Meier estimator to account for upper limits (left-censored data). Uncertainties are the standard deviation of the bootstrapped median (10,000 iterations), where each iteration resamples sources with replacement and perturbs each detected source's parameters within its asymmetric MCMC posterior uncertainties. Class I subsample: 7 sources with inclination $< 70^{\circ}$. Class II subsample: 12 sources with stellar mass $\le 0.7\,M_\odot$.}
\end{deluxetable}

\vspace{0.2cm}
\noindent\textbf{Excitation temperature:}  

\begin{figure*}[!htbp]
\centering
\vspace{-0.cm}
\includegraphics[width=0.9\textwidth]{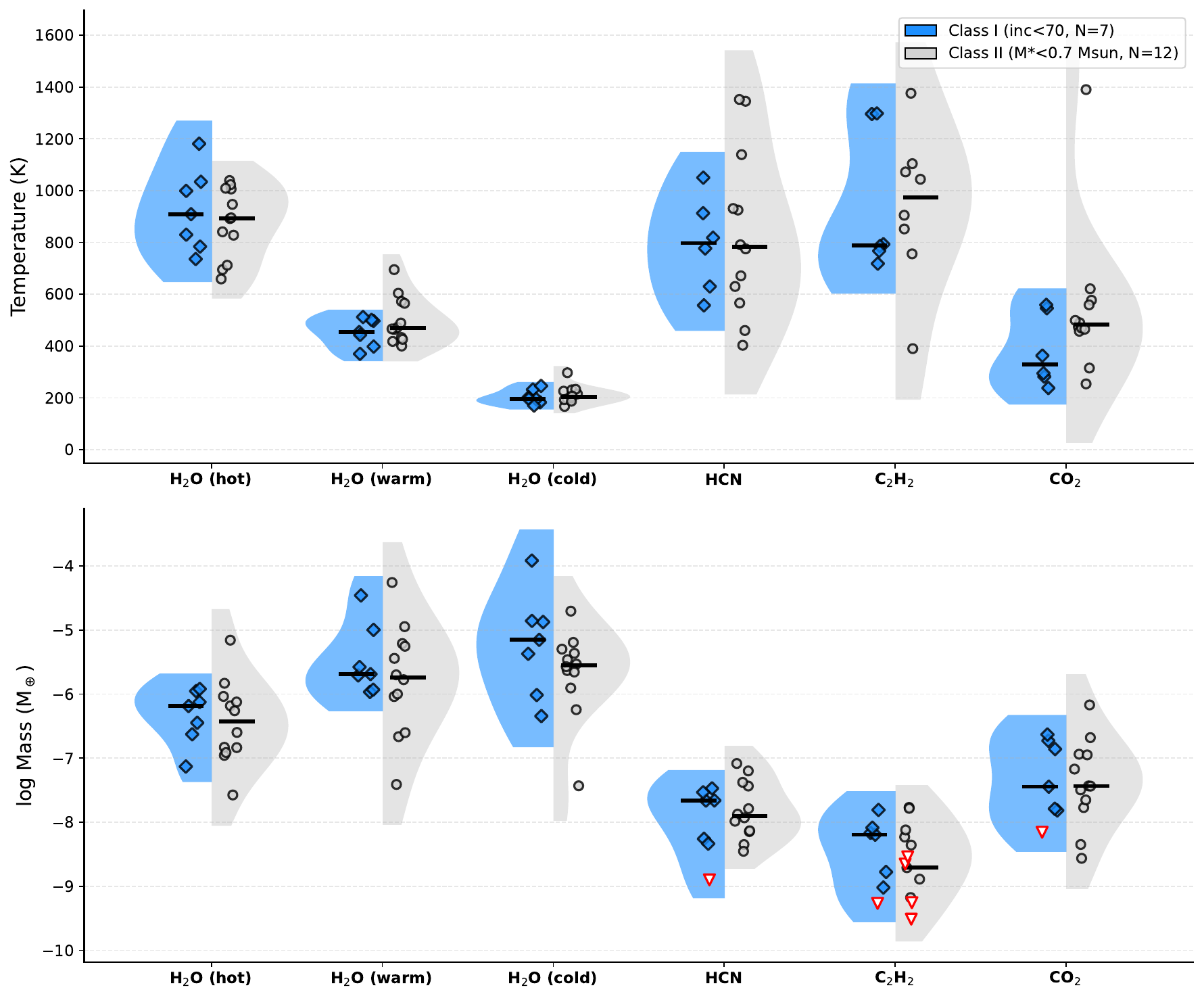}
\caption{Comparison of slab model parameters between Class I/FS ($i<70^{\circ}$, blue) and Class II (grey) disks. Top panel: excitation temperatures for the three water components (hot, warm, cold), HCN, C$_2$H$_2$, and CO$_2$. Bottom panel: log emitting masses for the same species. Individual sources are shown as points; violin plots show the kernel density estimate of each distribution, and horizontal bars mark the median. Upper limits are indicated by red downward arrows. The three water components have comparable temperatures in both classes, while median temperature of CO$_2$ is colder in Class I/FS (329 vs.\ 482\,K) with 1.5$\sigma$ separation.  \label{fig:slab_comparison}
}
\vspace{-0.cm}
\end{figure*}

Figure~\ref{fig:slab_comparison} compares the excitation temperatures and the observable molecular masses derived for six key components: hot, warm, and cold H$_2$O; HCN; C$_2$H$_2$; and CO$_2$. We exclude OH from this comparison because its emission is often in non-LTE (Section~\ref{sec:slab}). Furthermore, HCN and C$_2$H$_2$ emission can be optically thin at MIRI wavelengths \citep{Arulanantham2025}, introducing a degeneracy between emitting area and column density. However, several sources in our sample exceed $\log N \gtrsim 16$\,cm$^{-2}$, above which these lines become optically thick under thermal broadening (Xie et al., submitted), so the derived masses represent the observable emitting layer rather than the total reservoir. We therefore focus on excitation temperature and observable molecular mass (Appendix Figure~\ref{fig:slab_area_N} shows the full area--column density comparison).

Table~\ref{tab:slab_median} and Figure~\ref{fig:slab_comparison} show that the Class I/FS and Class II samples have similar median temperatures of $\sim$900, 460, and 200\,K for the hot, warm, and cold water components, respectively. 
HCN and C$_2$H$_2$ exhibit a broad range of excitation temperatures (200–1400\,K), but the median values are similar between the two samples. 

For CO$_2$, Class I/FS sources appear to be generally colder (median of 329\,K) than that of Class II disks (median of 482\,K). To estimate the sample size required to detect the CO$_2$ temperature difference at 2$\sigma$ significance, we performed a power analysis assuming the median difference and bootstrapped uncertainties scale as $1/\sqrt{n}$. Based on the current 6 Class~I/FS and 12 Class~II CO$_2$ detections (153\,K median difference; 1.5$\sigma$), we find that approximately 7 additional detections per class (totaling $\sim$13 Class~I/FS and $\sim$19 Class~II) are needed to reach 2$\sigma$, assuming the underlying distributions remain unchanged. The full Ophiuchus Class~I/FS sample from JWST Cycle~4, combined with existing JDISCS GO observations expanding the Class~II sample, will provide such a test.

\vspace{0.2cm}
\noindent\textbf{Emitting mass:} The median observable masses of three water components in Class I/FS disks are comparable to that of Class II disks, but the hot and cold water components have slightly higher masses than that of Class II. For HCN and C$_2$H$_2$, the median observable masses of Class I/FS disks are also similar to the Class II disks. The median observable CO$_2$ mass in Class I/FS disks is slightly lower than that of Class II. But given the large spread in individual values, none of the differences are statistically significant.  

\begin{figure*}[!htbp]
\centering
\vspace{-0.cm}
\includegraphics[width=0.9\textwidth]{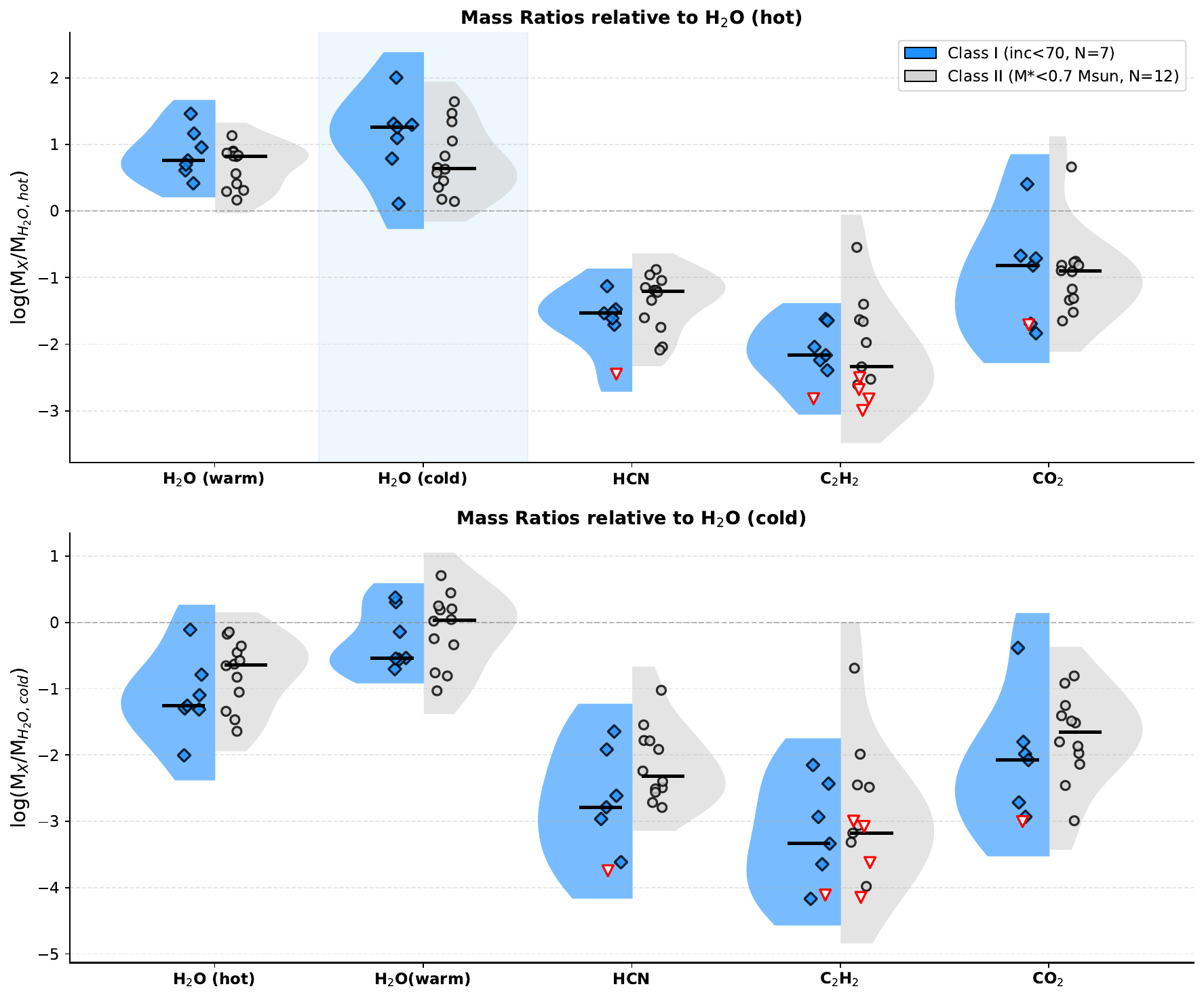}
\caption{Violin plots of molecular mass ratios for Class I/FS ($i<70^{\circ}$, blue) and Class II (grey) disks. Top panel: mass ratios normalized to hot water. Bottom panel: mass ratios normalized to cold water. Individual sources are shown as points, with upper limits indicated by downward arrows. When normalized to hot water, the ratios of HCN, C$_2$H$_2$, and CO$_2$ are similar between the two classes, but the cold-to-hot water ratio is elevated in Class I/FS (median $\log(M_{\rm cold}/M_{\rm hot})=1.30$ vs.\ 0.63). When normalized to cold water, all ratios are systematically lower in Class I/FS, indicating that the chemical difference is driven by a cold water excess rather than a carbon-species deficit.  \label{fig:mass_ratio}
}
\vspace{-0.cm}
\end{figure*}

\noindent\textbf{Mass ratios:} Figure~\ref{fig:mass_ratio} displays the mass ratios relative to both hot and cold water. The cold-to-hot water mass ratio is higher in Class I/FS sources (median $\log(M_{\rm cold}/M_{\rm hot})=1.30\pm0.42$) than in Class II sources (median $\log(M_{\rm cold}/M_{\rm hot})=0.63\pm0.32$), though the difference is only 1.4$\sigma$, primarily limited by small sample sizes. Simulating a larger population with the same mean and scatter as the observed sample, we estimate that approximately 8 additional detections per class would be needed for the median difference to reach 2$\sigma$ significance. When normalized to hot water, the mass ratios of HCN, C$_2$H$_2$, and CO$_2$ are similar between Class I/FS and Class II, indicating that the relative abundances of carbon-bearing species to hot water are not significantly different between the two evolutionary stages. In contrast, when normalized to cold water, the abundances of all other water components—as well as HCN, C$_2$H$_2$, and CO$_2$—are systematically lower in Class I/FS sources. This indicates that the chemical difference between the two classes is driven by an excess of cold water in Class I/FS disks, rather than by a deficit of carbon-bearing species.

\subsection{Correlations between molecular masses and disk properties}\label{sec:correlation_analysis}

Having established the molecular inventories of the two samples, we now ask which disk and stellar properties govern them. Because our samples are small, we cannot regress on all properties simultaneously without overfitting, so we proceed in two stages: we first use univariate censored correlation tests to identify which properties carry signal (\S\ref{sec:correlation}), then feed the dominant ones into a multivariable regression that isolates their independent contributions (\S\ref{sec:multivariate}).

\subsubsection{Univariate screening: censored Kendall $\tau$ tests}\label{sec:correlation}

For this screening we use a generalized Kendall $\tau$ test for censored data \citep{Brown1974,Isobe1986}, which incorporates upper limits rather than discarding non-detections, thereby retaining all sources and maximizing the available information (Appendix~\ref{sec:stat_methods}). We test dust disk size ($R_{\rm d,90\%}$), dust disk mass ($M_{\rm d}$), and accretion luminosity ($L_{\rm acc}$) for the Class~I/FS ($i<70^{\circ}$), Class~II, and combined samples; bolometric temperature ($T_{\rm bol}$) is tested for Class~I/FS only, and age and stellar luminosity ($L_\star$) for Class~II only, as these are the most reliable indicators available for each class.

\begin{figure*}
\centering
\vspace{-0.cm}
\includegraphics[width=\textwidth]{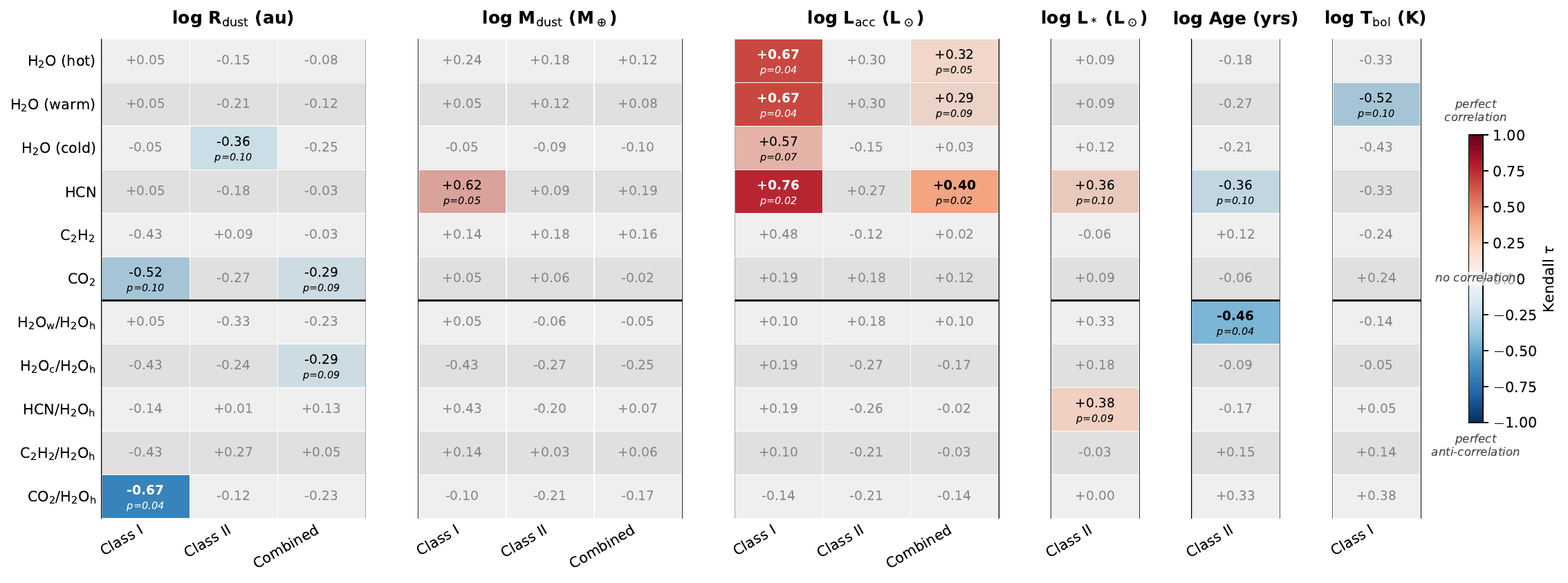}
\caption{ Generalized Kendall $\tau$ correlation coefficients between slab model molecular properties and disk/stellar properties. Each panel corresponds to a different disk/stellar property: dust radius ($R_{\rm dust}$), dust mass ($M_{\rm dust}$), accretion luminosity ($L_{\rm acc}$), stellar luminosity ($L_\star$), and stellar age. The first three panels show results for the Class I, Class II, and combined samples separately; the stellar luminosity and age panels show Class II only. Rows are divided into absolute slab model masses (top) and mass ratios relative to hot water (bottom). Cell color encodes the Kendall $\tau$ (blue = negative, red = positive), with the $p$-value annotated in each cell. Colored cells denote correlations significant at the $p < 0.05$ level (full saturation, bold text) or marginally significant at $p < 0.1$ (faded color); grey cells indicate no significant correlation. Upper limits on molecular masses and accretion luminosities are included as censored data. \label{fig:heatmap}
}
\vspace{-0.cm}
\end{figure*}

Given the small samples and the large number of property--molecule combinations tested, we control the false-discovery rate with a Benjamini--Hochberg (BH-FDR) correction \citep{Benjamini2018,Seabold2010}, labeling correlations that survive the correction ``robust'' and those that are only nominally significant ($p<0.1$) ``tentative'' (Appendix~\ref{sec:stat_methods}). All results below should be regarded as exploratory and sensitive to individual data points. Figure~\ref{fig:heatmap} summarizes the full set of censored Kendall $\tau$ results; we highlight the key findings here.

Among the properties tested, \textbf{accretion luminosity is the dominant positive correlate} of molecular mass. HCN, hot water, and warm water correlate significantly with $L_{\rm acc}$ in the Class~I/FS sample ($\tau = 0.67$--$0.76$, $p<0.04$) and remain the strongest correlations in the combined sample, while cold water, CO$_2$, and C$_2$H$_2$ show no significant $L_{\rm acc}$ dependence. None of the mass \emph{ratios} correlate with $L_{\rm acc}$, indicating that accretion luminosity scales the overall emission level rather than the relative composition, with no systematic offset between the two classes (Appendix Figure~\ref{fig:Mratio_Lacc}).

\textbf{Dust disk size is the dominant negative correlate.} CO$_2$ mass anti-correlates with $R_{\rm d}$ in both the Class~I/FS ($\tau=-0.52$, $p=0.1$) and combined ($\tau=-0.29$, $p=0.09$) samples, cold water mass is marginally anti-correlated with $R_{\rm d}$ in Class~II ($\tau=-0.36$, $p=0.1$), and the CO$_2$-to-hot-water ratio anti-correlates with disk size in Class~I/FS ($\tau=-0.67$, $p=0.04$). These trends echo earlier findings that compact disks show stronger water emission \citep{Banzatti2020,Banzatti2023,RomeroMirza2024}. Other properties yield only weak or isolated trends (e.g., $M_{\rm d}$ with HCN in Class~I/FS, and age and $L_\star$ in Class~II), shown in Appendix Figures~\ref{fig:M_Lacc}, \ref{fig:Tbol}, \ref{fig:water_mass_evolution}, and \ref{fig:M_Rd}. \emph{None} of the correlations survive BH-FDR correction, so all remain tentative.

This screening identifies $L_{\rm acc}$ and $R_{\rm d}$ as the two dominant properties, and---importantly---they act with \emph{opposite signs} ($L_{\rm acc}$ positive, $R_{\rm d}$ negative). Because the two predictors themselves co-vary, opposite-sign predictors can partially mask one another in univariate tests, biasing each apparent correlation toward zero. We therefore carry exactly these two properties into the multivariable regression of Section~\ref{sec:multivariate}, which isolates their independent contributions and, as shown below, sharpens the trends.

\subsubsection{Multivariate regression}\label{sec:multivariate}

The univariate tests in Section~\ref{sec:correlation} treat each property independently and, as noted, can understate opposite-sign predictors that co-vary. Guided by that screening, we carry the two dominant properties---disk size and accretion luminosity---into a multivariable weighted linear regression of the form
\begin{equation}\label{eq:multivar}
  \log M_{\rm slab} = a + b_{R_{\rm d}}\,\hat{z}_{\log R_{\rm d}}
                        + b_{L_{\rm acc}}\,\hat{z}_{\log L_{\rm acc}}\,,
\end{equation}
where $\hat{z}$ denotes standardized predictors. For a given property $x$ (e.g., $\log R_{\rm d}$), the standardized value for source $i$ is
\begin{equation}
  \hat{z}_{x,i} = \frac{x_i - \bar{x}}{\sigma_x}\,,
\end{equation}
where $\bar{x}$ and $\sigma_x$ are the sample mean and standard deviation of $x$, computed from the sources included in each fit. This standardization yields zero-mean, unit-variance predictors, so that the slopes $b_{R_{\rm d}}$ and $b_{L_{\rm acc}}$ are directly comparable in magnitude and each measures the partial effect of one predictor at a fixed value of the other. Standardization is performed independently for each sample (Class~II and Combined). We do not fit the Class~I/FS sample separately, as only four sources remain after excluding $L_{\rm acc}$ upper limits, too few for a two-predictor regression.

\begin{figure*}[!htbp]
\centering
\vspace{-0.cm}
\includegraphics[width=0.7\textwidth]{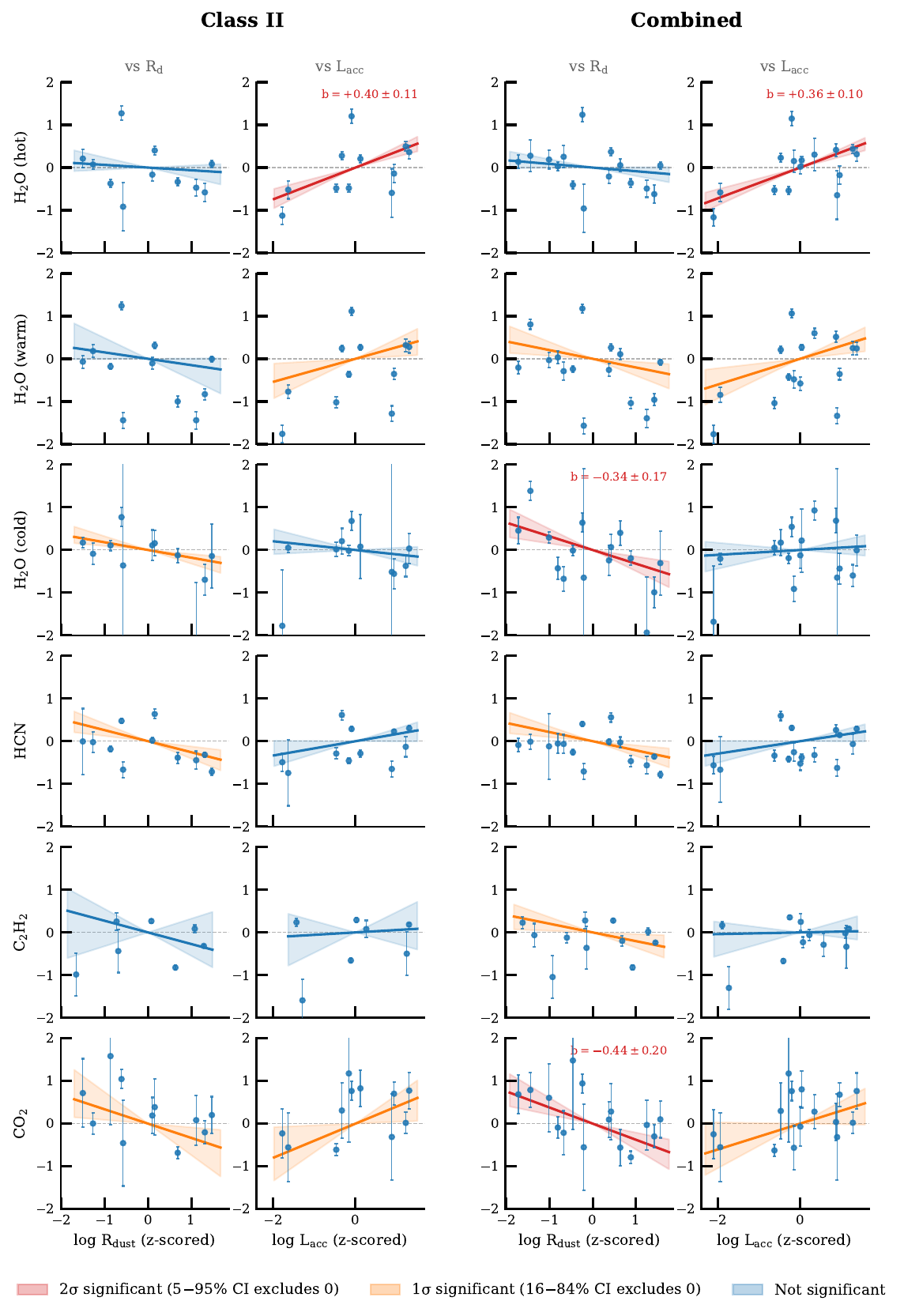}
\vspace{-0.cm}
\caption{Partial residuals from the multivariate regression of log slab mass on z-scored log $R_{\rm dust}$ and log $L_{\rm acc}$, for the Class~II (left) and Combined (right) samples. Rows correspond to the six molecular tracers and columns to the two predictors. The band color indicates the significance of the slope: red = strongly significant ($2\sigma$; 5--95\% CI excludes zero), orange = significant ($1\sigma$; 16--84\% CI excludes zero), blue = not significant. Uncertainties are estimated from 1500 bootstrap iterations with iterative weighted least squares incorporating measurement errors on both predictors and the response variable. The Class~I/FS sample is not shown separately because only four sources remain after excluding $L_{\rm acc}$ upper limits. \label{fig:partial_residuals}
}
\vspace{-0.cm}
\end{figure*}

We fit each molecule using iterative weighted least squares (WLS). In the first iteration the weights are $w_i = 1/\sigma_{y,i}^{2}$, where $\sigma_{y,i}$ is the symmetrized uncertainty on $\log M_{\rm slab}$. In subsequent iterations the effective uncertainty is inflated to account for predictor errors propagated through the current slopes:
\begin{equation}
  \sigma_{\rm eff,\,i}^{2} = \sigma_{y,i}^{2}
    + \bigl(b_{R_{\rm d}}\,\sigma_{\hat{z}_{R_{\rm d}},i}\bigr)^{2}
    + \bigl(b_{L_{\rm acc}}\,\sigma_{\hat{z}_{L_{\rm acc}},i}\bigr)^{2}\,.
\end{equation}
Convergence is reached after two iterations. Uncertainties on the slopes are estimated by bootstrap resampling (1500 draws with replacement); we report the 16th--84th percentile interval as the $1\sigma$ confidence interval (CI) and the 5th--95th percentile interval as the $\approx 2\sigma$ CI. A slope is considered significant if its $1\sigma$ CI excludes zero, and strongly significant if its $2\sigma$ CI excludes zero. Upper limits on both molecular masses and $L_{\rm acc}$ are excluded using the same detection flags as in the Kendall $\tau$ tests, and molecules with fewer than four detections in a given sample are omitted. Figure~\ref{fig:partial_residuals} summarizes the results for the Class~II and Combined samples (Appendix Figure~\ref{fig:heatmap_multivar} gives a compact heatmap view).

\paragraph{Class~II.}
In the Class~II sample, hot water mass is driven almost entirely by accretion luminosity ($b_{L_{\rm acc}} = 0.40$, $2\sigma$), with no significant dependence on disk size; warm water likewise depends only on $L_{\rm acc}$ ($b_{L_{\rm acc}} = 0.28$, $1\sigma$). CO$_2$ is the one species that depends on both predictors at the $1\sigma$ level ($b_{R_{\rm d}} = -0.39$, $b_{L_{\rm acc}} = 0.42$). In contrast, cold water ($b_{R_{\rm d}} = -0.16$) and HCN ($b_{R_{\rm d}} = -0.28$) anti-correlate with disk size while showing no significant $L_{\rm acc}$ dependence, and C$_2$H$_2$ is not significantly correlated with either predictor. The three water components thus separate cleanly: the hot and warm reservoirs are set by accretion luminosity, whereas cold water is governed by disk size rather than accretion rate, consistent with a supply mechanism linked to the radial extent of the dust disk.

\paragraph{Combined sample.}
With all 15 sources (after excluding $L_{\rm acc}$ upper limits), the trends sharpen. The dichotomy seen in Class~II persists: hot water is controlled by accretion luminosity ($b_{L_{\rm acc}} = 0.36$, $2\sigma$) and insensitive to disk size, while cold water ($b_{R_{\rm d}} = -0.34$, $2\sigma$) and CO$_2$ ($b_{R_{\rm d}} = -0.44$, $2\sigma$) are governed by disk size. Warm water, HCN, and C$_2$H$_2$ show weaker but significant ($1\sigma$) anti-correlations with disk radius, while warm water and CO$_2$ also retain a positive dependence on $L_{\rm acc}$ at the $1\sigma$ level.

\subsection{Summary of key results}
The main findings from our comparison of seven Class I/FS ($i<70^{\circ}$) and twelve Class II disks are:
\begin{enumerate}
    \item Disk inclination strongly affects molecular detections in Class I/FS sources: edge-on systems ($i>70^{\circ}$) show significantly suppressed line emission.
    \item After restricting to $i<70^{\circ}$, Class I/FS and Class II disks show similar detection rates for H$_2$O, HCN, C$_2$H$_2$, and CO$_2$, as well as similar line luminosity--$L_{\rm acc}$ trends.
    \item The three water components have comparable median excitation temperatures ($\sim$900, 460, 200\,K) in both classes. HCN and C$_2$H$_2$ also show similar median excitation temperatures between the two samples, while the median temperature of CO$_2$ is notably colder in Class I/FS (329 vs.\ 482\,K, 1.5$\sigma$).
    \item The cold-to-hot water mass ratio is higher in Class I/FS sources (median $\log(M_{\rm cold}/M_{\rm hot})=1.30$) than in Class II (median $\log(M_{\rm cold}/M_{\rm hot})=0.63$), though not statistically significant ($1.4\sigma$) given the small sample. When normalized to cold water, the mass ratios of all other water components as well as HCN, C$_2$H$_2$, and CO$_2$ are lower in Class I/FS than in Class II, suggesting that Class I/FS disks possess consistently higher cold water masses relative to their other molecular species.
    \item Censored Kendall $\tau$ tests, which properly incorporate upper limits on both molecular masses and accretion luminosity, identify accretion luminosity as the strongest correlate of molecular emitting mass, with HCN and hot water showing the strongest correlations in the combined sample, though none survive BH-FDR correction for multiple comparisons. Univariate tests also reveal a tentative anti-correlation between dust disk size and CO$_2$ mass in Class~I/FS. 
    \item Multivariate regression for the Class~II and combined samples confirms that compact disks tend to have higher cold water and CO$_2$ masses at a given accretion luminosity---these two species show the strongest ($2\sigma$) anti-correlations with disk size---whereas hot water mass is insensitive to disk size and is instead set by accretion luminosity. The remaining species fall between these poles: HCN also anti-correlates with disk size without an accretion dependence, while warm water and C$_2$H$_2$ show weaker ($1\sigma$) disk-size trends. This gradient points to different physical origins: the hot water component traces water vapor produced by gas-phase chemistry in the accretion-heated surface layers, while cold water and CO$_2$ may be largely replenished by sublimation of inward-drifting icy grains --- particularly in compact disks, where radial drift is expected to dominate the solid mass budget.
    
\end{enumerate}
\section{Discussion} \label{sec:discussion}

\subsection{Chemical Evolution from Embedded Disks to Class II Disks}

Given the growing evidence for early planet formation, it is critical to understand the chemical compositions of embedded disks and how they compare to the more evolved Class II disks. Keplerian rotating disks have been observed as early as the Class 0 stage and become common in Class I sources \citep{Murillo2013,Yen2014,Han2025}. Due to their generally higher accretion rates, embedded disks are expected to be warmer on average than Class II disks, with the mid-plane of the inner few au dominated by accretion heating.

Early observational evidence for hot gas in embedded disks came from broad $^{12}$CO rovibrational emission lines at 4.6\,$\mu$m, detected in 14 low-mass embedded sources, which was attributed to hot gas inside $\sim$0.5\,au \citep{Herczeg2011,Smith2015}. More recently, JWST/MIRI spectra of 15 Class 0 and 6 Class I sources from the JOYS and CORINOS programs have shown that water and CO$_2$ are often detected in absorption, while HCN and \ctht\ are rarely detected in either emission or absorption \citep[e.g.,][]{vanGelder2024,vanDishoeck2025}. When water emission is present, it is usually dominated by a cold $\sim$200\,K component, and CO$_2$ emission also has low excitation temperatures ($\sim$200\,K) \citep{vanDishoeck2025}. These characteristics suggest that the emission may primarily originate in relatively cold gas in the inner envelope or outflows, rather than from the hot inner disk.

Our Ophiuchus Class I/FS sample shows a markedly different pattern. Among the seven sources with $i<70^{\circ}$, water emission is detected in all of them (100\%), spanning three distinct temperature components, and the cold component does not dominate the emission. HCN, \ctht, and CO$_2$ are each detected in six out of seven (86\%). Notably, while the CO$_2$ detection rate is comparable to Class II, its excitation temperature is lower ($\sim$329 vs.\ 482\,K), suggesting different excitation conditions. HCN and \ctht\ show similar excitation temperatures to their Class II counterparts. A key distinction from earlier embedded-disk studies is that our Class I/FS sample has higher bolometric temperatures (200–750\,K), compared to $T_\mathrm{bol} < 200$\,K in those previous samples, suggesting that our targets are more evolved disks at the embedded stage.

Taken together, existing observations hint at a chemical evolution sequence, which is summarized in Table~\ref{tab:evolution}. Class 0 sources are generally molecular-line-poor (aside from H$_2$), possibly due to high dust opacity in the disk and/or surrounding envelope, as well as genuine molecular destruction at high temperatures. Some younger Class I sources begin to show more molecular emission, but with low excitation temperatures ($\sim$200\,K) and likely dominated by envelope or hot-corino contributions. In contrast, more evolved Class I/FS disks exhibit abundant water and organic emission with excitation temperatures and detection rates comparable to Class II disks, indicating that inner-disk chemistry is already well established at this stage. Together, these comparisons place the transition from envelope-dominated, molecule-poor (or cold-emission-dominated) spectra to Class~II-like inner-disk molecular emission at roughly $T_{\rm bol}\sim200$\,K. A key difference is that Class I/FS disks show elevated cold water content and colder CO$_2$ excitation temperatures relative to Class II, while the hot and warm water components and organic species are broadly similar between the two classes. We caution that this sequence is assembled from heterogeneous surveys (JOYS, CORINOS, this work, JDISCS) of different star-forming regions, so some variation may be environmental rather than evolutionary.

This trend could reflect a combination of factors: decreasing dust opacity and cooling temperatures in the inner disk surface, allowing molecules to survive and accumulate in the gas phase; enrichment from volatile-rich ices delivered by inward-drifting pebbles; and accretion-driven heating, as suggested by the strong positive correlations between molecular emitting masses and $L_{\rm acc}$. In the next section, we examine how predictions from pebble-drift models compare with the observed differences in water and organic content between our Class I/FS and Class II samples. 

\begin{deluxetable*}{lccc}
\tabletypesize{\footnotesize}
\tablecaption{Proposed chemical evolution of inner-disk molecular emission from embedded protostars to Class~II disks \label{tab:evolution}}
\tablewidth{0pt}
\tablehead{\colhead{Property} & \colhead{Class 0} & \colhead{Class I/FS (this work)} & \colhead{Class II}}
\startdata
$T_{\rm bol}$ & $<200$\,K & 200--750\,K & more evolved \\
Structure & envelope\,$+$\,outflow & disk established  & revealed disk \\
Inner-disk molecular emission & poor (mainly H$_2$) & H$_2$O, HCN, C$_2$H$_2$, CO$_2$ & H$_2$O, HCN, C$_2$H$_2$, CO$_2$ \& other organics \\
H$_2$O emission & Cold, envelope/outflow & 3 comp.(Hot, Warm, Cold) & 3 comp. \\
Cold water in the disk &  & elevated & reduced \\
CO$_2$ excitation & non-detection/absorption & $\sim$329\,K & $\sim$482\,K \\
Pebble drift to inner disk & not yet efficient & efficient & impeded by traps \\
\enddata
\tablecomments{Class~0 and youngest Class~I ($T_{\rm bol}<200$\,K) properties are compiled from \citet{vanGelder2024,vanDishoeck2025}; the Class~I/FS and Class~II columns are from this work. In Class~0, the cold-water emission might be dominated by the envelope/outflow rather than the disk. The sequence is assembled from heterogeneous surveys (JOYS, CORINOS, this work, JDISCS) and is qualitative; see Section~\ref{sec:discussion}.}
\end{deluxetable*}

\subsection{Excess of cold water and cold CO$_2$ in the context of pebble drift} 

Our comparison between the Class I/FS ($i<70^{\circ}$) and Class II samples reveals several notable trends:
(1) Class I/FS disks show elevated cold-to-hot water mass ratios, and relative to other species, the cold water mass in the Class I/FS sample is higher than that of Class II;
(2) CO$_2$ shows lower excitation temperatures compared to Class II disks, despite similar detection rates and temperatures of most species; and
(3) multivariate regression shows that cold water and CO$_2$ masses have the strongest anti-correlations with mm dust disk size, whereas hot water is insensitive to disk size and scales instead with accretion luminosity; cold water is governed by disk size alone, while CO$_2$ retains an additional positive dependence on $L_{\rm acc}$.
Here, we place these results in the context of current understanding of pebble drift during planet(esimal) formation.

Over the past few years, there has been growing interest in how the volatile compositions of the terrestrial planet-forming region (the inner few au) are shaped by the inward drift of icy pebbles. When these pebbles cross various snowlines, their icy mantles sublimate, releasing volatiles and enriching the inner disk with carbon- and oxygen-bearing species \citep[e.g.,][]{Oeberg2016,Booth2017,Booth2019}. Current pebble-evolution models predict that most inward drift occurs within the first Myr in the absence of strong pressure traps \citep[e.g.,][]{Birnstiel2012,Pinilla2012,Booth2019,Kalyaan2023,Houge2025,Williams2025,Krijt2025}. In such cases, water vapor abundances inside the snowline peak early, within the first Myr \citep[e.g.,][]{Kalyaan2023}. Because the CO$_2$ snowline lies farther out than the water snowline, enrichment of the innermost $\lesssim$1\,au region—where MIRI spectra are most sensitive—takes $\sim$1\,Myr, leading to a rapid evolution in the H$_2$O/CO$_2$ mass ratio and C/O elemental ratio inside the water snowline over this timescale \citep{Sellek2025}. These models therefore predict an early \textit{water-rich, CO$_2$-poor} phase in young disks, followed by gradual CO$_2$ enrichment and a rising C/O ratio. Consistent with the late-time end of this progression, JWST/MIRI observations of the evolved Upper Scorpius disks find water-depleted spectra with hints of elevated C/O \citep{Raul2026,Xie2026}.

Several of our findings are consistent with the predicted early water-rich phase. The elevated cold water mass in Class I/FS disks relative to other species compared to Class II sources is consistent with enhanced water ice delivery across the snowline during the earliest evolutionary stages. The lower CO$_2$ excitation temperatures in Class I/FS disks ($\sim$329 vs.\ 482\,K) suggest that CO$_2$ sublimation from drifting pebbles has not yet fully enriched the innermost disk. In this picture, the CO$_2$ emission in Class I/FS may be dominated by gas released near the CO$_2$ snowline at larger radii, where temperatures are lower, rather than from well within the water snowline as in Class II disks. As the disk evolves and pebble-delivered CO$_2$ accumulates closer in, the CO$_2$-emitting region would shift inward and the excitation temperature would rise, consistent with the higher CO$_2$ temperatures observed in Class II sources. The compact Class~II disk CX~Tau, in which CO$_2$ is abundant while H$_2$O is comparatively weak, may represent an advanced stage of this progression \citep{Vlasblom_2025_CXTau}. Furthermore, the mass ratios of carbon-bearing species (HCN, C$_2$H$_2$) to water are lower in Class I/FS sources when normalized to cold water, qualitatively consistent with a lower C/O ratio during this water-dominated phase \citep[e.g.,][]{Oeberg2016,Booth2019,Sellek2025}. However, translating slab model masses into robust abundances or C/O constraints requires detailed thermo-chemical modeling that is beyond the scope of this work \citep{Anderson2021,Arabhavi2026, Esteve2026}. Recent analyses that apply LTE H$_2$O slab retrievals to detailed thermochemical disk models---using different disk codes---provide a useful benchmark for how the slab-derived water abundances relate to the underlying disk values \citep{Kaeufer_2024,Vlasblom_2025b}.

\begin{figure*}[!htbp]
\centering
\vspace{-0.cm}
\includegraphics[width=0.9\textwidth]{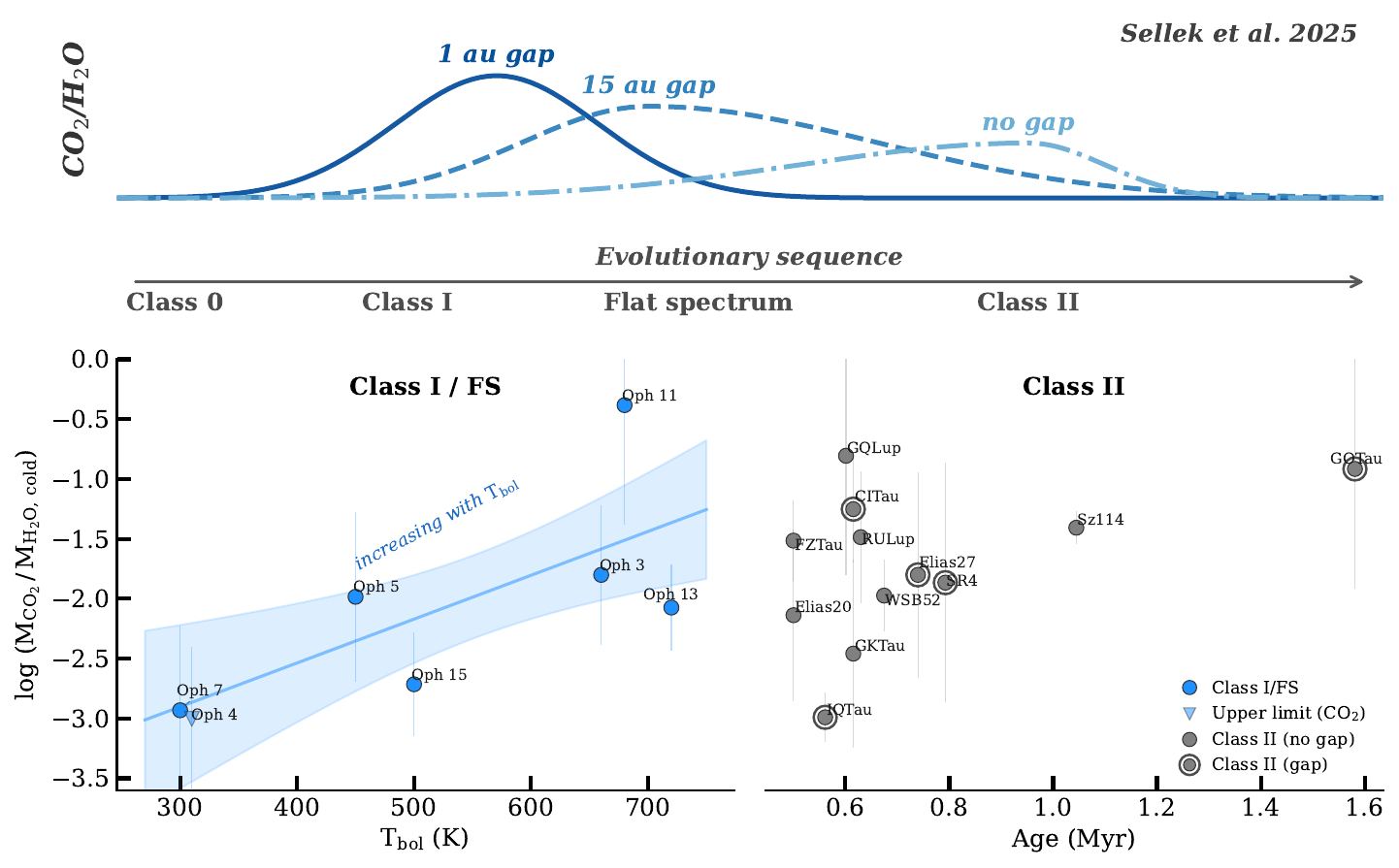}
\vspace{-0.2cm}
\caption{The ratio of observable masses of CO$_2$ and H$_2$O(cold) as a function of bolometric temperature, an age indicator for Class I/FS and isochrone age for Class II. Blue points are Class I/FS detections; downward triangles indicate CO$_2$ upper limits. Grey points are Class II sources, with circled symbols indicating disks with known dust gaps. The top panel shows the predicted CO$_2$/H$_2$O mass ratio evolution from the pebble drift models of \citet{Sellek2025} for disks with no gap, a 15\,au gap, and a 1\,au gap. The Class I/FS sources show a tentative increasing trend of CO$_2$/H$_2$O ratios with bolometric temperature, consistent with the predicted early water-rich phase. \label{fig:CO2_H2O_ratio}
}
\vspace{-0.cm}
\end{figure*}

Figure~\ref{fig:CO2_H2O_ratio} compares the observed CO$_2$/H$_2$O(cold) mass ratio against bolometric temperature, alongside the evolutionary tracks predicted by the pebble drift models of \citet{Sellek2025}. The Class~I/FS sources exhibit a tentative upward trend in CO$_2$/H$_2$O(cold) with T$_{\rm bol}$, qualitatively consistent with the progressive CO$_2$ enrichment expected as icy pebbles drift inward and sequentially cross the H$_2$O and CO$_2$ snowlines. In the gap-free model, this enrichment sets in early, whereas the presence of a large gap (e.g., 15\,au) delays CO$_2$ delivery by trapping pebbles beyond the gap. The current data are not sufficient to distinguish between these scenarios. A larger sample of Class~I/FS detections will be needed to distinguish whether the observed trend reflects unimpeded drift or delayed transport past a gap, and to confirm the predicted early water-rich phase in young disks.

These Class~I/FS trends connect to a growing body of evidence in Class~II disks. Spitzer/IRS studies first showed that compact Class~II disks tend to have stronger water line fluxes \citep{Banzatti2020}, and JWST/MIRI has since revealed that this excess is concentrated in the cold ($<$400\,K) water component \citep[e.g.,][]{Banzatti2023,RomeroMirza2024,Banzatti2025}, qualitatively consistent with pebble drift delivering ices more efficiently in compact, drift-dominated disks. The picture is not universal, however: some compact disks lack strong cold-water emission, and dust gaps do not always suppress H$_2$O \citep{Gasman2025,Temmink2025,Arulanantham2025}. Theoretical models likewise show that the observable signature of pebble drift depends on disk thermal structure, the location and depth of substructures, and disk age \citep{Kalyaan2021,Kalyaan2023,Easterwood2024,Sellek2025,Krijt2025}.

The relationship between CO$_2$ mass and dust disk size provides additional support. The univariate censored Kendall $\tau$ test shows a tentative anti-correlation in the Class~I/FS sample, and in the combined sample the multivariate regression confirms that CO$_2$ has the strongest disk-size dependence of all species ($b_{R_{\rm d}} = -0.44$, $2\sigma$), even after controlling for accretion luminosity. This is consistent with compact disks experiencing more efficient pebble drift and thus greater CO$_2$ delivery to the inner disk.

Similarly, the multivariate regression reveals that cold water mass in Class~II disks is governed by dust disk radius rather than accretion luminosity ($b_{R_{\rm d}}$ significant, $b_{L_{\rm acc}}$ not). This is consistent with pebble drift expectations: the cold water reservoir inside the snowline is replenished by sublimation of inward-drifting icy pebbles, and compact disks---which are less likely to harbor pressure traps capable of intercepting the pebble flux---are expected to deliver more ice to the inner disk. In contrast, hot water remains insensitive to disk size in both classes, consistent with these components tracing thermally produced gas in the innermost disk whose mass scales primarily with accretion luminosity.

\begin{figure*}[!htbp]
\centering
\vspace{-0.cm}
\includegraphics[width=0.8\textwidth]{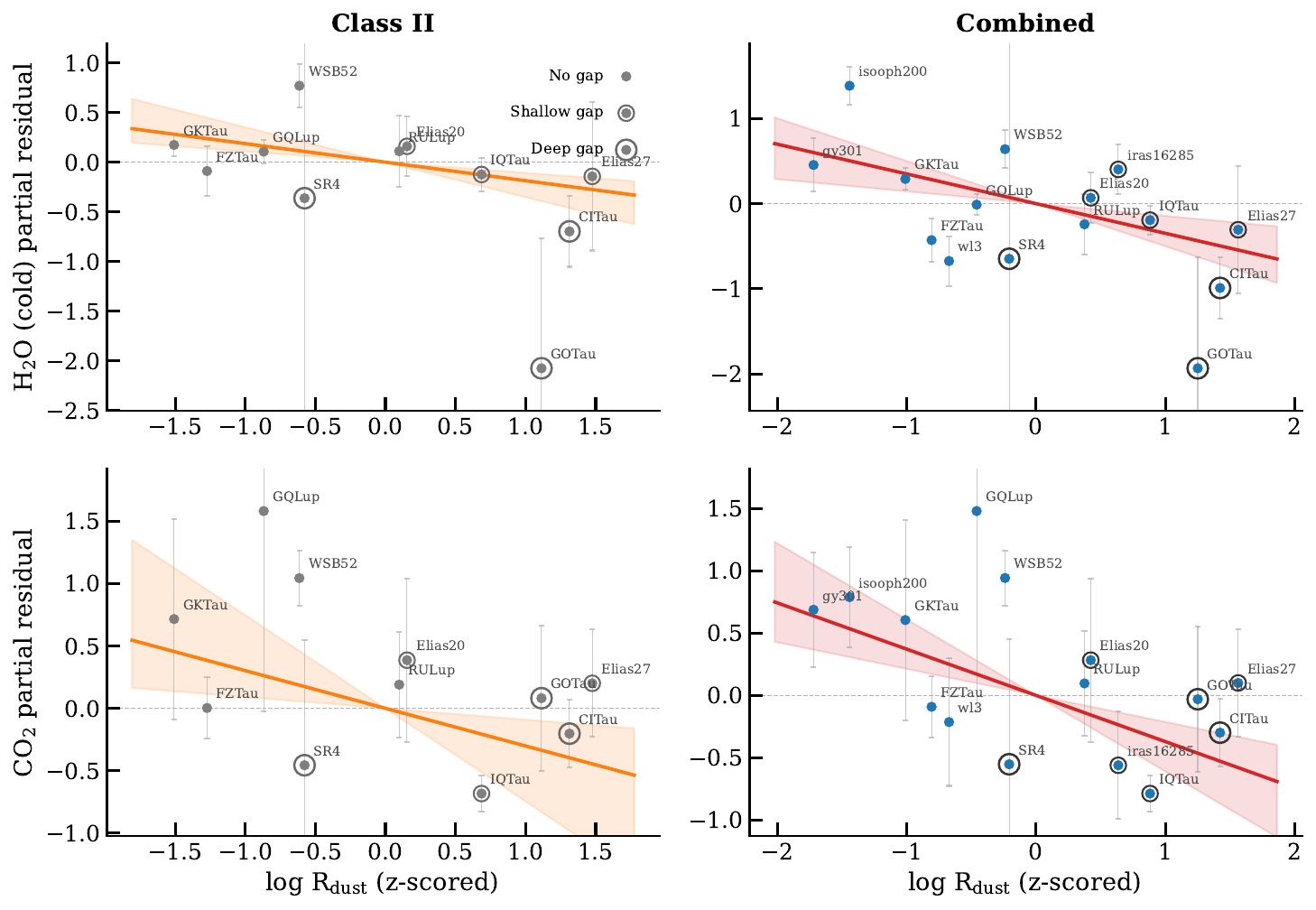}
\vspace{-0.2cm}
\caption{Partial residuals of cold water (top) and CO$_2$ (bottom) slab mass against z-scored log $R_{\rm dust}$, after removing the $L_{\rm acc}$ dependence from the multivariate regression. Panels show  Class~II (left) and the combined sample (right). Symbol styles indicate disk substructure: filled circles for disks with no known gaps, single-ring symbols for shallow gaps, and double-ring symbols for deep gaps. In the Class~II cold water panel, the three deep-gap sources (GO~Tau, CI~Tau, SR~4) fall below the general trend, suggesting that dust gaps may suppress cold water delivery. This suppression is not evident in the CO$_2$ residuals. \label{fig:partial_Rd}
}
\vspace{-0.cm}
\end{figure*}

Figure~\ref{fig:partial_Rd} shows the partial residuals of cold water and CO$_2$ mass against dust disk radius (after removing the $L_{\rm acc}$ dependence) for Class~II and the combined sample. Among the Class~II sources, the three disks with deep gaps (SR 4, CI~Tau, and GO~Tau) fall notably below the cold water trend \citep{Huang2018,Clarke2018_CITau,Zhang2023}, suggesting that their dust gaps may be trapping pebbles and suppressing water ice delivery to the inner disk \citep{Banzatti2023,Mah2024,Sellek2025,Krijt2025}. Intriguingly, this suppression is not seen in the CO$_2$ partial residuals, where the deep-gap sources scatter around the trend. This difference may reflect the distinct snowline locations: if the dust gaps lie between the water and CO$_2$ snowlines, they would preferentially block water ice delivery while allowing CO$_2$-bearing pebbles from beyond the gap to continue drifting inward. Notably, all three deep-gap sources have close-in inner gaps (at $\sim$11, 11.8, and 17\,au for SR~4, CI~Tau, and GO~Tau, respectively; \citealt{Krijt2025}), consistent with the prediction that gaps must be located close to the water snowline to efficiently suppress pebble-delivered cold water \citep{Krijt2025}. The different behavior of H$_2$O and CO$_2$ in disks with gaps needs to be confirmed in a larger sample.

The role of dust gaps in modulating pebble drift connects to the broader question of when and how disk substructures emerge during protostellar evolution. Recent high-resolution millimeter surveys paint a consistent picture: substructures such as rings and gaps are rare in the most deeply embedded Class~0 sources, begin to appear during the Class~I phase, and become prevalent in Class~II disks \citep[e.g.,][]{Ohashi2023,Andrews2018a,Cieza2021,Hsieh2025}. This evolutionary trend suggests that the transition from Class~I to Class~II coincides with the emergence of pressure traps that can increasingly impede pebble drift. In this picture, the elevated cold water content we observe in Class~I/FS disks may reflect a brief evolutionary window in which pebble drift is still efficient---before substructures have fully developed to intercept the inward flow of icy solids.

\subsection{Alternative explanations for the cold water excess}
While the pebble drift framework provides a coherent explanation for several of our findings, alternative explanations for the cold water excess should be considered. Class~I/FS disks are expected to retain more vertically extended, flared geometries than settled Class~II disks, which would expose a larger emitting surface area at the radii where cold water lines originate ($\gtrsim$1\,au). This geometric effect could enhance the observed cold water flux without requiring a higher column of water. Its magnitude relative to the hot and warm components remains to be quantified with radiative transfer modeling. More generally, because the slab masses are \emph{observable} masses, a more extended or flared emitting layer would raise the inferred mass even at fixed true column density; recovering the true gas mass requires thermochemical modeling \citep{Kaeufer_2024,Vlasblom_2025b}. More importantly, the flaring explanation would predict stronger cold water excess in larger, more flared disks, whereas we observe the opposite trend: compact disks show the highest cold water masses.

A different thermal structure in embedded disks---e.g., higher accretion rates pushing the snowline outward---could also create a larger cold water reservoir. However, the accretion luminosities of our Class~I/FS sample are comparable to, and in some cases lower than, those of the Class~II sample, suggesting similar thermal conditions in the inner disk and arguing against a purely thermal origin. Moreover, the multivariate regression shows that cold water mass does not correlate with $L_{\rm acc}$ in Class~II but does anti-correlate with dust disk radius, which is difficult to reconcile with a purely thermal explanation but naturally expected from pebble drift.

Finally, some cold water emission could originate from the envelope or outflow rather than the disk. As described in Section~\ref{subsec:miri_obs}, our continuum-subtracted line images (Appendix Figure~\ref{fig:line_images}) show that the water emission is spatially unresolved in all sources, while some sources exhibit extended H$_2$ and jet-like [Ne\,II] emission indicative of active outflows. Crucially, the strength of the cold water excess does not correlate with the presence or morphology of these outflow tracers, arguing against a significant outflow contribution to the water emission. While this is consistent with a disk origin, we cannot rule out contributions from a compact (sub-arcsecond) outflow on scales below the MIRI/MRS PSF. However, an envelope or outflow origin would not naturally explain the observed anti-correlation between cold water mass and mm dust disk size, which instead points to a disk-based process such as pebble drift.

Perhaps the most important caveat is that the disk-size distributions of the two samples differ significantly (Section~\ref{subsec:sample}): the Class~II sample from JDISCS GO~1 is biased toward mm-bright, large disks (median $R_{\rm dust}\approx54$\,au), whereas our Class~I/FS sample, selected to span a broader range of mm luminosities, has systematically smaller dust disks (median $R_{\rm dust}\approx13$\,au). Beyond this selection effect, Class~0/I disks are known to be intrinsically smaller than Class~II disks \citep{Tobin2020}, although whether this reflects the timescale for dust growth at outer disk radii or gradual disk spreading remains uncertain. Because the multivariate regression shows that cold water mass anti-correlates with $R_{\rm dust}$ ($b_{R_{\rm dust}}=-0.34$ at 2$\sigma$ in the combined sample; Section~\ref{sec:multivariate}), some or all of the apparent cold water excess in Class~I/FS could simply reflect their smaller disk sizes rather than a distinct chemical evolutionary stage. Disentangling disk-size effects from evolutionary effects will require better overlap in disk-size distributions between the two samples: a Class~II comparison sample that includes compact disks comparable in size to the Class~I/FS population, and a Class~I/FS sample that extends to larger, more massive disks.

\subsection{Limitations and future directions}

Our conclusions are subject to several limitations. The samples are small---seven Class~I/FS sources with $i<70^{\circ}$ and twelve Class~II disks, drawn from a single star-forming region (Ophiuchus) and the public JDISCS GO~1 data, respectively---which limits the statistical power and makes it difficult to separate evolutionary effects from initial conditions. As already noted (Section~\ref{subsec:classII_sample}), the Class~II sample is also biased toward mm-bright, larger disks and younger ages ($<$1\,Myr), so we cannot yet probe the older ($>$3\,Myr) regime where pebble-drift reduction may be most pronounced. Recent JWST/MIRI surveys of the more evolved ($\sim$5--10\,Myr) Upper Scorpius disks find fading molecular emission and water depletion relative to younger regions \citep{Raul2026,Xie2026}, consistent with a declining pebble flux at late times; extending our comparison to such ages is a natural next step.

The stellar mass distributions of the two samples may also not be fully matched. While the ten AGE-PRO sources (Oph~1--10) span 0.3--0.65\,M$_\odot$, stellar spectral types for Oph~11--15 suggest they are lower-mass objects (M4--M5, $M_\star\sim$0.07--0.18\,M$_\odot$; \citealt{McClure2010}, Ruiz-Rodriguez et al.\ in prep), three of which (Oph~11, 13, 15) are in the $i<70^{\circ}$ subsample used for the Class~I/FS vs.\ Class~II comparison. The Class~II sample does include one comparably low-mass source (Sz~114, M5, $M_\star = 0.17$\,M$_\odot$), but the Class~I/FS subsample contains a higher fraction of very low-mass stars. If confirmed, this stellar mass imbalance could affect the comparison, as lower-mass stars may have systematically different accretion luminosities and disk thermal structures. However, the robust detection of water and organics in these low-mass Class~I/FS sources is itself noteworthy.

Finally, key stellar properties such as stellar mass and luminosity are not reliably measured for most Class~I/FS sources, preventing us from testing correlations that have proven important in Class~II studies \citep[e.g.,][]{Banzatti2025,Grant2025}, and the HI-based accretion luminosities carry additional systematic uncertainty, as the conversion has not been calibrated for embedded systems.

Future observational progress will require expanding the sample of Class I/FS disks observed with JWST/MIRI to cover a broader range of bolometric temperatures and star-forming environments, and extending the Class II sample to older ages ($>$3\,Myr) and a wider range of millimeter continuum luminosities. Surveys across multiple star-forming regions will be essential to average over environmental effects and test whether the trends identified here are universal. Spatially resolved observations (e.g., with ALMA at high angular resolution and longer wavelengths) of the Class I/FS disk structures would help establish whether substructures are already present and potentially trapping pebbles at these early stages. 

On the modeling side, coupling time-dependent pebble drift with thermo-chemical evolution and radiative transfer will enable quantitative predictions of line fluxes and excitation temperatures, allowing direct comparison with the slab model parameters measured here. Such efforts will help distinguish between pebble-drift-driven and opacity-driven explanations for the observed trends in water and CO$_2$, and test whether the predicted C/O ratio evolution is consistent with the observed differences in carbon-bearing species between Class I/FS and Class II disks.

\section{Conclusion}

We present the first detailed chemical survey of Class I/FS disks with JWST/MIRI. We analyze MIRI spectra of 16 Class I/FS disks in the Ophiuchus star-forming region using both empirical measurements and slab modeling, and compare their molecular properties with twelve Class II disks within a similar stellar mass and accretion luminosity range. Our main findings are:

\begin{itemize}
    \item Disk inclination strongly affects molecular detections in Class I/FS sources: edge-on systems ($i>70^{\circ}$) show markedly suppressed line emission, likely due to obscuration by the optically thick outer disk and residual envelope. This geometric effect must be accounted for when interpreting molecular inventories of disks.

    \item Class I/FS sources with favorable viewing angle ($i<70^{\circ}$) demonstrate that inner-disk chemistry is already well established: water, HCN, C$_2$H$_2$, and CO$_2$ are commonly detected at rates and excitation conditions broadly comparable to Class II disks. This places the onset of chemically rich inner-disk environments earlier than previously characterized, bridging the gap between the molecular-emission-poor Class 0 stage and the molecule-rich Class II phase.

    \item Class I/FS disks show tentatively elevated cold water relative to other molecular species and to Class II disks, though the small sample sizes preclude statistical significance. CO$_2$ similarly exhibits lower excitation temperatures, consistent with its emitting region not yet having shifted inward to the warm innermost disk, though this trend also requires confirmation with larger samples.

  \item Correlation analyses identify accretion luminosity and dust disk size as the two dominant drivers of molecular mass, acting with opposite signs. The two predictors separate the species along a gradient: hot water scales with accretion luminosity and is insensitive to disk size, whereas cold water and CO$_2$ are most strongly governed by disk size, with compact disks hosting more of both; HCN likewise anti-correlates with disk size, and warm water and C$_2$H$_2$ show weaker disk-size trends. This points to a dichotomy in physical origins: the hot water reservoir is produced by gas-phase chemistry in the accretion-heated inner disk atmosphere, while cold water and CO$_2$ are replenished by the inward drift and sublimation of icy pebbles.

    \item These patterns can be qualitatively understood in the framework of pebble drift models \citep[e.g.,][]{Kalyaan2023,Sellek2025}, in which inward-drifting icy pebbles first enrich the inner disk in water as they cross the snow line, while CO$_2$ delivery lags because its snow line lies further out. The observed CO$_2$/H$_2$O mass ratio shows a tentative increasing trend with bolometric temperature among Class I/FS sources, in line with the predicted progressive CO$_2$ enrichment as the system evolves. Among Class II disks, sources with deep dust gaps appear to have suppressed cold water delivery while CO$_2$ remains unaffected, possibly because the gaps lie between the two snow lines---though this remains tentative given the small number of gap sources.
\end{itemize}

Expanding the Class I/FS sample across multiple star-forming regions, extending the Class II comparison to older ages, and coupling pebble drift models with thermo-chemical evolution will be essential to test whether this evolutionary picture is universal.

\begin{acknowledgments}
This work is based on observations made with the
NASA/ESA/CSA James Webb Space Telescope. The
data were obtained from the Mikulski Archive for Space
Telescopes at the Space Telescope Science Institute,
which is operated by the Association of Universities
for Research in Astronomy, Inc., under NASA contract
NAS 5-03127 for JWST. These observations are associated with JWST GO program ID 3034 and 7135 (PI: K. Zhang). Supports for K.Z., A.W., and E.R. through these two programs were provided by NASA through a grant from
the Space Telescope Science Institute, which is operated by the Association of Universities for Research in Astronomy, Inc., under NASA contract NAS 5-03127. SK and TK acknowledge support from STFC Grant ST/Y002415/1. JW is funded by the UK Science and Technology Facilities Council (STFC), grant code ST/Y509383/1. L.A.C is funded by ANID through the  Millennium Science Initiative Program - Center Code NCN2024\_001 and the Fondecyt grant 1241056.

The JWST data presented in this article were obtained from the Mikulski Archive for Space Telescopes (MAST) at the Space Telescope Science Institute. The specific observations analyzed can be accessed via \dataset[doi: 10.17909/21qc-9951]{https://doi.org/10.17909/21qc-9951}.

\end{acknowledgments}

\vspace{5mm}
\facilities{JWST(MIRI)}

\software{
  astropy \citep{AstropyCollaboration2022,AstropyCollaboration2018,AstropyCollaboration2013},
  emcee \citep{ForemanMackey2013},
  spectools\_ir \citep{Salyk2022},
  ctool \citep{Pontoppidan2026},
  iSLAT \citep{Jellison2024},
  ysoisochrone \citep{Deng2025a},
  lifelines \citep{DavidsonPilon2024},
  statsmodels \citep{Seabold2010},
  matplotlib \citep{Hunter2007},
}

\appendix

\section{Observational Log}

\begin{deluxetable*}{lcccr}
\tablecaption{Observational Log for Ophiuchus Sources \label{tab:obs_log}}
\tablehead{
\colhead{Source} & \colhead{Program} & \colhead{DATE-BEG} & \colhead{Visit ID} & \colhead{Exposure (s)}}
\startdata
Oph 1 & 3034 & 2024-03-16 & 03034001001 & 455 \\
Oph 2 & 3034 & 2024-03-16 & 03034002001 & 1520 \\
Oph 3 & 3034 & 2024-04-01 & 03034003001 & 677 \\
Oph 4 & 3034 & 2024-04-01 & 03034004001 & 688 \\
Oph 5 & 3034 & 2024-04-01 & 03034005001 & 688 \\
Oph 6 & 3034 & 2024-04-01 & 03034006001 & 688 \\
Oph 7 & 3034 & 2024-04-01 & 03034007001 & 355 \\
Oph 8 & 3034 & 2024-03-16 & 03034008001 & 1354 \\
Oph 9 & 3034 & 2023-08-16 & 03034009001 & 677 \\
Oph 10 & 3034 & 2024-04-01 & 03034010001 & 677 \\
Oph 11 & 7135 & 2025-08-18 & 07135003001 & 1354 \\
Oph 12 & 7135 & 2025-08-18 & 07135001001 & 1354 \\
Oph 13 & 7135 & 2025-08-05 & 07135017001 & 677 \\
Oph 14 & 7135 & 2025-08-18 & 07135006001 & 1187 \\
Oph 15 & 7135 & 2025-08-18 & 07135031001 & 455 \\
Oph 16 & 7135 & 2025-08-18 & 07135032001 & 455
\enddata
\end{deluxetable*}

\section{Line luminosities}\label{sec:lineflux_appendix}

Table~\ref{tab:lineflux} lists the line luminosities of the Class~I/FS sample, and Table~\ref{tab:lineflux_classII} lists those of the Class~II comparison sample used in Figure~\ref{fig:flux_Lacc}.

\begin{deluxetable*}{lccccccccc}
\tablecaption{Line luminosities of Class I/FS sample\label{tab:lineflux}}
\tablewidth{0pt}
\tablehead{
\colhead{Source} & \colhead{$L_{\rm H_2O,Hot}$} & \colhead{$L_{\rm H_2O,Warm}$} & \colhead{$L_{\rm H_2O,Cold}$} & \colhead{$L_{\rm HCN}$} & \colhead{$L_{\rm C_2H_2}$} & \colhead{$L_{\rm CO_2}$} & \colhead{$L_{\rm HI(10-7)}$} & \colhead{$L_{\rm acc}$} & \colhead{$n_{13-26}$} \\
\colhead{} & \colhead{$17.3\mu$m} & \colhead{$17.5\mu$m} & \colhead{$23.8$--$23.9\mu$m} & \colhead{$13.5$--$13.7\mu$m} & \colhead{$13.8$--$14.0\mu$m} & \colhead{$14.8$--$15.0\mu$m} & \colhead{$8.76\mu$m} & \colhead{} & \colhead{} \\
\colhead{} & \colhead{(log$L_\odot$)} & \colhead{(log$L_\odot$)} & \colhead{(log$L_\odot$)} & \colhead{(log$L_\odot$)} & \colhead{(log$L_\odot$)} & \colhead{(log$L_\odot$)} & \colhead{(log$L_\odot$)} & \colhead{(log$L_\odot$)} & \colhead{}
}
\startdata
\multicolumn{10}{c}{\textit{$i < 70^{\circ}$}} \\
Oph 3& $-5.59 \pm 0.01$& $-5.29 \pm 0.01$& $-4.99 \pm 0.04$& $-4.17$& $-4.32$& $-5.01$& $-5.51 \pm 0.06$& $-0.74 \pm 0.08$& $-0.35$ \\
Oph 4& $-6.28 \pm 0.12$& $-5.74 \pm 0.03$& $-5.20 \pm 0.08$& $< -5.44$& $< -5.37$& $< -5.77$& $< -5.80$& $< -1.01$& $-1.27$ \\
Oph 5& $-5.41 \pm 0.03$& $-5.22 \pm 0.03$& $-4.49 \pm 0.01$& $-4.36$& $-4.01$& $-4.75$& $-5.42 \pm 0.10$& $-0.66 \pm 0.11$& $0.08$ \\
Oph 7& $-5.39 \pm 0.08$& $-5.18 \pm 0.03$& $-4.90 \pm 0.08$& $-4.03$& $-4.29$& $-4.95$& $-4.96 \pm 0.06$& $-0.24 \pm 0.08$& $0.12$ \\
Oph 11& $-6.30 \pm 0.04$& $-5.85 \pm 0.01$& $-5.35 \pm 0.04$& $-4.80$& $-4.92$& $-5.78$& $< -6.63$& $< -1.78$& $-0.68$ \\
Oph 13& $-6.32 \pm 0.08$& $-5.99 \pm 0.06$& $-5.08 \pm 0.02$& $-4.90$& $-5.21$& $-5.32$& $< -7.08$& $< -2.19$& $-0.36$ \\
Oph 15& $-5.60 \pm 0.08$& $-5.12 \pm 0.03$& $-4.30 \pm 0.01$& $-4.34$& $-4.24$& $-4.50$& $-5.24 \pm 0.07$& $-0.50 \pm 0.08$& $-0.22$ \\
\hline
\multicolumn{10}{c}{\textit{$i > 70^{\circ}$}} \\
Oph 1& $-5.80 \pm 0.04$& $-5.15 \pm 0.02$& $-4.28 \pm 0.01$& absorption& absorption& absorption& $< -5.53$& $< -0.76$& $-0.67$ \\
Oph 2& $-7.39 \pm 0.08$& $< -7.73$& $< -7.14$& $< -7.07$& $< -7.01$& $< -7.40$& $< -7.15$& $< -2.25$& $0.90$ \\
Oph 6& $< -7.21$& $-6.49 \pm 0.06$& $-5.40 \pm 0.03$& $-5.43$& $-5.43$& $< -6.67$& $< -7.09$& $< -2.20$& $1.13$ \\
Oph 8& $< -7.82$& $< -12.97$& $< -6.40$& $< -6.91$& $< -6.85$& $< -7.25$& $< -7.41$& $< -2.49$& $1.61$ \\
Oph 9& $< -7.39$& $< -7.09$& $<-5.96$& $< -6.42$& $< -6.36$& $< -6.75$& $< -7.11$& $< -2.22$& $1.02$ \\
Oph 10& $-5.82 \pm 0.05$& $-5.60 \pm 0.02$& $-5.20 \pm 0.10$& $-4.56$& $-4.66$& $-5.32$& $-5.27 \pm 0.03$& $-0.52 \pm 0.06$& $-0.45$ \\
Oph 16& $<-5.32 $& $<-5.23$& $<-4.54$& $< -4.68$& $< -4.62$& absorption& $< -5.46$& $< -0.70$& $0.22$ \\
\hline
\multicolumn{10}{c}{\textit{Inclination unknown}} \\
Oph 12& $-6.57 \pm 0.06$& $-5.96 \pm 0.03$& $-5.22 \pm 0.06$& $< -6.47$& $< -6.41$& $-5.45$& $< -6.47$& $< -1.62$& $-0.84$ \\
Oph 14& $-5.86 \pm 0.02$& $-5.53 \pm 0.01$& $-5.13 \pm 0.03$& $< -5.50$& $< -5.44$& $-5.40$& $-6.21 \pm 0.09$& $-1.38 \pm 0.12$& $-0.71$ \\
\enddata
\tablecomments{Upper limits are indicated with $<$. Sources showing ``absorption'' have absorption features instead of emission (Oph 1: HCN, C$_2$H$_2$, CO$_2$; Oph 16: CO$_2$). HCN, C$_2$H$_2$, and CO$_2$ luminosities are derived from slab models and do not have individual measurement uncertainties. $L_{\rm HI(10-7)}$ is the luminosity of the H\,\textsc{i} (10--7) line at 8.76\,$\mu$m, measured after subtracting the best-fitting water slab model to remove blended H$_2$O emission at this wavelength. $L_{\rm acc}$ is derived from $L_{\rm HI(10-7)}$ (Section~\ref{sec:slab}); we tabulate the line luminosity so that the accretion luminosities can be recomputed with any adopted H\,\textsc{i}--$L_{\rm acc}$ calibration.}
\end{deluxetable*}

\begin{deluxetable*}{lcccccccc}
\tablecaption{Line luminosities of the Class II sample used in Figure~\ref{fig:flux_Lacc}\label{tab:lineflux_classII}}
\tablewidth{0pt}
\tablehead{
\colhead{Source} & \colhead{$L_{\rm H_2O,Hot}$} & \colhead{$L_{\rm H_2O,Warm}$} & \colhead{$L_{\rm H_2O,Cold}$} & \colhead{$L_{\rm HCN}$} & \colhead{$L_{\rm C_2H_2}$} & \colhead{$L_{\rm CO_2}$} & \colhead{$L_{\rm HI(10-7)}$} & \colhead{$L_{\rm acc}$} \\
\colhead{} & \colhead{$17.3\mu$m} & \colhead{$17.5\mu$m} & \colhead{$23.8$--$23.9\mu$m} & \colhead{$13.5$--$13.7\mu$m} & \colhead{$13.8$--$14.0\mu$m} & \colhead{$14.8$--$15.0\mu$m} & \colhead{$8.76\mu$m} & \colhead{} \\
\colhead{} & \colhead{(log$L_\odot$)} & \colhead{(log$L_\odot$)} & \colhead{(log$L_\odot$)} & \colhead{(log$L_\odot$)} & \colhead{(log$L_\odot$)} & \colhead{(log$L_\odot$)} & \colhead{(log$L_\odot$)} & \colhead{(log$L_\odot$)}
}
\startdata
CI Tau& $-5.37 \pm 0.02$& $-5.25 \pm 0.01$& $-5.03 \pm 0.01$& $-4.25$& $-4.30$& $-4.85$& $-4.91 \pm 0.01$& $-0.19 \pm 0.08$ \\
GQ Lup& $-5.68 \pm 0.02$& $-5.25 \pm 0.01$& $-4.59 \pm 0.01$& $-4.56$& $< -5.62$& $-5.22$& $-5.58 \pm 0.06$& $-0.80 \pm 0.09$ \\
IQ Tau& $-5.75 \pm 0.02$& $-5.65 \pm 0.01$& $-5.23 \pm 0.01$& $-4.57$& $-4.99$& $-5.38$& $-5.76 \pm 0.04$& $-0.97 \pm 0.09$ \\
RU Lup& $-4.92 \pm 0.03$& $-4.69 \pm 0.02$& $-4.46 \pm 0.02$& $-3.84$& $< -4.65$& $-4.44$& $-4.69 \pm 0.03$& $0.01 \pm 0.10$ \\
FZ Tau& $-5.09 \pm 0.02$& $-4.82 \pm 0.02$& $-4.58 \pm 0.01$& $-4.08$& $< -4.76$& $-4.36$& $-4.74 \pm 0.03$& $-0.04 \pm 0.10$ \\
Elias 27& $-5.60 \pm 0.01$& $-5.28 \pm 0.01$& $-4.93 \pm 0.01$& $-4.71$& $< -5.36$& $-4.86$& $-5.41 \pm 0.03$& $-0.65 \pm 0.08$ \\
GO Tau& $-6.98 \pm 0.03$& $-6.53 \pm 0.02$& $-6.18 \pm 0.06$& $-5.63$& $-5.07$& $-6.17$& $-6.55 \pm 0.06$& $-1.70 \pm 0.16$ \\
Elias 20& $-5.40 \pm 0.01$& $-5.00 \pm 0.01$& $-4.49 \pm 0.01$& $-3.92$& $-3.90$& $-4.80$& $-5.68 \pm 0.08$& $-0.90 \pm 0.11$ \\
WSB 52& $-5.11 \pm 0.02$& $-4.71 \pm 0.01$& $-4.36 \pm 0.01$& $-3.84$& $-4.01$& $-4.11$& $-5.54 \pm 0.08$& $-0.77 \pm 0.11$ \\
GK Tau& $-5.71 \pm 0.04$& $-5.43 \pm 0.02$& $-4.87 \pm 0.01$& $-5.21$& $-5.24$& $-5.38$& $-6.46 \pm 0.22$& $-1.62 \pm 0.25$ \\
SR 4& $-5.56 \pm 0.02$& $-5.46 \pm 0.02$& $-5.29 \pm 0.08$& $-4.75$& $-4.65$& $-5.29$& $-4.95 \pm 0.03$& $-0.22 \pm 0.08$ \\
Sz 114& $-5.98 \pm 0.02$& $-5.58 \pm 0.01$& $-5.02 \pm 0.02$& $-4.67$& $-4.78$& $-4.52$& $< -7.10$& $< -2.21$ \\
\enddata
\tablecomments{Line luminosities for the Class~II comparison sample plotted in Figure~\ref{fig:flux_Lacc}. Water luminosities are measured from single-Gaussian fits to the diagnostic lines listed in the column headers; HCN, C$_2$H$_2$, and CO$_2$ luminosities are derived from the best-fit slab models and do not have individual measurement uncertainties. Upper limits are indicated with $<$. $L_{\rm HI(10-7)}$ is the luminosity of the H\,\textsc{i} (10--7) line at 8.76\,$\mu$m, measured after subtracting the best-fitting water slab model to remove blended H$_2$O emission at this wavelength. $L_{\rm acc}$ is derived from $L_{\rm HI(10-7)}$ (Section~\ref{sec:slab}); we tabulate the line luminosity so that the accretion luminosities can be recomputed with any adopted H\,\textsc{i}--$L_{\rm acc}$ calibration.}
\end{deluxetable*}

\section{Continuum-subtracted line images}\label{sec:line_images_appendix}

\begin{figure*}[!htbp]
\centering
\includegraphics[height=0.88\textheight]{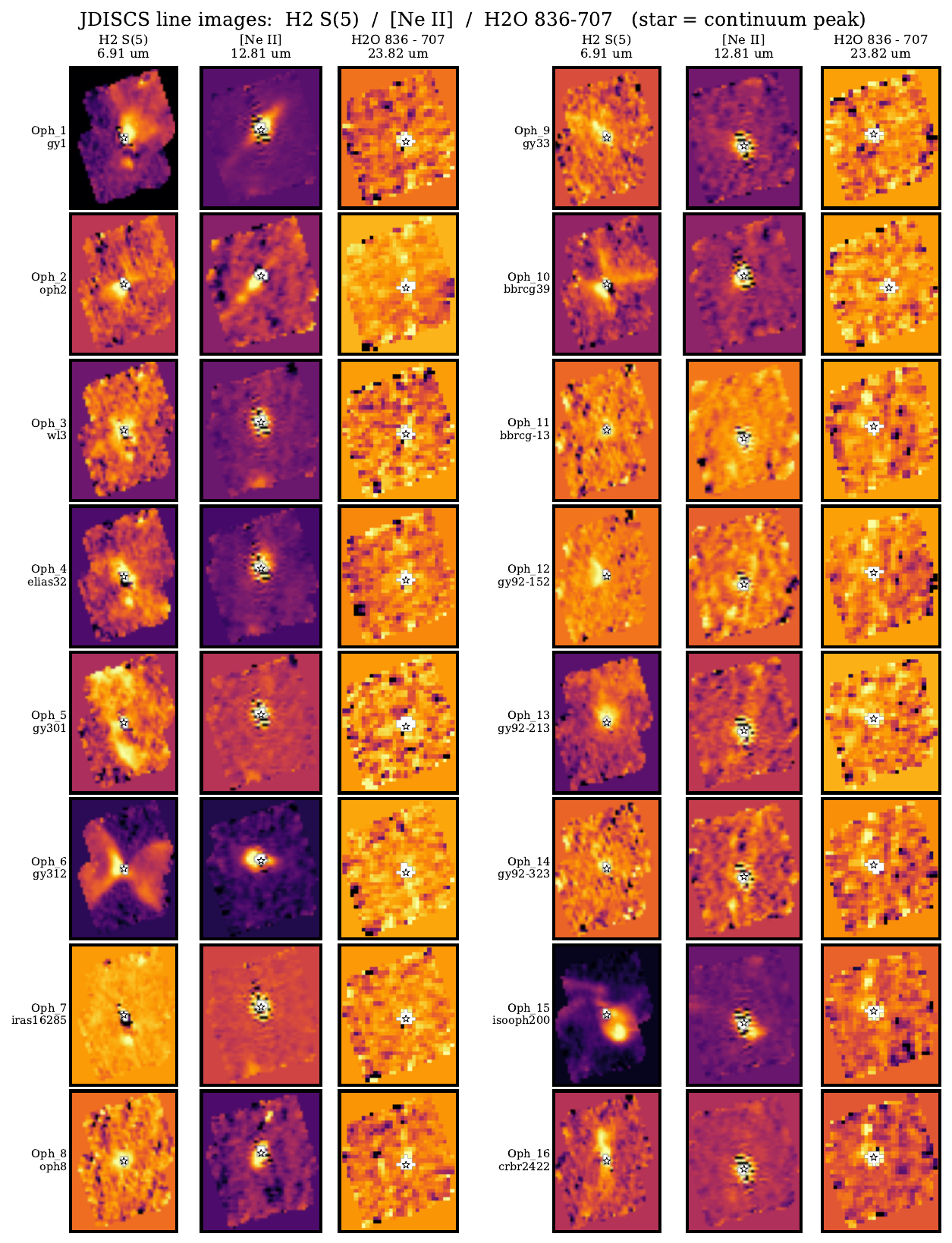}
\caption{Continuum-subtracted line images of H$_2$ S(5) (6.91\,$\mu$m), [Ne\,II]
(12.81\,$\mu$m), and H$_2$O (23.8\,$\mu$m) for all 16 Class~I/FS sources.
Each row shows one source; columns are the three lines. The white star in each panel marks the peak of the normalized continuum
image measured near the wavelength of each line, i.e., the location of the
central source, which is not necessarily at the center of each image. Each panel is
stretched independently, so surface brightness is not comparable between
panels. The water emission is point-like and coincident with the continuum
peak in all sources, consistent with an inner-disk origin, whereas the H$_2$
and [Ne\,II] images reveal extended or jet-like morphologies in several
sources, indicating active outflows. The image construction follows the
procedure of \citet{Narang_2026}. \label{fig:line_images}}
\end{figure*}

To distinguish disk emission from extended envelope/outflow contributions, we constructed continuum-subtracted line images for each Class~I/FS source following the procedure of \citet{Narang_2026}, in which the local continuum is fit per spaxel and subtracted before integrating over the line wavelength range. Figure~\ref{fig:line_images} shows the resulting images of H$_2$O at 23.8\,$\mu$m, H$_2$ S(5) at 6.91\,$\mu$m, and [Ne\,II] at 12.81\,$\mu$m for all 16 Class~I/FS sources. The water emission is spatially unresolved in all cases. Some sources show extended H$_2$ and/or jet-like [Ne\,II] morphology, indicating active outflows, but the morphology of these outflow tracers does not correlate with the strength of the cold water excess.

\section{Extinction correction}\label{sec:extinction_appendix}

Here we expand on the extinction treatment summarized in Section~\ref{subsec:extinction}.

We adopt the \citet{McClure2009} extinction curve because it is appropriate for the moderate extinctions of our relatively evolved, less deeply embedded Class~I/FS sources ($T_{\rm bol}=200$--750\,K). More deeply embedded Class~0 and young Class~I sources show deeper 10\,$\mu$m silicate and ice absorption and are often better described by the KP5 curve \citep{Pontoppidan2024}, which includes stronger silicate and ice features; the most appropriate curve depends on the degree of embedding \citep[see the discussion in][]{Assani_2025}. For the Class~II comparison sample we apply no extinction correction, as $A_V<1$ and the correction is negligible across the MIRI range.

The $A_V$ values (Table~\ref{table:stars}) are adopted from the literature and works in preparation; where multiple estimates exist, we preferentially adopt those for which the spectral type and $A_V$ were derived simultaneously from UV and/or optical stellar spectra, for consistency across the sample. Because the values originate from several studies rather than a single uniform analysis, residual systematic differences in $A_V$ contribute to the uncertainty on the absolute molecular fluxes.

In principle, the extinction can be constrained directly from the JWST spectra using H$_2$ pure-rotational lines, which exploit the differential reddening across the 10\,$\mu$m silicate feature in the H$_2$ excitation diagram \citep[e.g.,][]{Salyk2024,Francis_2025_D2Hratio,Assani_2025}. In our sample, the lower-J lines (S(1), S(2), and S(3)) are commonly detected, but the higher-J lines (S(5)--S(7)) needed to provide a long enough upper-level energy baseline to break the degeneracy between excitation temperature and extinction are not clearly detected in the majority of our sources. We therefore do not attempt an H$_2$-based extinction determination and instead rely on the literature $A_V$ values.

\begin{figure*}[!htbp]
\centering
\vspace{-0.cm}
\includegraphics[width=0.8\textwidth]{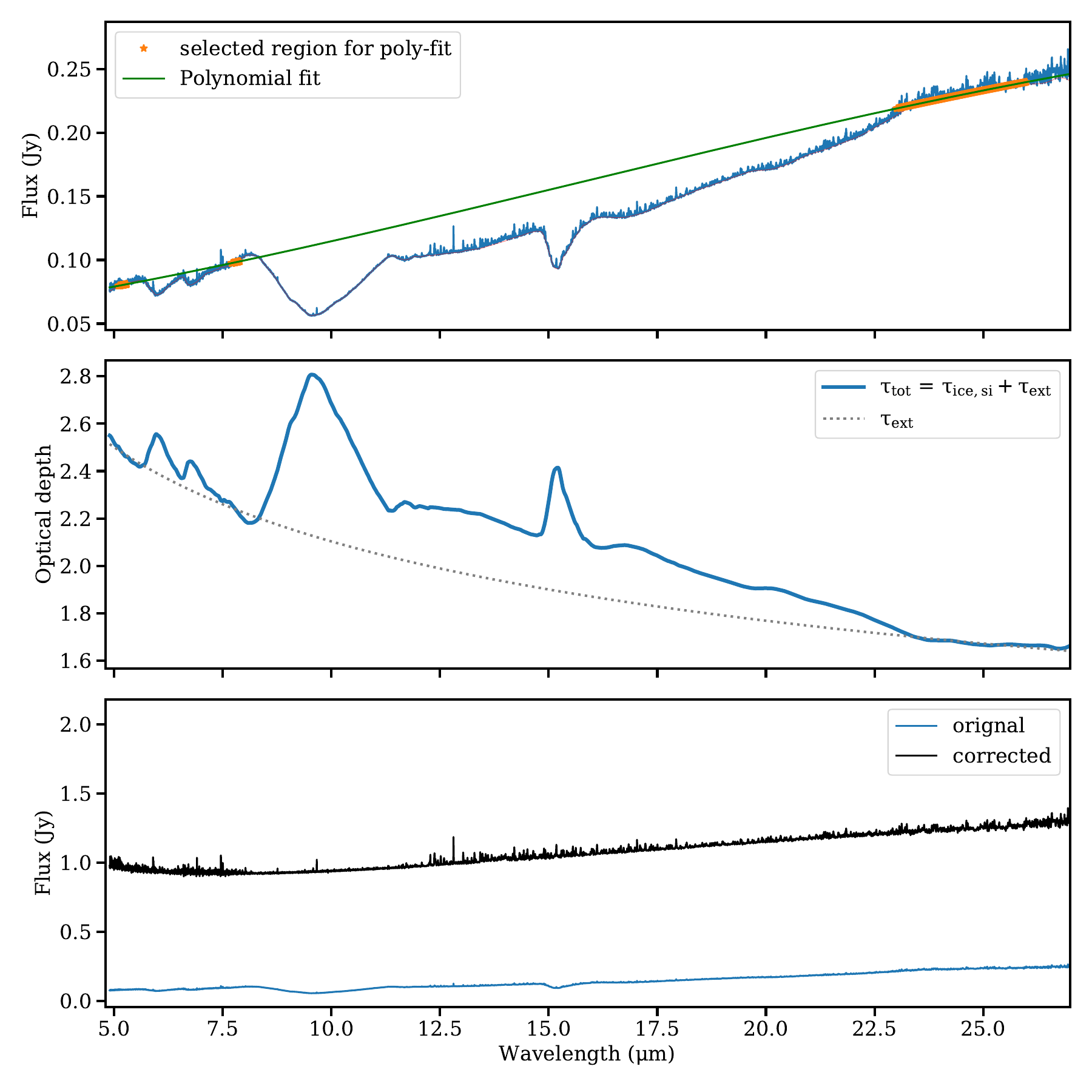}
\vspace{-.7cm}
\caption{Extinction correction example for Oph 10.  \label{fig:extinction_example}
}
\vspace{-0.cm}
\end{figure*}

\section{Censored Kendall $\tau$ correlation}\label{sec:stat_methods}

This appendix details the censored rank-correlation method used for the univariate screening in Section~\ref{sec:correlation}. To test correlations in the presence of upper limits, we use the generalized Kendall $\tau$ rank-correlation statistic for censored data \citep{Brown1974,Isobe1986}. Unlike the standard Kendall $\tau$, which requires excluding non-detections, the censored statistic incorporates left-censored points (upper limits) so they contribute information without being discarded (see \citealt{Brown1974,Isobe1986} for details). For the $L_{\rm acc}$ correlations, three Class~I/FS sources and one Class~II source with only upper limits on $L_{\rm acc}$ are included as left-censored points. For the linear fits shown in the scatter plots, we use the Bayesian \texttt{linmix} regression \citep{Kelly_2007}, which incorporates left-censored data (upper limits on $L_{\rm acc}$ are handled by swapping the regression axes); significance is assessed from the censored Kendall $\tau$ test at $p<0.1$.

Because we test many property--molecule combinations simultaneously, the probability of obtaining spurious significant results increases---the multiple-comparisons problem \citep[e.g.,][]{Benjamini2018}. For example, testing 66 independent hypotheses at $p<0.10$ would yield $\sim$7 false positives even if no true correlations existed. To control for this we apply the Benjamini--Hochberg false-discovery-rate correction using the \texttt{multipletests} function of the Python \texttt{statsmodels} package \citep{Seabold2010} within each property group (i.e., all molecule--sample combinations tested against a given property are corrected together). Correlations that survive the correction are termed ``robust'' and those that are only nominally significant ``tentative''; as noted in Section~\ref{sec:correlation}, none of the reported correlations survive the correction.

\begin{figure*}[!htbp]
\centering
\vspace{-0.cm}
\includegraphics[width=0.7\textwidth]{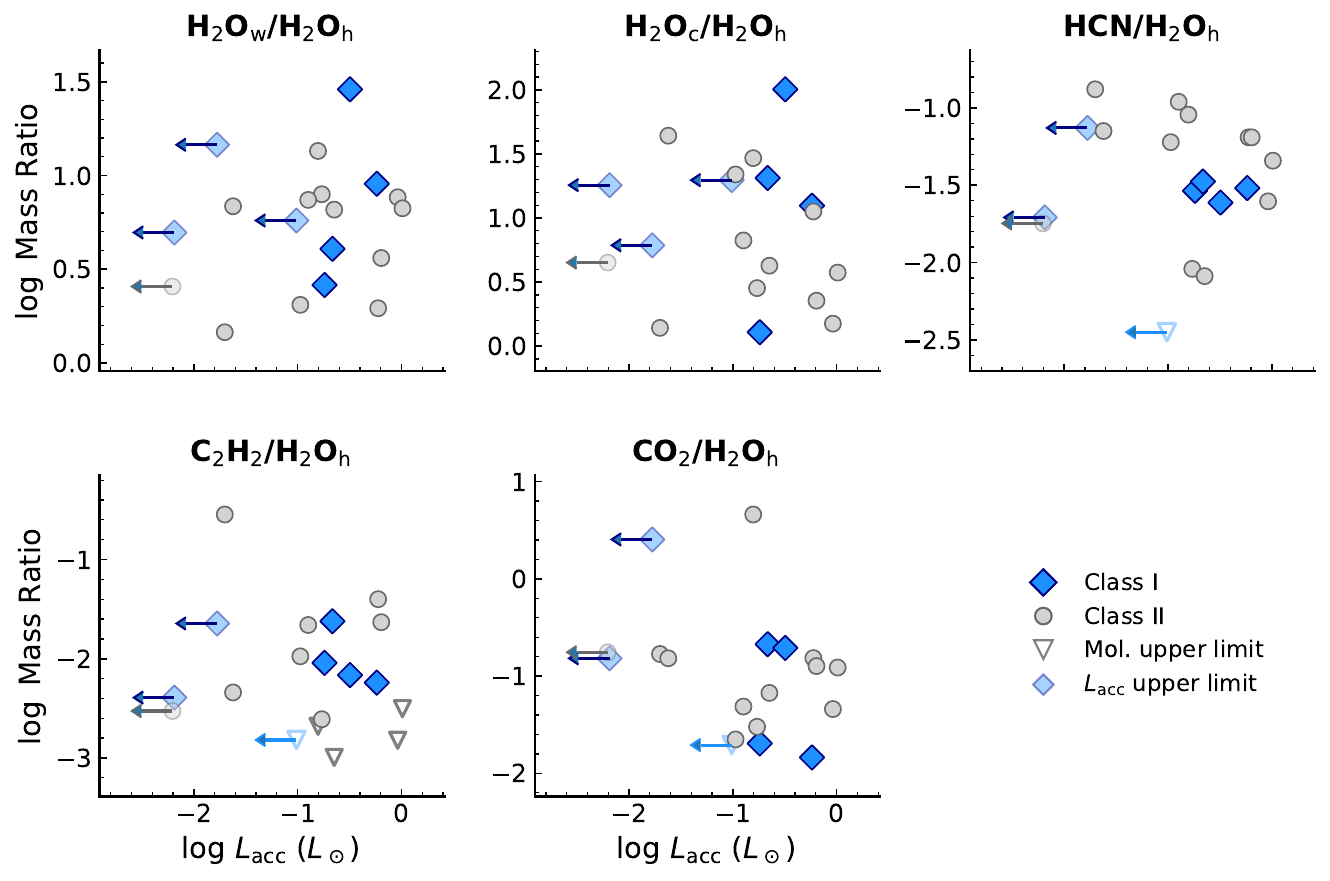}
\vspace{-.0cm}
\caption{Mass ratios relative to hot water as a function of accretion luminosity. Blue points are Class I/FS sources and grey points are Class II sources. No statistically significant correlations are found. \label{fig:Mratio_Lacc}
}
\vspace{-0.cm}
\end{figure*}

\begin{figure*}[!htbp]
\centering
\vspace{-0.cm}
\includegraphics[width=\textwidth]{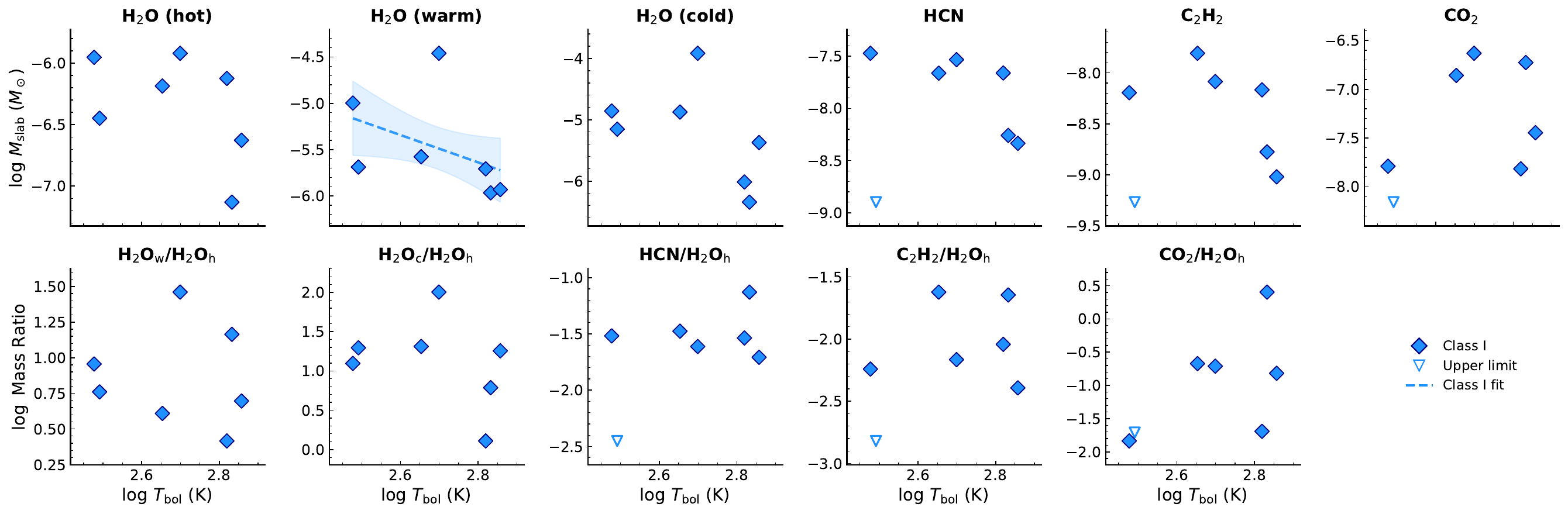}
\vspace{-.0cm}
\caption{Slab model masses and mass ratios relative to bolometric temperature.  Blue points are Class I/FS sources and grey points are Class II sources. \label{fig:Tbol}
}
\vspace{-0.cm}
\end{figure*}

\begin{figure*}[!htbp]
\centering
\vspace{-0.cm}
\includegraphics[width=0.9\textwidth]{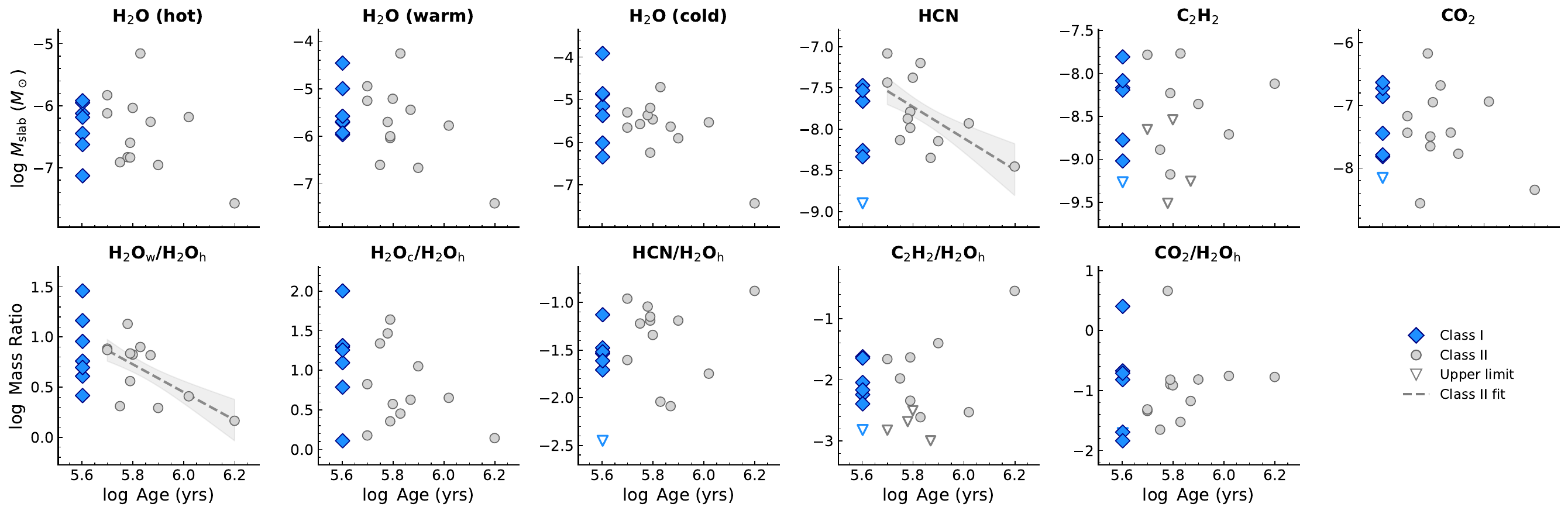}
\vspace{-0.cm}
\caption{Slab model masses and mass ratios as a function of estimated age for the Class II sample. \label{fig:water_mass_evolution}
}
\vspace{-0.cm}
\end{figure*}

\begin{figure*}[!htbp]
\centering
\vspace{-0.cm}
\includegraphics[width=0.8\textwidth]{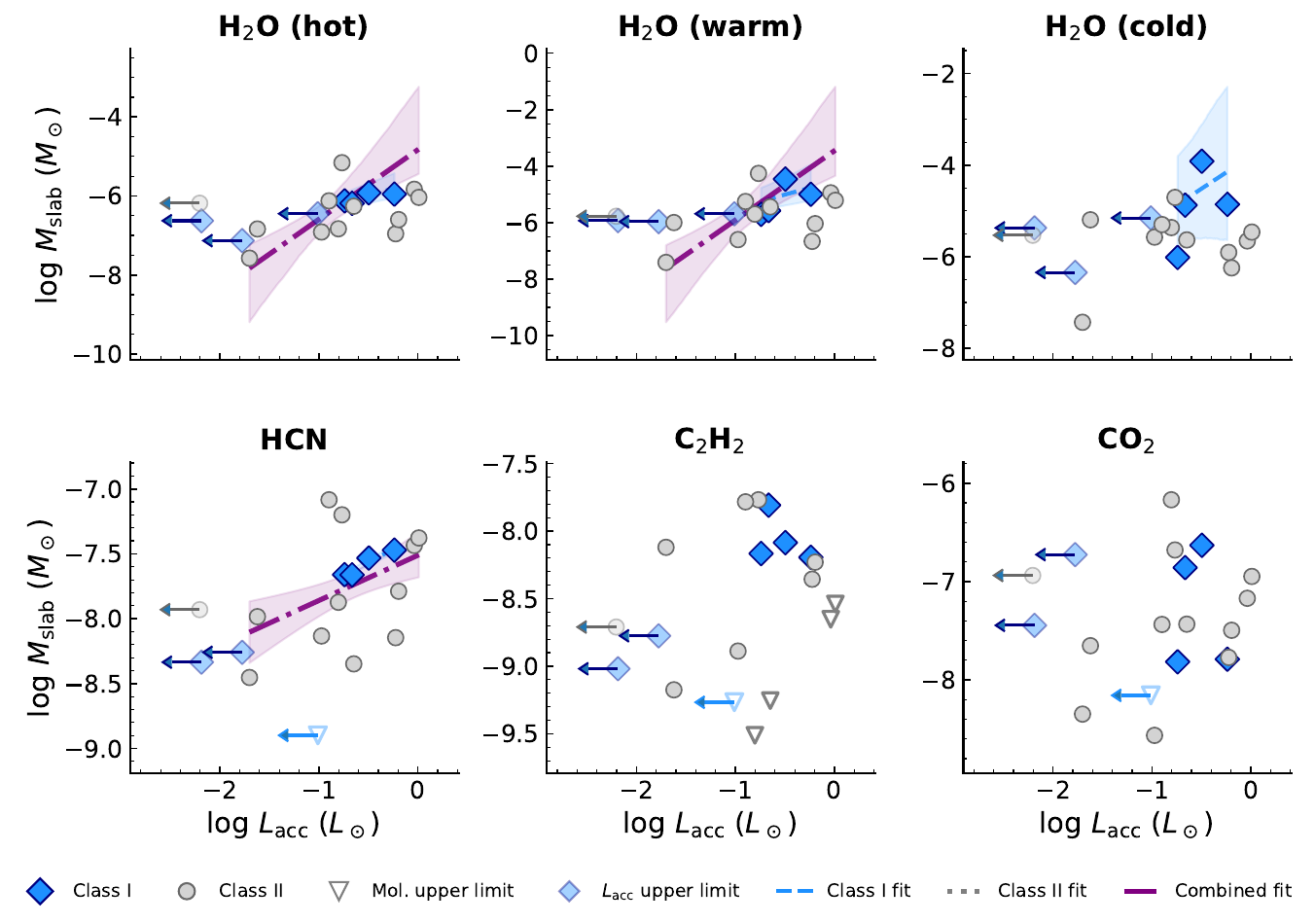}
\vspace{-0.cm}
\caption{Slab model masses of water and organic species as a function of accretion luminosity. Blue points are Class I/FS sources and grey points are Class II sources. Left-pointing arrows indicate sources with upper limits on $L_{\rm acc}$. Only statistically significant correlations (censored Kendall $\tau$ $p<0.1$) are shown; linear fits are computed with \texttt{linmix} \citep{Kelly_2007}, which incorporates upper limits on either axis as left-censored data. \label{fig:M_Lacc}
}
\vspace{-0.cm}
\end{figure*}

\begin{figure*}[!htbp]
\centering
\vspace{-0.cm}
\includegraphics[width=0.9\textwidth]{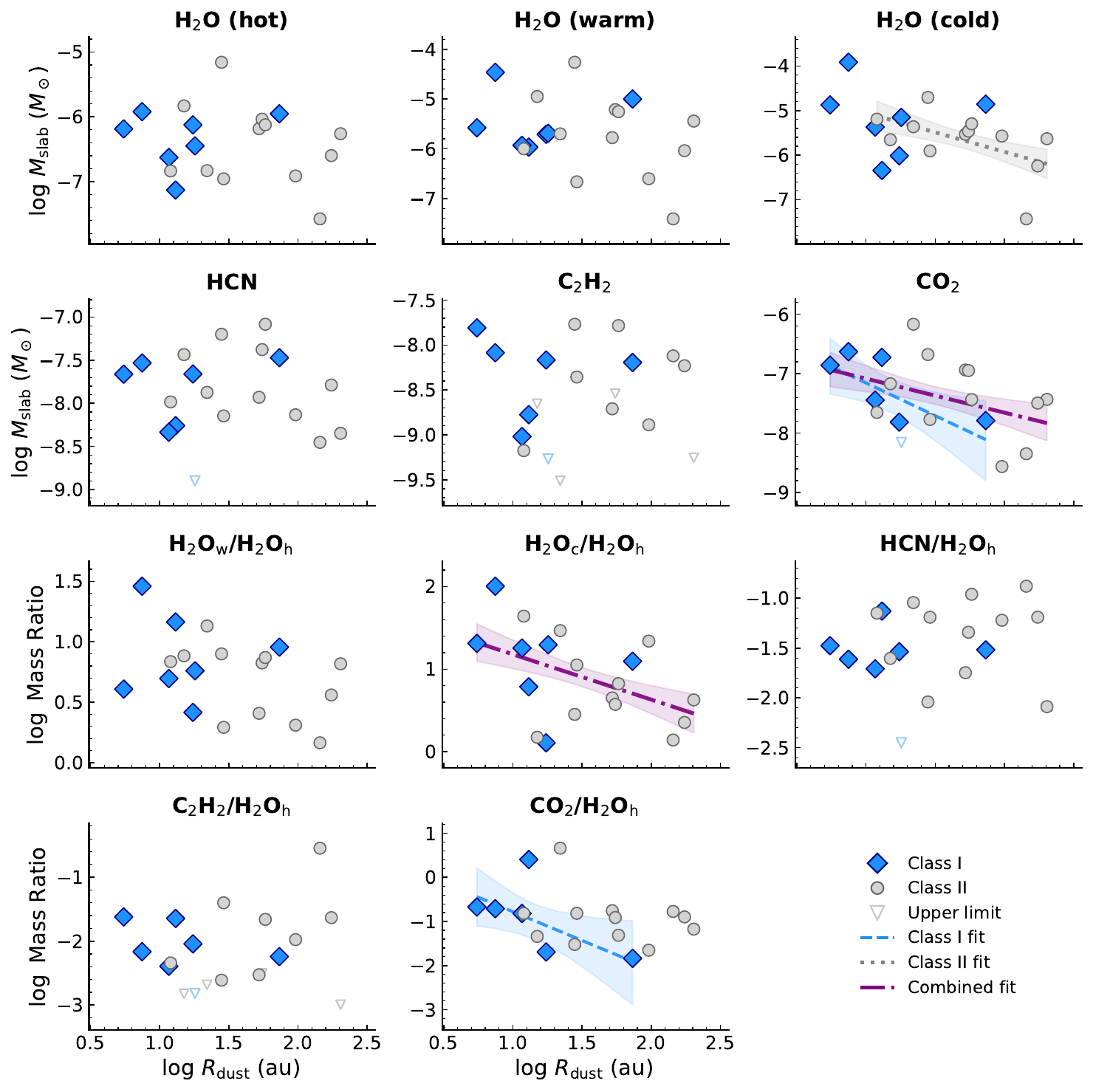}
\vspace{-0.cm}
\caption{Slab model masses and mass ratios as a function of mm dust disk size. Blue points are Class I/FS sources and grey points are Class II sources. \label{fig:M_Rd}
}
\vspace{-0.cm}
\end{figure*}

\begin{figure*}[!htbp]
\centering
\vspace{-0.cm}
\includegraphics[width=0.9\textwidth]{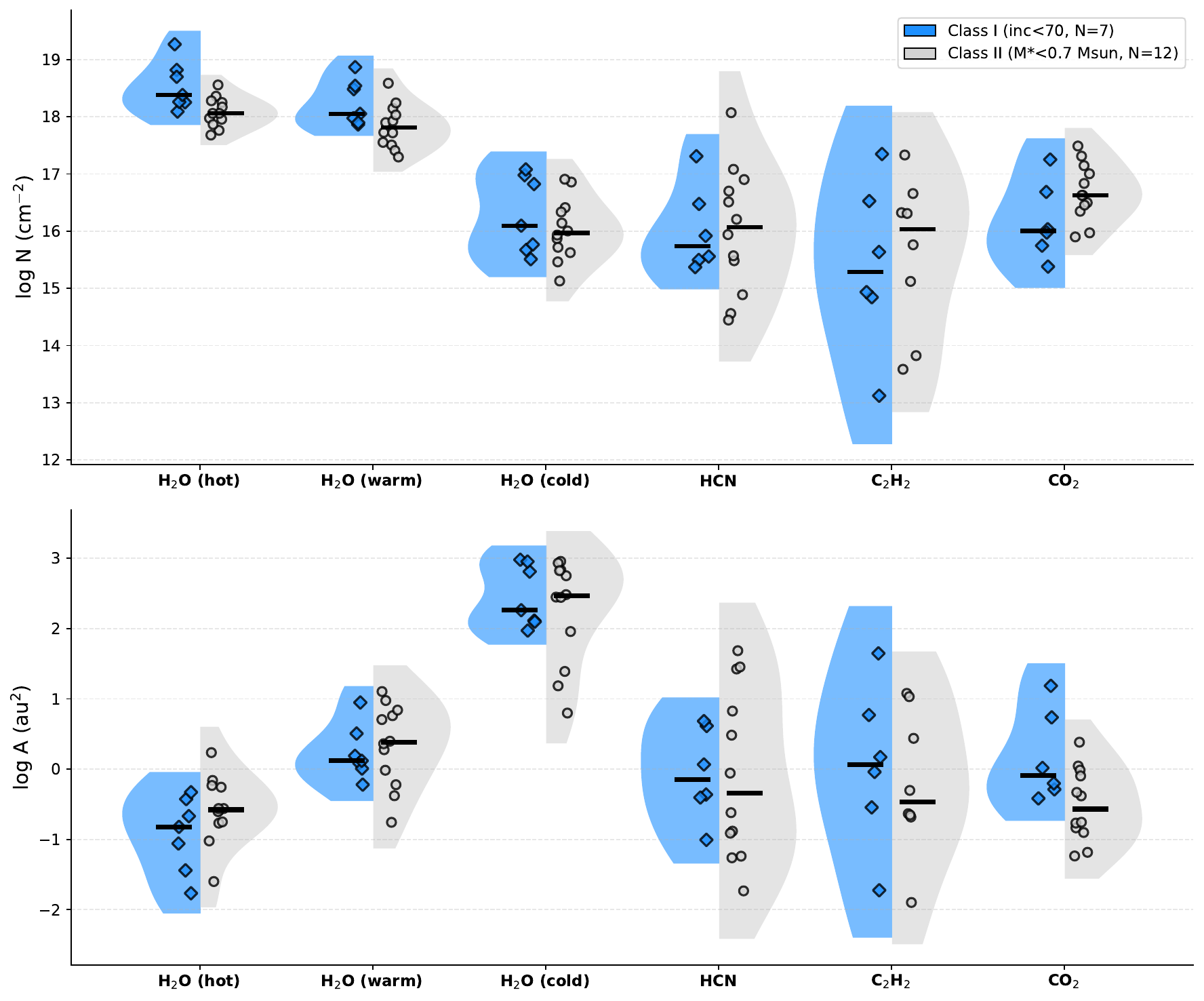}
\caption{Comparison of slab model emitting areas and column densities between Class I/FS ($i<70^{\circ}$, blue) and Class II (grey) disks. Format is the same as Figure~\ref{fig:slab_comparison}.  \label{fig:slab_area_N}
}
\end{figure*}

\begin{figure}[!htbp]
\centering
\vspace{-0.cm}
\includegraphics[width=0.5\textwidth]{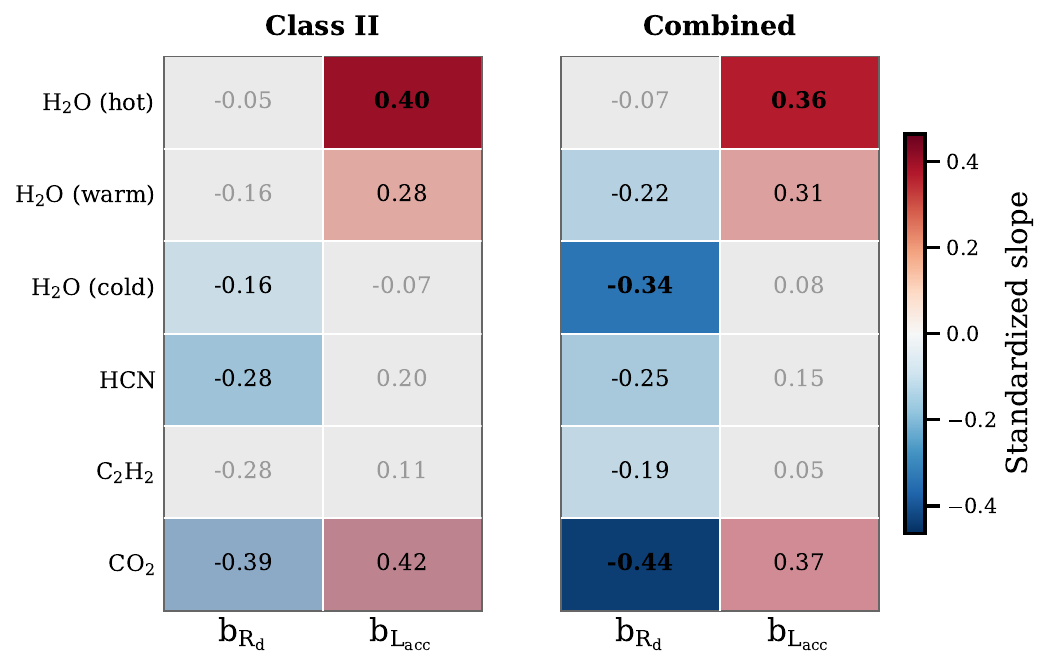}
\vspace{-0.cm}
\caption{Standardized partial regression slopes of log slab mass on z-scored log $R_{\rm dust}$ ($b_{R_{\rm d}}$) and log $L_{\rm acc}$ ($b_{L_{\rm acc}}$) for the Class I, Class II, and Combined samples. Each slope quantifies the partial effect of one predictor while controlling for the other. Red and blue indicate positive and negative slopes, respectively. Colored cells are statistically significant: full opacity for the 5–95\% bootstrap confidence interval excluding zero, semi-transparent for 16–84\%. Grey cells are consistent with zero. Confidence intervals are derived from 1500 bootstrap iterations. \label{fig:heatmap_multivar}
}
\vspace{-0.cm}
\end{figure}

\section{Age estimation}
\begin{figure*}[!htbp]
\centering
\vspace{-0.cm}
\includegraphics[width=\textwidth]{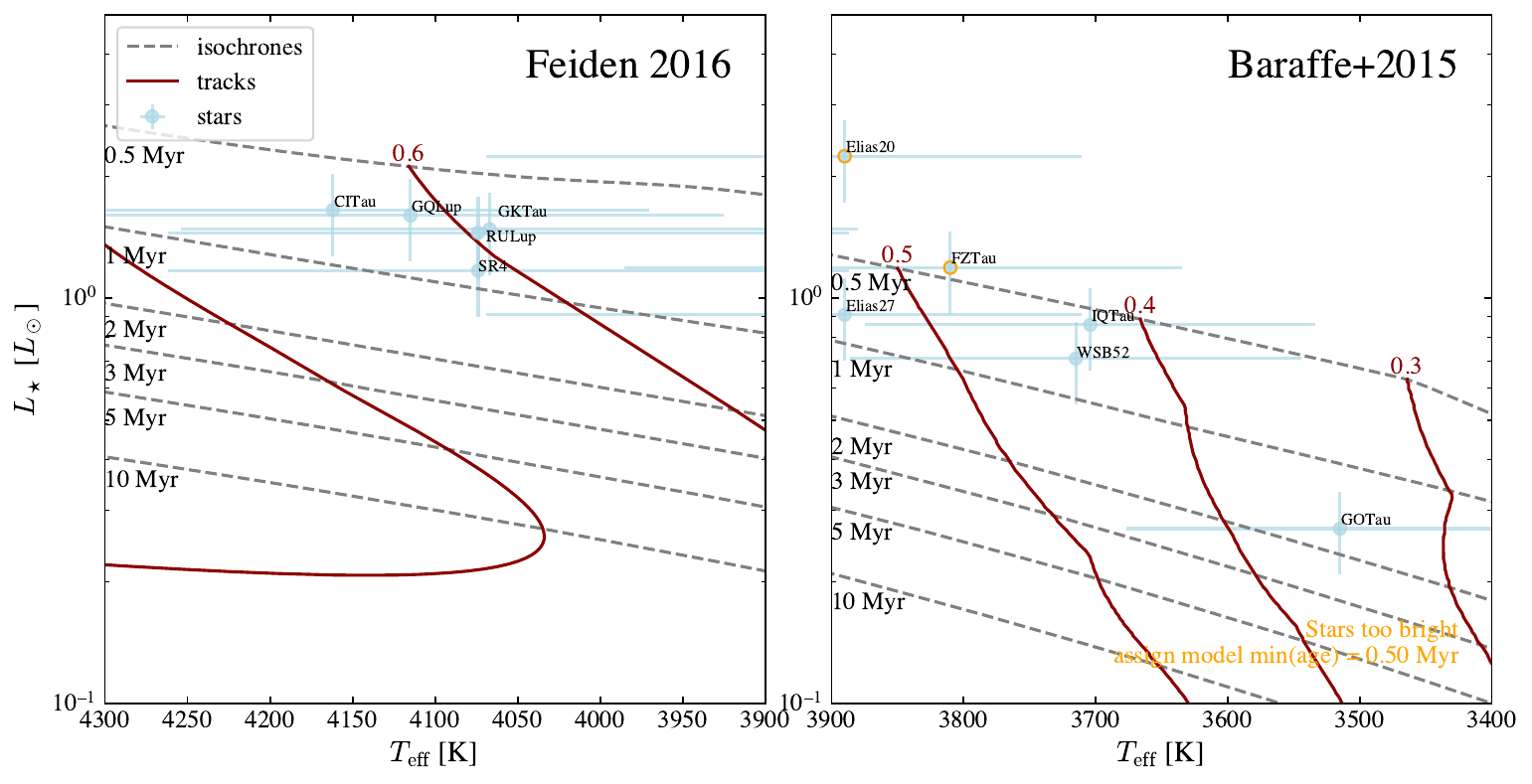}
\vspace{-.0cm}
\caption{H-R diagram of Class II sources  \label{fig:age}
}
\vspace{-0.cm}
\end{figure*}

\section{Slab model parameters of Class I/FS disks}
\begin{deluxetable*}{l|cccc|cccc|cccc}
\tablecaption{Slab Model Parameters for Class I/FS sources: H$_2$O \label{tab:slab_h2o}}
\tablewidth{0pt}
\tablehead{
 & \multicolumn{4}{c|}{Hot Water} & \multicolumn{4}{c|}{Warm Water} & \multicolumn{4}{c}{Cold Water} \\
\colhead{Source} & \colhead{T (K)} & \colhead{logN} & \colhead{logA} & \colhead{logM} & \colhead{T (K)} & \colhead{logN} & \colhead{logA} & \colhead{logM} & \colhead{T (K)} & \colhead{logN} & \colhead{logA} & \colhead{logM}
}
\startdata
Oph 1 & \ldots & \ldots & \ldots & $<-6.48$ & $431^{+8}_{-132}$ & $18.08^{+0.05}_{-0.08}$ & $0.90^{+0.46}_{-0.04}$ & $-4.98^{+0.10}_{-0.06}$ & $194^{+3}_{-4}$ & $16.53^{+0.10}_{-0.22}$ & $2.98^{+0.02}_{-0.06}$ & $-4.44^{+0.06}_{-0.11}$ \\
Oph 2 & \ldots & \ldots & \ldots & $<-7.89$ & \ldots & \ldots & \ldots & $<-6.99$ & \ldots & \ldots & \ldots & $<-6.56$ \\
Oph 3 & $783^{+31}_{-382}$ & $18.25^{+0.10}_{-0.07}$ & $-0.43^{+0.26}_{-0.08}$ & $-6.12^{+0.39}_{-0.12}$ & $496^{+22}_{-18}$ & $18.05^{+0.09}_{-0.38}$ & $0.19^{+0.03}_{-0.23}$ & $-5.71^{+0.10}_{-0.33}$ & $231^{+27}_{-30}$ & $15.67^{+0.54}_{-1.50}$ & $2.26^{+0.24}_{-0.48}$ & $-6.02^{+0.33}_{-0.35}$ \\
Oph 4 & $1033^{+272}_{-226}$ & $19.27^{+1.11}_{-0.69}$ & $-1.77^{+0.28}_{-0.17}$ & $-6.45^{+0.83}_{-0.72}$ & $501^{+37}_{-45}$ & $18.48^{+3.51}_{-0.21}$ & $-0.22^{+0.07}_{-0.11}$ & $-5.69^{+0.52}_{-0.23}$ & $182^{+13}_{-17}$ & $16.82^{+0.55}_{-0.40}$ & $1.97^{+0.28}_{-0.33}$ & $-5.15^{+0.40}_{-0.25}$ \\
Oph 5 & $909^{+85}_{-34}$ & $18.09^{+0.07}_{-0.23}$ & $-0.33^{+0.09}_{-0.09}$ & $-6.19^{+0.08}_{-0.40}$ & $454^{+26}_{-170}$ & $17.86^{+0.12}_{-0.51}$ & $0.51^{+0.28}_{-0.08}$ & $-5.58^{+0.16}_{-0.14}$ & $201^{+6}_{-185}$ & $16.09^{+0.20}_{-0.31}$ & $2.98^{+0.02}_{-0.14}$ & $-4.87^{+0.12}_{-0.12}$ \\
Oph 6 & \ldots & \ldots & \ldots & $<-7.74$ & $538^{+21}_{-28}$ & $18.31^{+0.12}_{-0.15}$ & $-0.83^{+0.09}_{-0.06}$ & $-6.46^{+0.14}_{-0.13}$ & $171^{+8}_{-11}$ & $16.84^{+0.69}_{-0.24}$ & $2.04^{+0.20}_{-0.53}$ & $-5.07^{+0.69}_{-0.14}$ \\
Oph 7 & $1180^{+125}_{-122}$ & $18.82^{+0.22}_{-0.15}$ & $-0.82^{+0.13}_{-0.14}$ & $-5.95^{+0.21}_{-0.18}$ & $512^{+75}_{-47}$ & $18.87^{+0.69}_{-0.22}$ & $0.09^{+0.11}_{-0.19}$ & $-5.00^{+0.46}_{-0.27}$ & $195^{+32}_{-38}$ & $16.98^{+1.04}_{-0.86}$ & $2.11^{+0.66}_{-0.48}$ & $-4.86^{+1.80}_{-0.60}$ \\
Oph 8 & \ldots & \ldots & \ldots & $<-8.15$ & \ldots & \ldots & \ldots & $<-7.25$ & \ldots & \ldots & \ldots & $<-6.81$ \\
Oph 9 & \ldots & \ldots & \ldots & $<-7.81$ & \ldots & \ldots & \ldots & $<-6.91$ & \ldots & \ldots & \ldots & $<-6.34$ \\
Oph 10 & $1025^{+107}_{-248}$ & $18.42^{+0.66}_{-0.13}$ & $-1.09^{+0.45}_{-0.22}$ & $-6.61^{+0.49}_{-0.22}$ & $595^{+60}_{-227}$ & $18.41^{+0.17}_{-0.12}$ & $-0.50^{+0.06}_{-0.05}$ & $-6.03^{+0.20}_{-0.14}$ & $256^{+57}_{-30}$ & $16.63^{+0.49}_{-0.62}$ & $1.05^{+0.38}_{-2.81}$ & $-6.26^{+0.37}_{-0.45}$ \\
Oph 11 & $999^{+61}_{-26}$ & $18.26^{+0.06}_{-0.06}$ & $-1.44^{+0.04}_{-0.16}$ & $-7.13^{+0.07}_{-0.11}$ & $443^{+232}_{-6}$ & $17.97^{+0.04}_{-0.15}$ & $0.01^{+0.02}_{-0.54}$ & $-5.97^{+0.04}_{-0.18}$ & $246^{+30}_{-8}$ & $15.51^{+0.50}_{-1.11}$ & $2.10^{+0.90}_{-0.44}$ & $-6.34^{+0.10}_{-0.31}$ \\
Oph 12 & \ldots & \ldots & \ldots & $<-7.52$ & $468^{+7}_{-4}$ & $17.88^{+0.03}_{-0.05}$ & $-0.02^{+0.02}_{-0.02}$ & $-6.09^{+0.03}_{-0.05}$ & $263^{+9}_{-7}$ & $14.98^{+0.52}_{-0.57}$ & $2.65^{+0.35}_{-0.51}$ & $-6.31^{+0.07}_{-0.11}$ \\
Oph 13 & $735^{+29}_{-41}$ & $18.38^{+0.11}_{-0.54}$ & $-1.06^{+0.21}_{-0.06}$ & $-6.63^{+0.14}_{-0.15}$ & $369^{+19}_{-51}$ & $17.90^{+0.13}_{-0.11}$ & $0.12^{+0.28}_{-0.08}$ & $-5.93^{+0.50}_{-0.12}$ & $195^{+9}_{-4}$ & $15.77^{+0.46}_{-0.36}$ & $2.81^{+0.19}_{-0.40}$ & $-5.37^{+0.08}_{-0.21}$ \\
Oph 14 & $857^{+344}_{-18}$ & $18.27^{+0.04}_{-0.32}$ & $-0.80^{+0.04}_{-0.52}$ & $-6.47^{+0.06}_{-0.80}$ & $425^{+8}_{-10}$ & $18.26^{+0.20}_{-0.05}$ & $0.22^{+0.02}_{-0.03}$ & $-5.46^{+0.22}_{-0.05}$ & $228^{+11}_{-8}$ & $15.89^{+0.66}_{-0.53}$ & $2.05^{+0.52}_{-0.70}$ & $-6.01^{+0.10}_{-0.13}$ \\
Oph 15 & $830^{+76}_{-171}$ & $18.70^{+0.91}_{-0.15}$ & $-0.67^{+0.46}_{-0.14}$ & $-5.92^{+0.69}_{-0.23}$ & $397^{+9}_{-16}$ & $18.54^{+0.06}_{-0.09}$ & $0.95^{+0.28}_{-0.03}$ & $-4.46^{+0.08}_{-0.08}$ & $170^{+3}_{-246}$ & $17.08^{+0.18}_{-0.11}$ & $2.95^{+0.04}_{-0.05}$ & $-3.92^{+0.19}_{-0.09}$ \\
Oph 16 & \ldots & \ldots & \ldots & $<-6.17$ & \ldots & \ldots & \ldots & $<-5.27$ & \ldots & \ldots & \ldots & $<-4.95$ \\
\enddata
\tablecomments{Temperatures (T) are in Kelvin (K). Column densities (logN) are in cm$^{-2}$, emitting areas (logA) are in au$^2$, and masses (logM) are in $M_\oplus$. Uncertainties are shown as subscripts and superscripts. Upper limits are indicated with ``$<$''.}
\end{deluxetable*}
\begin{deluxetable*}{l|cccc|cccc|cccc}
\tablecaption{Slab Model Parameters for Class I/FS sources: HCN, C$_2$H$_2$, CO$_2$ \label{tab:slab_organics}}
\tablewidth{0pt}
\tablehead{
 & \multicolumn{4}{c|}{HCN} & \multicolumn{4}{c|}{C$_2$H$_2$} & \multicolumn{4}{c}{CO$_2$} \\
\colhead{Source} & \colhead{T} & \colhead{logN} & \colhead{logA} & \colhead{logM} & \colhead{T} & \colhead{logN} & \colhead{logA} & \colhead{logM} & \colhead{T} & \colhead{logN} & \colhead{logA} & \colhead{logM}
}
\startdata
Oph 1 & \ldots & \ldots & \ldots & $<-8.47$ & \ldots & \ldots & \ldots & $<-8.84$ & \ldots & \ldots & \ldots & $<-7.73$ \\
Oph 2 & \ldots & \ldots & \ldots & $<-10.53$ & \ldots & \ldots & \ldots & $<-10.90$ & \ldots & \ldots & \ldots & $<-9.79$ \\
Oph 3 & $912^{+49}_{-321}$ & $16.47^{+0.96}_{-0.49}$ & $-0.36^{+0.46}_{-0.77}$ & $-7.66^{+0.39}_{-0.04}$ & $1295^{+63}_{-147}$ & $17.35^{+0.82}_{-2.03}$ & $-1.72^{+1.97}_{-0.65}$ & $-8.17^{+0.19}_{-0.05}$ & $545^{+536}_{-95}$ & $16.03^{+0.98}_{-2.66}$ & $-0.29^{+1.54}_{-0.48}$ & $-7.82^{+0.38}_{-1.02}$ \\
Oph 4 & \ldots & \ldots & \ldots & $<-8.90$ & \ldots & \ldots & \ldots & $<-9.27$ & \ldots & \ldots & \ldots & $<-8.16$ \\
Oph 5 & $630^{+70}_{-79}$ & $15.50^{+0.55}_{-1.57}$ & $0.61^{+1.39}_{-0.54}$ & $-7.66^{+0.11}_{-0.08}$ & $792^{+61}_{-136}$ & $16.52^{+0.93}_{-0.74}$ & $-0.55^{+0.62}_{-0.80}$ & $-7.81^{+0.24}_{-0.05}$ & $363^{+114}_{-135}$ & $16.68^{+1.00}_{-0.91}$ & $0.02^{+0.77}_{-0.70}$ & $-6.86^{+2.67}_{-0.39}$ \\
Oph 6 & $873^{+118}_{-414}$ & $17.00^{+1.24}_{-1.28}$ & $-2.04^{+1.00}_{-0.92}$ & $-8.81^{+0.94}_{-0.13}$ & $1047^{+233}_{-236}$ & $17.52^{+0.49}_{-0.86}$ & $-2.93^{+0.35}_{-0.89}$ & $-9.20^{+0.27}_{-0.14}$ & \ldots & \ldots & \ldots & $<-9.06$ \\
Oph 7 & $1050^{+143}_{-510}$ & $17.31^{+1.41}_{-0.99}$ & $-1.01^{+0.86}_{-1.01}$ & $-7.47^{+0.46}_{-0.11}$ & $1297^{+347}_{-155}$ & $15.63^{+1.61}_{-1.33}$ & $-0.04^{+1.34}_{-1.53}$ & $-8.19^{+0.12}_{-0.17}$ & $559^{+413}_{-298}$ & $15.97^{+2.60}_{-1.78}$ & $-0.20^{+1.46}_{-1.06}$ & $-7.79^{+3.78}_{-0.80}$ \\
Oph 8 & \ldots & \ldots & \ldots & $<-10.38$ & \ldots & \ldots & \ldots & $<-10.74$ & \ldots & \ldots & \ldots & $<-9.63$ \\
Oph 9 & \ldots & \ldots & \ldots & $<-9.88$ & \ldots & \ldots & \ldots & $<-10.25$ & \ldots & \ldots & \ldots & $<-9.14$ \\
Oph 10 & $1125^{+123}_{-398}$ & $17.75^{+0.26}_{-2.07}$ & $-1.89^{+1.88}_{-0.11}$ & $-7.91^{+0.16}_{-0.21}$ & $1352^{+44}_{-71}$ & $17.08^{+0.43}_{-1.17}$ & $-1.81^{+1.11}_{-0.18}$ & $-8.53^{+0.06}_{-0.07}$ & $648^{+418}_{-290}$ & $17.73^{+3.73}_{-1.69}$ & $-1.91^{+0.97}_{-0.09}$ & $-7.73^{+3.73}_{-0.68}$ \\
Oph 11 & $817^{+35}_{-51}$ & $15.92^{+0.89}_{-1.03}$ & $-0.40^{+0.99}_{-0.82}$ & $-8.26^{+0.13}_{-0.02}$ & $786^{+106}_{-52}$ & $14.84^{+0.94}_{-1.85}$ & $0.17^{+1.83}_{-0.89}$ & $-8.77^{+0.03}_{-0.07}$ & $238^{+68}_{-38}$ & $17.25^{+2.19}_{-0.69}$ & $-0.42^{+2.41}_{-0.37}$ & $-6.73^{+1.86}_{-0.88}$ \\
Oph 12 & \ldots & \ldots & \ldots & $<-9.63$ & \ldots & \ldots & \ldots & $<-10.00$ & $488^{+180}_{-46}$ & $15.03^{+1.03}_{-1.82}$ & $0.30^{+1.70}_{-0.84}$ & $-8.23^{+0.15}_{-0.23}$ \\
Oph 13 & $776^{+50}_{-49}$ & $15.37^{+1.04}_{-0.94}$ & $0.07^{+0.94}_{-0.97}$ & $-8.34^{+0.04}_{-0.06}$ & $717^{+147}_{-112}$ & $13.12^{+1.33}_{-0.56}$ & $1.65^{+0.35}_{-1.32}$ & $-9.02^{+0.15}_{-0.07}$ & $281^{+38}_{-33}$ & $15.38^{+0.47}_{-1.56}$ & $0.74^{+1.26}_{-0.28}$ & $-7.44^{+0.26}_{-0.43}$ \\
Oph 14 & \ldots & \ldots & \ldots & $<-8.96$ & \ldots & \ldots & \ldots & $<-9.33$ & $246^{+84}_{-240}$ & $17.55^{+3.75}_{-0.99}$ & $-0.22^{+0.55}_{-0.44}$ & $-6.23^{+3.14}_{-1.10}$ \\
Oph 15 & $556^{+82}_{-222}$ & $15.56^{+2.33}_{-0.87}$ & $0.68^{+0.78}_{-2.42}$ & $-7.53^{+0.52}_{-0.10}$ & $767^{+111}_{-191}$ & $14.93^{+1.00}_{-1.02}$ & $0.77^{+1.06}_{-0.93}$ & $-8.09^{+0.10}_{-0.05}$ & $294^{+35}_{-73}$ & $15.74^{+0.41}_{-0.57}$ & $1.18^{+0.42}_{-0.35}$ & $-6.63^{+0.28}_{-0.19}$ \\
Oph 16 & \ldots & \ldots & \ldots & $<-8.14$ & \ldots & \ldots & \ldots & $<-8.51$ & \ldots & \ldots & \ldots & $<-7.40$ \\
\enddata
\tablecomments{Temperatures (T) are in Kelvin (K). Column densities (logN) are in cm$^{-2}$, emitting areas (logA) are in au$^2$, and masses (logM) are in $M_\oplus$. Uncertainties are shown as subscripts and superscripts. Upper limits are indicated with ``$<$''.}
\end{deluxetable*}

\section{Slab models for additional Class I/FS disks}

\begin{figure*}[!htbp]
\centering
\vspace{-0.cm}
\includegraphics[width=\textwidth]{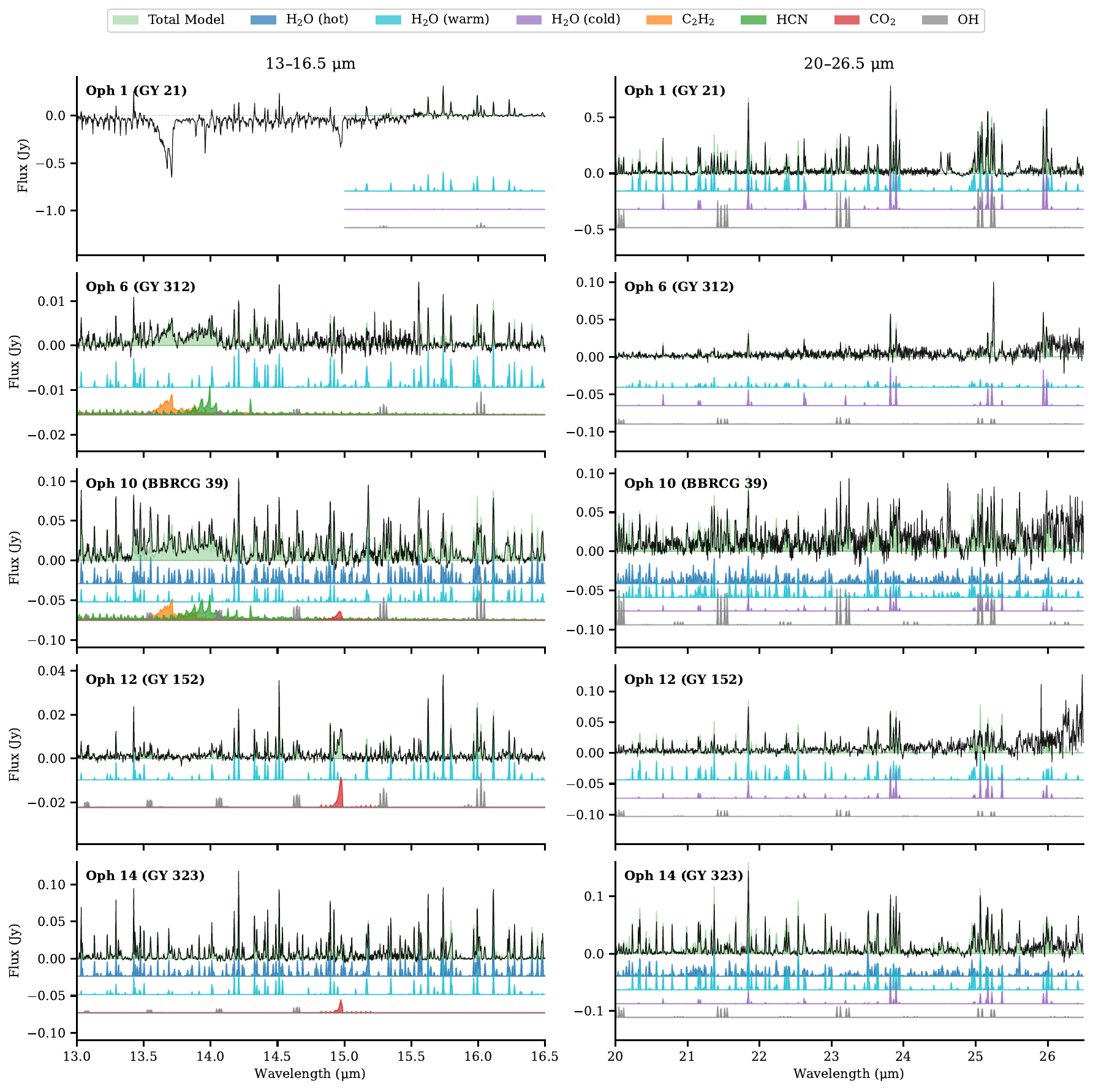}
\vspace{-.0cm}
\caption{Best-fitting slab models for the highly inclined ($i>70^{\circ}$) or unknown-inclination Class I/FS sources, shown over the 13--16.5 and 20--26.5\,$\mu$m wavelength ranges. Format is the same as Figure~\ref{fig:slab_model_classI}. These sources are excluded from the statistical comparison due to inclination-driven suppression of line emission.  \label{fig:ClassI_slab_inclined}
}
\vspace{-0.cm}
\end{figure*}

\section{Slab model parameters of Class II disks}
\begin{deluxetable*}{l|cccc|cccc|cccc}
\tablecaption{Slab Model Parameters for Class II sources: H$_2$O \label{tab:slab_classII_h2o}}
\tablewidth{0pt}
\tablehead{
 & \multicolumn{4}{c|}{Hot Water} & \multicolumn{4}{c|}{Warm Water} & \multicolumn{4}{c}{Cold Water} \\
\colhead{Source} & \colhead{T (K)} & \colhead{logN} & \colhead{logA} & \colhead{logM} & \colhead{T (K)} & \colhead{logN} & \colhead{logA} & \colhead{logM} & \colhead{T (K)} & \colhead{logN} & \colhead{logA} & \colhead{logM}
}
\startdata
SR 4 & $1006^{+383}_{-76}$ & $17.76^{+0.10}_{-0.17}$ & $-0.77^{+0.22}_{-0.54}$ & $-6.96^{+0.20}_{-0.92}$ & $694^{+68}_{-70}$ & $17.51^{+0.14}_{-0.19}$ & $-0.22^{+0.06}_{-0.07}$ & $-6.66^{+0.16}_{-0.19}$ & $197^{+173}_{-34}$ & $16.86^{+1.48}_{-2.73}$ & $1.19^{+0.91}_{-1.51}$ & $-5.91^{+1.30}_{-3.80}$ \\
CI Tau & $1039^{+154}_{-30}$ & $17.96^{+0.03}_{-0.07}$ & $-0.61^{+0.06}_{-0.17}$ & $-6.60^{+0.07}_{-0.34}$ & $603^{+22}_{-21}$ & $17.93^{+0.20}_{-0.07}$ & $-0.02^{+0.03}_{-0.02}$ & $-6.04^{+0.19}_{-0.06}$ & $297^{+92}_{-29}$ & $16.91^{+0.63}_{-0.27}$ & $0.80^{+0.12}_{-0.16}$ & $-6.24^{+0.42}_{-0.29}$ \\
GQ Lup & $895^{+26}_{-21}$ & $17.68^{+0.05}_{-0.06}$ & $-0.56^{+0.05}_{-0.06}$ & $-6.83^{+0.06}_{-0.07}$ & $432^{+14}_{-8}$ & $17.41^{+0.06}_{-0.04}$ & $0.84^{+0.03}_{-0.05}$ & $-5.70^{+0.05}_{-0.05}$ & $215^{+7}_{-8}$ & $16.14^{+0.15}_{-0.17}$ & $2.44^{+0.06}_{-0.10}$ & $-5.36^{+0.14}_{-0.09}$ \\
IQ Tau & $1009^{+33}_{-25}$ & $18.06^{+0.05}_{-0.06}$ & $-1.02^{+0.05}_{-0.06}$ & $-6.91^{+0.06}_{-0.08}$ & $565^{+25}_{-32}$ & $17.72^{+0.12}_{-0.07}$ & $-0.38^{+0.04}_{-0.04}$ & $-6.60^{+0.16}_{-0.08}$ & $192^{+17}_{-5}$ & $15.62^{+0.29}_{-0.46}$ & $2.75^{+0.25}_{-0.29}$ & $-5.57^{+0.07}_{-0.26}$ \\
RU Lup & $1023^{+53}_{-52}$ & $18.17^{+0.20}_{-0.06}$ & $-0.26^{+0.09}_{-0.12}$ & $-6.04^{+0.18}_{-0.10}$ & $571^{+28}_{-16}$ & $18.03^{+0.05}_{-0.18}$ & $0.71^{+0.02}_{-0.04}$ & $-5.21^{+0.05}_{-0.20}$ & $231^{+25}_{-26}$ & $16.00^{+0.39}_{-3.20}$ & $2.48^{+0.02}_{-0.31}$ & $-5.46^{+0.42}_{-0.29}$ \\
FZ Tau & $891^{+22}_{-104}$ & $18.28^{+0.05}_{-0.06}$ & $-0.16^{+0.14}_{-0.05}$ & $-5.83^{+0.09}_{-0.07}$ & $472^{+10}_{-14}$ & $18.24^{+0.05}_{-0.19}$ & $0.76^{+0.03}_{-0.02}$ & $-4.95^{+0.05}_{-0.22}$ & $225^{+22}_{-18}$ & $16.33^{+1.56}_{-0.49}$ & $1.96^{+0.46}_{-1.12}$ & $-5.66^{+0.26}_{-0.24}$ \\
Elias 27 & $840^{+19}_{-21}$ & $18.25^{+0.04}_{-0.16}$ & $-0.56^{+0.12}_{-0.03}$ & $-6.26^{+0.06}_{-0.08}$ & $463^{+6}_{-121}$ & $18.14^{+0.05}_{-0.04}$ & $0.36^{+0.02}_{-0.02}$ & $-5.44^{+0.05}_{-0.04}$ & $201^{+12}_{-325}$ & $15.87^{+1.58}_{-0.45}$ & $2.45^{+0.42}_{-0.52}$ & $-5.63^{+1.30}_{-0.19}$ \\
GO Tau & $712^{+25}_{-246}$ & $17.97^{+0.19}_{-0.12}$ & $-1.60^{+0.30}_{-0.13}$ & $-7.57^{+0.22}_{-0.10}$ & $489^{+22}_{-71}$ & $17.30^{+0.10}_{-0.39}$ & $-0.76^{+0.07}_{-0.07}$ & $-7.41^{+0.08}_{-0.30}$ & $233^{+80}_{-43}$ & $15.13^{+1.60}_{-3.12}$ & $1.39^{+1.11}_{-1.11}$ & $-7.43^{+0.74}_{-1.87}$ \\
Elias 20 & $828^{+14}_{-53}$ & $18.06^{+0.04}_{-0.13}$ & $-0.23^{+0.11}_{-0.03}$ & $-6.12^{+0.05}_{-0.05}$ & $425^{+6}_{-25}$ & $17.72^{+0.03}_{-0.05}$ & $0.98^{+0.05}_{-0.02}$ & $-5.25^{+0.04}_{-0.04}$ & $206^{+8}_{-273}$ & $15.72^{+0.24}_{-0.61}$ & $2.93^{+0.07}_{-0.21}$ & $-5.30^{+0.49}_{-0.11}$ \\
WSB 52 & $659^{+12}_{-117}$ & $18.55^{+0.07}_{-0.04}$ & $0.23^{+0.12}_{-0.04}$ & $-5.16^{+0.23}_{-0.06}$ & $400^{+5}_{-7}$ & $18.59^{+0.04}_{-0.12}$ & $1.10^{+0.01}_{-0.01}$ & $-4.26^{+0.04}_{-0.10}$ & $167^{+8}_{-13}$ & $16.41^{+0.32}_{-0.24}$ & $2.83^{+0.14}_{-0.22}$ & $-4.71^{+0.27}_{-0.17}$ \\
GK Tau & $946^{+26}_{-42}$ & $17.86^{+0.09}_{-0.05}$ & $-0.75^{+0.07}_{-0.06}$ & $-6.83^{+0.16}_{-0.06}$ & $466^{+16}_{-24}$ & $17.55^{+0.05}_{-0.10}$ & $0.40^{+0.15}_{-0.04}$ & $-6.00^{+0.08}_{-0.07}$ & $187^{+6}_{-8}$ & $15.93^{+0.27}_{-0.31}$ & $2.82^{+0.18}_{-0.25}$ & $-5.19^{+0.11}_{-0.10}$ \\
Sz 114 & $695^{+17}_{-20}$ & $18.36^{+0.06}_{-0.25}$ & $-0.59^{+0.15}_{-0.05}$ & $-6.18^{+0.08}_{-0.09}$ & $418^{+11}_{-40}$ & $17.90^{+0.06}_{-0.06}$ & $0.28^{+0.05}_{-0.03}$ & $-5.78^{+0.09}_{-0.06}$ & $197^{+7}_{-5}$ & $15.46^{+0.19}_{-0.32}$ & $2.96^{+0.04}_{-0.17}$ & $-5.53^{+0.12}_{-0.08}$ \\
\enddata
\tablecomments{Temperatures (T) are in Kelvin (K). Column densities (logN) are in cm$^{-2}$, emitting areas (logA) are in au$^2$, and masses (logM) are in $M_\oplus$. Uncertainties are shown as subscripts and superscripts. Upper limits are indicated with ``$<$''.}
\end{deluxetable*}
\begin{deluxetable*}{l|cccc|cccc|cccc}
\tablecaption{Slab Model Parameters for Class II sources: HCN, C$_2$H$_2$, CO$_2$ \label{tab:slab_classII_organics}}
\tablewidth{0pt}
\tablehead{
 & \multicolumn{4}{c|}{HCN} & \multicolumn{4}{c|}{C$_2$H$_2$} & \multicolumn{4}{c}{CO$_2$} \\
\colhead{Source} & \colhead{T} & \colhead{logN} & \colhead{logA} & \colhead{logM} & \colhead{T} & \colhead{logN} & \colhead{logA} & \colhead{logM} & \colhead{T} & \colhead{logN} & \colhead{logA} & \colhead{logM}
}
\startdata
SR 4 & $790^{+97}_{-350}$ & $16.51^{+0.57}_{-0.80}$ & $-0.88^{+0.74}_{-0.45}$ & $-8.15^{+0.29}_{-0.08}$ & $851^{+154}_{-394}$ & $17.33^{+0.67}_{-0.25}$ & $-1.90^{+0.16}_{-0.20}$ & $-8.36^{+0.87}_{-0.14}$ & $457^{+132}_{-117}$ & $16.62^{+2.58}_{-0.73}$ & $-0.84^{+0.58}_{-1.14}$ & $-7.77^{+1.67}_{-0.33}$ \\
CI Tau & $924^{+81}_{-23}$ & $14.56^{+1.06}_{-0.60}$ & $1.43^{+0.57}_{-1.05}$ & $-7.79^{+0.02}_{-0.06}$ & $1072^{+54}_{-43}$ & $15.12^{+0.98}_{-1.57}$ & $0.44^{+1.56}_{-0.97}$ & $-8.23^{+0.01}_{-0.05}$ & $559^{+119}_{-64}$ & $16.83^{+0.33}_{-1.03}$ & $-0.77^{+0.16}_{-0.11}$ & $-7.49^{+0.26}_{-0.27}$ \\
GQ Lup & $630^{+40}_{-42}$ & $14.44^{+0.77}_{-0.67}$ & $1.45^{+0.55}_{-0.76}$ & $-7.87^{+0.06}_{-0.04}$ & \ldots & \ldots & \ldots & $<-9.51$ & $254^{+47}_{-45}$ & $17.49^{+1.55}_{-1.97}$ & $-0.10^{+0.90}_{-0.43}$ & $-6.17^{+1.24}_{-1.97}$ \\
IQ Tau & $1138^{+57}_{-47}$ & $16.90^{+0.26}_{-1.23}$ & $-1.26^{+1.15}_{-0.21}$ & $-8.13^{+0.05}_{-0.20}$ & $1375^{+24}_{-40}$ & $13.82^{+1.03}_{-1.07}$ & $1.08^{+0.92}_{-0.99}$ & $-8.89^{+0.07}_{-0.04}$ & $1390^{+9}_{-59}$ & $15.90^{+1.10}_{-1.01}$ & $-0.90^{+1.00}_{-1.10}$ & $-8.56^{+0.20}_{-0.05}$ \\
RU Lup & $930^{+71}_{-154}$ & $15.57^{+0.80}_{-1.28}$ & $0.83^{+1.17}_{-0.78}$ & $-7.38^{+0.07}_{-0.04}$ & \ldots & \ldots & \ldots & $<-8.54$ & $468^{+161}_{-93}$ & $16.63^{+0.48}_{-0.90}$ & $-0.01^{+0.41}_{-0.67}$ & $-6.95^{+0.47}_{-0.36}$ \\
FZ Tau & $1352^{+47}_{-75}$ & $18.07^{+0.49}_{-1.25}$ & $-1.73^{+1.06}_{-0.26}$ & $-7.43^{+0.23}_{-0.23}$ & \ldots & \ldots & \ldots & $<-8.65$ & $577^{+297}_{-66}$ & $16.35^{+0.38}_{-1.93}$ & $0.04^{+0.54}_{-0.18}$ & $-7.17^{+0.19}_{-0.26}$ \\
Elias 27 & $1344^{+55}_{-124}$ & $15.48^{+0.63}_{-2.13}$ & $-0.06^{+2.06}_{-0.58}$ & $-8.35^{+0.04}_{-0.11}$ & \ldots & \ldots & \ldots & $<-9.26$ & $621^{+104}_{-183}$ & $17.31^{+0.76}_{-0.37}$ & $-1.18^{+0.32}_{-0.37}$ & $-7.43^{+0.59}_{-0.27}$ \\
GO Tau & $403^{+99}_{-33}$ & $15.94^{+0.25}_{-1.18}$ & $-0.62^{+0.97}_{-0.11}$ & $-8.45^{+0.15}_{-0.25}$ & $389^{+17}_{-24}$ & $16.32^{+0.06}_{-0.07}$ & $-0.65^{+0.17}_{-0.05}$ & $-8.12^{+0.09}_{-0.07}$ & $315^{+183}_{-56}$ & $15.97^{+0.88}_{-1.01}$ & $-0.76^{+1.00}_{-0.42}$ & $-8.35^{+0.67}_{-0.47}$ \\
Elias 20 & $565^{+16}_{-23}$ & $16.21^{+0.33}_{-0.13}$ & $0.48^{+0.10}_{-0.22}$ & $-7.08^{+0.17}_{-0.04}$ & $905^{+31}_{-51}$ & $16.31^{+0.31}_{-1.69}$ & $-0.30^{+1.67}_{-0.28}$ & $-7.78^{+0.03}_{-0.06}$ & $498^{+75}_{-74}$ & $16.45^{+1.37}_{-0.59}$ & $-0.33^{+0.43}_{-0.73}$ & $-7.44^{+1.14}_{-0.16}$ \\
WSB 52 & $670^{+47}_{-23}$ & $14.89^{+0.60}_{-0.38}$ & $1.69^{+0.31}_{-0.59}$ & $-7.20^{+0.03}_{-0.03}$ & $756^{+62}_{-69}$ & $16.66^{+0.65}_{-0.57}$ & $-0.64^{+0.47}_{-0.80}$ & $-7.77^{+0.32}_{-0.07}$ & $473^{+109}_{-30}$ & $16.50^{+0.16}_{-0.18}$ & $0.38^{+0.10}_{-0.38}$ & $-6.68^{+0.11}_{-0.30}$ \\
GK Tau & $459^{+123}_{-176}$ & $16.70^{+0.53}_{-0.71}$ & $-0.91^{+0.63}_{-0.32}$ & $-7.98^{+1.17}_{-0.36}$ & $1044^{+253}_{-338}$ & $13.58^{+1.30}_{-0.82}$ & $1.03^{+0.96}_{-0.96}$ & $-9.17^{+0.85}_{-0.14}$ & $465^{+150}_{-193}$ & $17.14^{+1.00}_{-0.49}$ & $-1.24^{+0.65}_{-0.51}$ & $-7.65^{+1.13}_{-0.43}$ \\
Sz 114 & $775^{+60}_{-52}$ & $17.08^{+0.61}_{-0.14}$ & $-1.24^{+0.10}_{-0.34}$ & $-7.93^{+0.13}_{-0.08}$ & $1104^{+294}_{-87}$ & $15.76^{+1.43}_{-0.90}$ & $-0.68^{+0.90}_{-1.32}$ & $-8.71^{+0.04}_{-0.03}$ & $490^{+24}_{-23}$ & $17.00^{+0.08}_{-0.12}$ & $-0.38^{+0.04}_{-0.04}$ & $-6.94^{+0.09}_{-0.09}$ \\
\enddata
\tablecomments{Temperatures (T) are in Kelvin (K). Column densities (logN) are in cm$^{-2}$, emitting areas (logA) are in au$^2$, and masses (logM) are in $M_\oplus$. Uncertainties are shown as subscripts and superscripts. Upper limits are indicated with ``$<$''.}
\end{deluxetable*}
\begin{figure*}[!htbp]
\centering
\vspace{-0.cm}
\includegraphics[width=\textwidth]{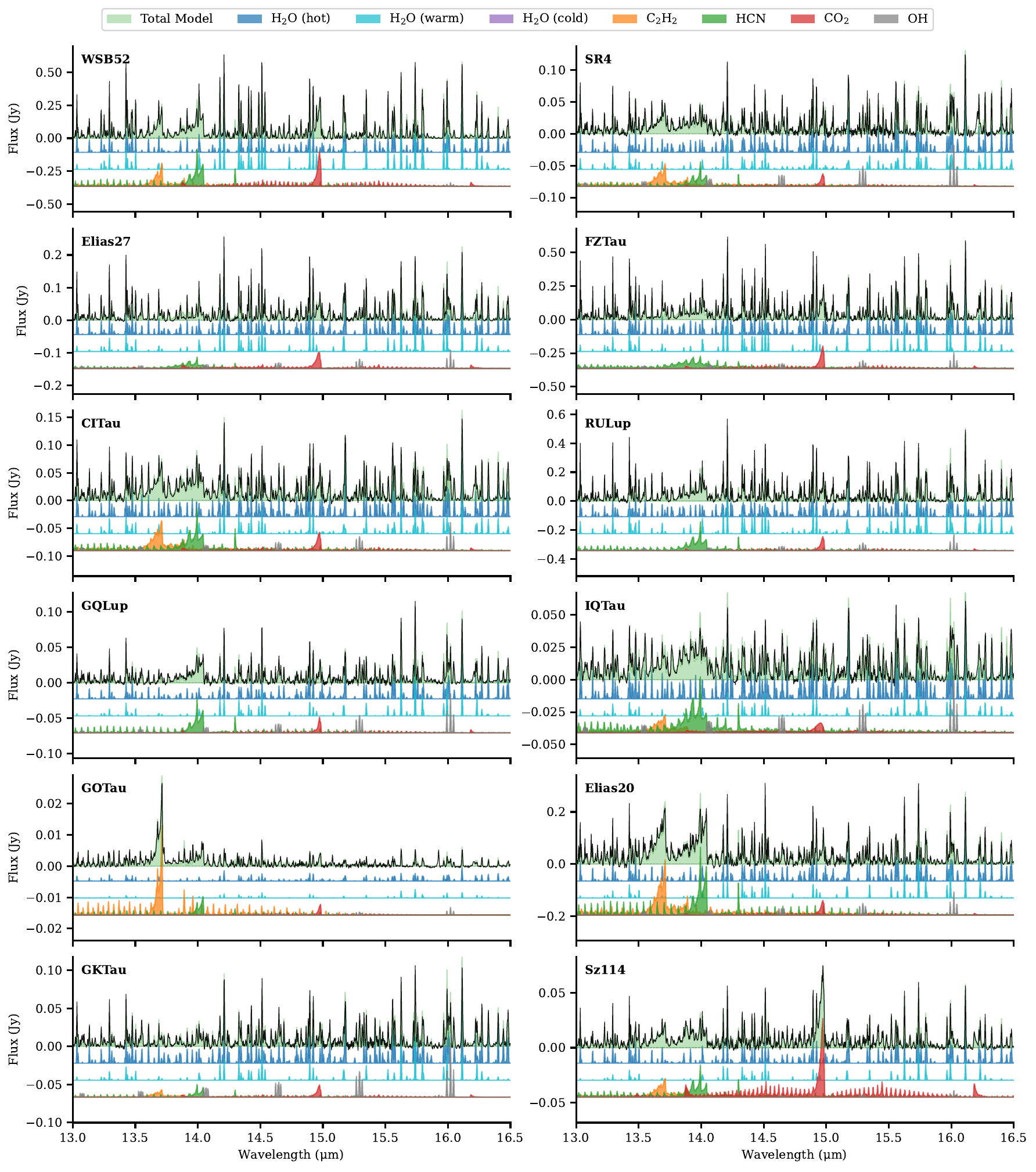}
\vspace{-.0cm}
\caption{Best-fitting slab models for the Class II disk sample over the 13--16.5\,$\mu$m wavelength range, covering hot water, HCN, C$_2$H$_2$, and CO$_2$ features. Format is the same as Figure~\ref{fig:slab_model_classI}.  \label{fig:ClassII_13um_slab}
}
\vspace{-0.cm}
\end{figure*}

\begin{figure*}[!htbp]
\centering
\vspace{-0.cm}
\includegraphics[width=\textwidth]{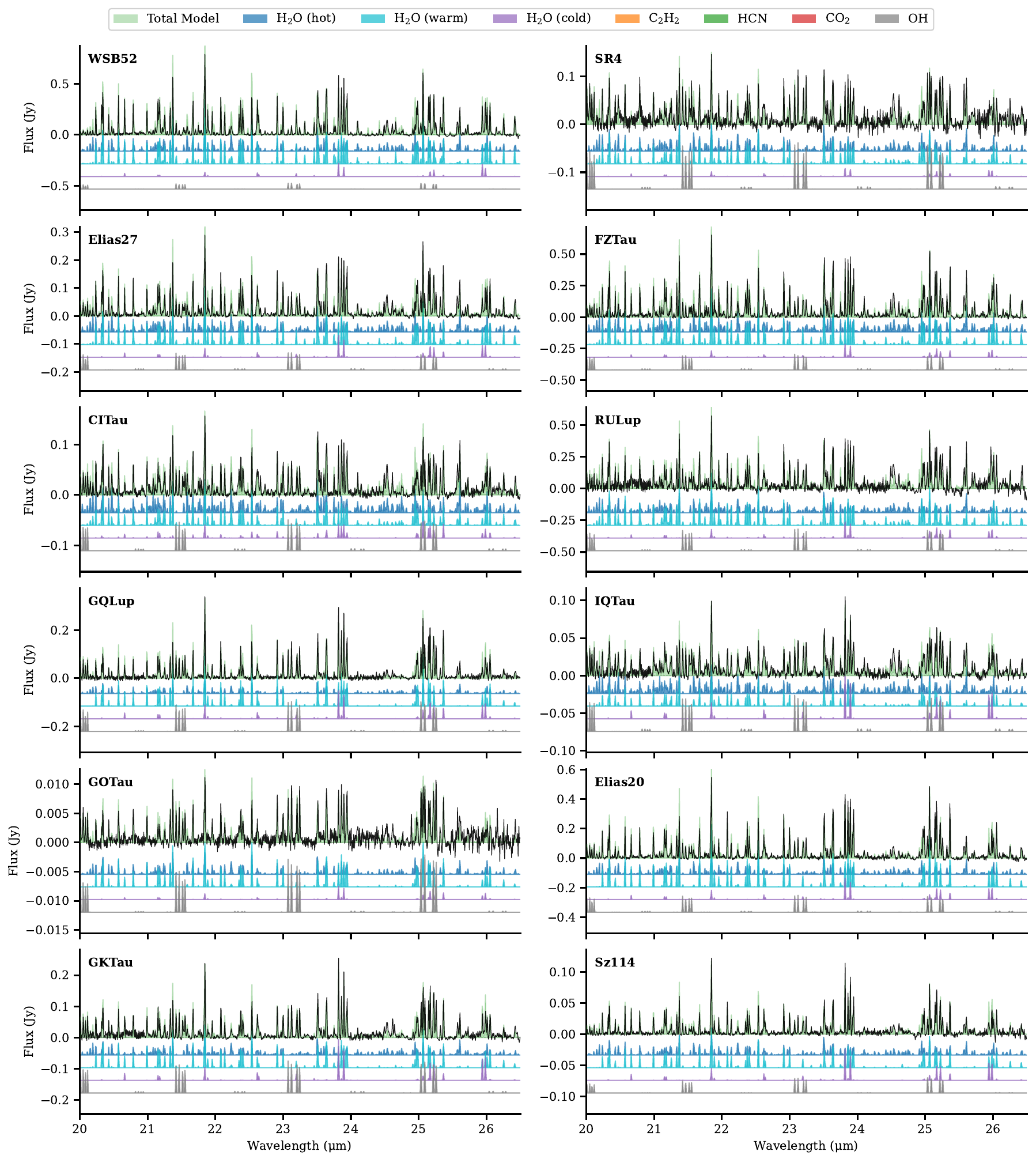}
\vspace{-.0cm}
\caption{Best-fitting slab models for the Class II disk sample over the 20--26.5\,$\mu$m wavelength range, covering warm water, cold water, and OH features. Format is the same as Figure~\ref{fig:slab_model_classI}.  \label{fig:ClassII_20um_slab}
}
\vspace{-0.cm}
\end{figure*}

\bibliography{ref_2026}{}
\bibliographystyle{aasjournalv7}

\end{document}